\documentclass[iop,apj,revtex4]{emulateapj}
\usepackage[]{natbib}
\bibpunct{(}{)}{;}{a}{}{,}
\usepackage{epstopdf}
\usepackage{url,hyperref}
\usepackage{amsmath,gensymb,latexsym,amssymb}

\newcommand\blankfootnote[1]{%
  \begingroup
  \renewcommand\thefootnote{}\footnote{#1}%
  \addtocounter{footnote}{-1}%
  \endgroup
}

\shorttitle{Excitation Conditions in J00266}
\shortauthors{Sharon et al.}
\submitted{ApJ Accepted 10/30/2014}
\begin{document}
\title{Excitation Conditions in the Multi-component Submillimeter Galaxy SMM\,J00266+1708}

\author{Chelsea E. Sharon\altaffilmark{1}, Andrew J. Baker}
\affil{Department of Physics and Astronomy, Rutgers, the State University of New Jersey, Piscataway, NJ, 08854-8019}
\altaffiltext{1}{Current affiliation: Center for Radiophysics and Space Research, Cornell University, Ithaca, NY, 14853}
\author{Andrew I. Harris}
\affil{Department of Astronomy, University of Maryland, College Park, MD 20742}
\author{Linda J. Tacconi, Dieter Lutz}
\affil{Max-Planck-Institut f\"{u}r extraterrestrische Physik (MPE), Giessenbachstr. 1, 85748 Garching, Germany}
\and
\author{Steven N. Longmore}
\affil{Astrophysics Research Institute, Liverpool John Moores University, Twelve Quays House, Egerton Warf, Birkenhead CH41 1LD, UK}

\begin{abstract} 
We present multiline CO observations of the complex submillimeter galaxy SMM\,J00266+1708. Using the Zpectrometer on the Green Bank Telescope, we provide the first precise spectroscopic measurement of its redshift ($z=2.742$). Based on followup \mbox{CO(1--0)}, \mbox{CO(3--2)}, and \mbox{CO(5--4)} mapping, SMM\,J00266+1708 appears to have two distinct components separated by $\sim 500\,{\rm km\,s^{-1}}$ that are nearly coincident along our line of sight. The two components show hints of different kinematics, with the blue-shifted component dispersion-dominated and the red-shifted component showing a clear velocity gradient. CO line ratios differ slightly between the two components, indicating that the physical conditions in their molecular gas may not be alike. We tentatively infer that SMM\,J00266+1708 is an ongoing merger with a mass ratio of $(7.8\pm4.0)/\sin^2(i)$, with its overall size and surface brightness closely resembling that of other merging systems. We perform large velocity gradient modeling of the CO emission from both components and find that each component's properties are consistent with a single phase of molecular gas (i.\/e.\/, a single temperatures and density); additional multi-phase modeling of the red-shifted component, although motivated by a \mbox{CO(1--0)} size larger than the \mbox{CO(3--2)} size, is inconclusive. SMM\,J00266+1708 provides evidence of early stage mergers within the submillimeter galaxy population. Continuum observations of J00266 at the $\sim1^{\prime\prime}$ resolution of our observations could not have distinguished between the two components due to their separation ($0.^{\prime\prime}73\pm0.^{\prime\prime}06$), illustrating that the additional velocity information provided by spectral line studies is important for addressing the prevalence of unresolved galaxy pairs in low-resolution submillimeter surveys.
\end{abstract}

\keywords{galaxies: high-redshift---galaxies: individual (SMM\,J00266+1708)---galaxies: ISM---galaxies: starburst---ISM: molecules}

\section{Introduction}

\blankfootnote{Based on observations carried out with the IRAM Plateau de Bure Interferometer. IRAM is supported by INSU/CNRS (France), MPG (Germany), and IGN (Spain).} 

Submillimeter galaxies (SMGs) have provided crucial insights to our understanding of the cosmic star formation history. First discovered as a population of dusty luminous galaxies in cluster and blank-field surveys \citep{smail1997, barger1998, hughes1998}, SMGs have been revealed to be substantial contributors to the far-infrared and submillimeter backgrounds \citep{puget1996, fixsen1998, devlin2009,zemcov2010,berta2010,berta2011,bethermin2012}. The faintness of SMGs' X-ray counterparts \citep{alexander2003,lindner2012} and the presence of polycyclic aromatic hydrocarbon (PAH) emission features atop otherwise weak mid-infrared continua \citep[e.\/g.\/,][]{lutz2005, valiante2007, pope2008,menendez2009} indicate that most of their luminosity is likely due to star formation rather than accretion in obscured active galactic nuclei (AGN; although there are exceptions, e.\/g.\/, \citealp{coppin2010}). Given that these properties are analogous to those of many local ultra/luminous infrared galaxies \citep[U/LIRGs;][]{sanders1996,tacconi2006,tacconi2008}, and that SMGs have large halo and/or baryonic masses \citep[e.\/g.\/,][]{genzel2003, blain2004, tecza2004, swinbank2008, ivison2010a, hainline2011}, SMGs have typically been characterized as the progenitors of present-day elliptical galaxies caught in their merger-driven starburst formation phase \citep[e.\/g.\/,][]{conselice2003, thomas2005, narayanan2010, engel2010}. However, some cosmological simulations suggest that large SFRs at high redshift are not necessarily caused by major mergers \citep[e.\/g.\/,][]{dave2010}, consistent with findings that SMGs include extended disks \citep[e.\/g.\/,][]{swinbank2011,hodge2012}, minor mergers \citep{aguirre2013}, and multiple galaxies caught within the same large beams of single-dish submillimeter surveys, with perhaps only a small fraction undergoing mergers \citep[e.\/g.\/,][]{hayward2011,hayward2013,hodge2013}.

Understanding how SMGs fit into the picture of early galaxy evolution and their relation to local U/LIRGs has included determining the physical conditions of the gas reservoirs that fuel their vigorous starbursts \citep[e.\/g.\/,][]{weiss2007, danielson2011}. One fruitful approach has been the application of radiative transfer models to observed CO spectral line energy distributions \citep[CO SLEDs; e.\/g.\/,][]{wild1992, guesten1993, mao2000, ward2003, danielson2011}. Early observations were restricted to mid-$J$ lines \citep[e.\/g.\/,][and references therein]{neri2003, greve2005, tacconi2006, tacconi2008}, leading to CO SLEDs that suggested single-phase molecular ISMs \citep[e.\/g.\/,][]{weiss2007}. However, with the increased frequency coverage provided by Ka band receivers at the Robert C. Byrd Green Bank Telescope (GBT) and Karl G. Jansky Very Large Array (VLA), observations of the \mbox{CO(1--0)} line, the best cold gas tracer, have revealed line fluxes in excess of the predictions for single-phase models (e.\/g.\/, \citealt{swinbank2010b, harris2010, ivison2010a, ivison2011, frayer2011}; see also \citealt{papadopoulos2001,carilli2002,greve2003,greve2004,klamer2005,hainline2006,riechers2006,dannerbauer2009,carilli2010,aravena2010a}). The detection of substantial cold gas reservoirs in a large fraction of the population suggests that SMGs possess multiple molecular phases with varying filling factors \citep{harris2010, ivison2011}, similar to those seen in other star-forming galaxies \citep[e.\/g.\/,][]{ward2003}.

Of particular use is the \mbox{CO(3--2)}/\mbox{CO(1--0)} line ratio ($r_{3,1}$, taken to be in brightness temperature units), which provides a rough probe of SMGs' molecular gas excitation as it traces both cold quiescent molecular gas and warm gas that is more actively involved in star formation. In these units, $r_{3,1}\sim1$ is measured in systems with a single phase of molecular gas (i.\/e.\/, a single fairly high temperature and density) while $r_{3,1}<1$ is measured for (a) multi-phase molecular gas, (b) subthermally excited gas, or (c) material at low temperature where there is a difference between the Planck and Rayleigh-Jeans temperatures. While a sub-unity $r_{3,1}$ is not \emph{strictly} indicative of multi-phase molecular gas, it is common for SMGs to have strong \mbox{CO(5--4)} and even higher-$J$ CO emission (including SMGs with observed $r_{3,1}<1$) which is difficult to reconcile with low $r_{3,1}$ originating from a single gas phase with either low temperature or subthermal CO excitation \citep[e.\/g.\/,][]{harris2010}. Observations of SMGs indicate that individual lines are still optically thick, with global averages clustering in a narrow range of $r_{3,1}\sim0.6$ (e.\/g.\/, \citealt{swinbank2010b, harris2010, ivison2011,danielson2011,thomson2012,bothwell2013}; cf. \citealt{sharon2013}). While a sub-unity $r_{3,1}$ is common for star-forming galaxies, the narrow range detected for $r_{3,1}$ is peculiar for two reasons: (1) U/LIRGs, the low-$z$ analogues of SMGs, seem to have a much broader range in $r_{3,1}$ \citep{yao2003, iono2009}, and (2) quasar host galaxies, plausible descendants of SMGs given quasars' potential ability to quench rapid starbursts \citep[e.\/g.\/,][]{granato2001,somerville2008}, seem to have \mbox{CO(3--2)/CO(1--0)} line ratios tightly clustered near an entirely different value, $r_{3,1}\sim1.0$ \citep{riechers2011f,thomson2012}, which is consistent with a single-phase molecular ISM. 

Since detailed studies of SMGs' molecular gas content have required precise redshifts, most of which have been determined using a combination of radio and optical counterparts \citep[e.\/g.\/,][]{chapman2005}, it may be that current results are biased towards lower-redshift and less-obscured members of the population even though \citet{harris2012} Zpectrometer observations of a {\it Herschel} sample have validated the \citet{chapman2005} redshift distribution \citep[cf.][]{weiss2013}. Therefore, it may be that the bulk of existing measurements of $r_{3,1}$ are biased towards a narrow range of U/LIRG-like states, such as intermediate-stage mergers where dust screens have begun to clear \citep[e.\/g.\/,][]{kormendy1992, hayward2011} and gas excitation is more settled \citep[e.\/g.\/,][and references therein]{narayanan2010}. For U/LIRGs per se, there is no observed correlation between $r_{3,1}$ and dust properties ($L_{\rm FIR}$, $T_{\rm dust}$, and $M_{\rm dust}$), molecular gas mass, star formation rate, or radio/infrared colors and luminosities \citep{mauersberger1999, yao2003}. In order to understand SMGs' tight range in $r_{3,1}$ and whether current samples are in fact biased, it is essential to find systems with a wider range of excitation conditions and investigate their potential evolutionary connections to other members of the population. 

In this paper, we report observations of SMM\,J00266+1708 (J00266 hereafter), whose multiple components have line ratios differing from the standard $r_{3,1}\sim0.6$. J00266 is the second brightest SMG in the SCUBA Lens Survey \citep[SLS;][]{smail2002} at $850\,{\rm \mu m}$ and has unusually high obscuration: it is the faintest of the detected SMGs in the \citet{frayer2004} near-IR followup of the SLS galaxies, with $K=22.36\pm0.16$ and $J>24.27$. Difficulties in counterpart identification led to an incorrect initial redshift \citep{frayer2000}. However, \citet{valiante2007} identified strong PAH emission features using the {\it Spitzer Space Telescope} and estimated the source's redshift to be $z_{\rm PAH}=2.73\pm0.02$. Based on this estimate, we observed J00266 with the Zpectrometer \citep{harris2007} on the GBT and successfully detected the \mbox{CO(1--0)} line at $z_{\rm CO}=2.742$. We then used this more precise redshift to make interferometric observations of \mbox{CO(1--0)} at the VLA, \mbox{CO(3--2)} and \mbox{CO(5--4)} at the IRAM Plateau de Bure Interferometer (PdBI), and \mbox{CO(7--6)} at the Submillimeter Array (SMA).

In Section\,\ref{sec:obs}, we describe the observations and data reduction for J00266. In Section\,\ref{sec:results}, we present our results, which are analyzed in Section\,\ref{sec:analysis}. In Section\,\ref{sec:discuss}, we discuss possible origins for the unusual properties of J00266 in the context of the broader population of SMGs. Our results and conclusions are summarized in Section\,\ref{sec:summary}. We assume the flat WMAP7+BAO+$H_0$ mean $\Lambda$CDM cosmology throughout this paper, with $\Omega_\Lambda=0.725$ and $H_0=70.2\,{\rm km\,s^{-1}\,Mpc^{-1}}$ \citep{komatsu2011}.

\section{Observations}
\label{sec:obs}

\subsection{Robert C. Byrd Green Bank Telescope}

J00266 was observed at the GBT on 2007 November 8, 9, 19, 25, 28, and 29, and on 2008 March 6 (project ID 05C-035), using the Zpectrometer \citep{harris2007}. The Zpectrometer is a fixed-bandwidth instrument covering a $25.6-36\,{\rm GHz}$ range within the Ka band that allows for blind detections of the \mbox{CO(1--0)} line for sources with $2.2<z<3.5$, and higher-$J$ CO lines for correspondingly higher redshifts. Observations were carried out with the standard combination of position switching and sub-reflector nodding described in \citet{harris2010}, using SMM\,J00265+1710 \citep{smail2002} as a position-switching ``partner." We observed 3C48 as a flux calibrator, adopting a flux density of $S_\nu=0.80\,{\rm Jy}$ at $32\,{\rm GHz}$ \citep{astroalmanac}. We assume a flux calibration uncertainty of 20\%. The spectrum was processed using a custom set of GBTIDL and R scripts.

\subsection{Plateau de Bure Interferometer}

\mbox{CO(3--2)} and \mbox{CO(5--4)} emission from J00266 were observed at the PdBI \citep{guilloteau1992}. The \mbox{CO(3--2)} line was observed on 2009 October 24 (project ID T03C) with the array in a five-antenna version of the C configuration (maximum baseline 144.5\,m) and on 2011 January 20 and 31 (project ID U0BB) with the full array in the A configuration (maximum baseline $760\,{\rm m}$), all in good to excellent weather conditions (precipitable water vapor between $0.5$--$10\,{\rm mm}$). The \mbox{CO(5--4)} line was observed on 2010 May 15 (project ID T{-}{-}4), in excellent weather conditions  ($\sim3.25\,{\rm mm}$ precipitable water vapor), but with the full array in the D configuration (maximum baseline 97.0\,m). For all observations we used the available narrow-band correlator mode with $5\,{\rm MHz}$ channels and a total bandwidth of $1\,{\rm GHz}$, obtaining both horizontal and vertical polarizations. For the \mbox{CO(3--2)} line, the correlator was tuned to $92.707\,{\rm GHz}$ for the 2009 observations and to $92.505\,{\rm GHz}$ for the 2011 observations (the change in tuning was informed by the line profile of the earlier observations). For the \mbox{CO(5--4)} observations, the correlator was tuned to $154.104\,{\rm GHz}$. The \mbox{CO(3--2)} and \mbox{CO(5--4)} lines of J00266 were observed for $10.5\,{\rm hours}$ in the $3\,{\rm mm}$ band and $7.7\,{\rm hours}$ in the $2\,{\rm mm}$ band, respectively. 

Bandpass and flux calibration were obtained by observing MWC\,349 or 3C454.3 near the beginning of each session. The model flux density for MWC349 in the 2009 \mbox{CO(3--2)} observations is $1.14\,{\rm Jy}$ at $92.7\,{\rm GHz}$. The adopted flux densities for 3C454.3 is $42.2\,{\rm Jy}$ and $37.6\,{\rm Jy}$ at $92.5\,{\rm GHz}$ for the \mbox{CO(3--2)} observations on 2011 January 20 and 31, respectively. We assume a 10\% flux calibration uncertainty for our \mbox{CO(3--2)} measurements. The model flux density for MWC349 is $1.55\,{\rm Jy}$ at $154.1\,{\rm GHz}$ for our 2010 \mbox{CO(5--4)} observations, for which we assume a 15\% flux calibration uncertainty. Phase and amplitude variation were tracked by alternating observations of the target and nearby bright quasars. Calibration and flagging for data quality were carried out using the CLIC program \citep[part of the IRAM GILDAS package;][]{guilloteau2000}. The resulting $uv$ datasets were smoothed spectrally by factors of three and four, resulting in $48.66\,{\rm km\,s^{-1}}$ and $38.93\,{\rm km\,s^{-1}}$ resolution, for the \mbox{CO(3--2)} and \mbox{CO(5--4)} observations, respectively. The GILDAS package was also used to produce the naturally weighted maps, with synthesized beams of $1^{\prime\prime}.36\times0^{\prime\prime}.81$ at position angle $31.07^{\degree}$ for the \mbox{CO(3--2)} observations and $3^{\prime\prime}.70\times3^{\prime\prime}.04$ at position angle $120.24^{\degree}$ for the \mbox{CO(5--4)} observations. The average RMS noise is $0.323\,{\rm mJy\,beam^{-1}}$ per $15\,{\rm MHz}$ channel for the \mbox{CO(3--2)} data cube and $0.715\,{\rm mJy\,beam^{-1}}$ per $40\,{\rm MHz}$ channel for the \mbox{CO(5--4)} data cube. Analysis of the resulting data cubes used a custom set of IDL scripts.

\subsection{Submillimeter Array}

We observed the \mbox{CO(7--6)} line in J00266 at the SMA on 2010 October 3 and 21 (proposal ID 2010A-S034), in good to excellent weather conditions ($225\,{\rm GHz}$ opacities between $0.08$--$0.3$). The tracks were both taken with the array in its compact configuration (maximum baseline $78\,{\rm m}$) using seven antennas, and together resulted in a total of 14.1 hours on source. The correlator was in the standard mode delivering the maximum bandwidth ($4\,{\rm GHz}$ per sideband) for single-receiver observations, giving a channel width of $812.5\,{\rm kHz}$. The correlator was tuned to $215.567\,{\rm GHz}$ at ``chunk s14" so that the \mbox{CO(7--6)} line would fall in the upper sideband, well away from the sideband's edges and the sideband's $32\,{\rm MHz}$ coverage gap between the two intermediate frequency (IF) bands. The C\,{\sc i} $^3{\rm P}_2\rightarrow {^3}{\rm P}_1$ fine structure line also falls within the upper sideband, but overlaps with the $32\,{\rm MHz}$ coverage gap.

Flux calibration used Uranus (October 21) and Neptune (October 3). Our adopted flux densities for Uranus ($33.4\,{\rm Jy}$ at $216.4\,{\rm GHz}$) and Neptune ($12.4\,{\rm Jy}$ at $204.2\,{\rm GHz}$) are determined from the Common Astronomy Software Application (CASA)\footnote{\href{http://casa.nrao.edu}{http://casa.nrao.edu}} package's ``Butler-JPL-Horizons 2010" standards\footnote{\href{https://science.nrao.edu/facilities/alma/aboutALMA/Technology/ALMA_Memo_Series/alma594/abs594}{https://science.nrao.edu/facilities/alma/aboutALMA/Technology/ALMA\_Memo\_Series/alma594/abs594}}. We assume a 20\% flux calibration uncertainty. Phase and amplitude variation were tracked by alternating observations of the target and several nearby bright quasars, which were also used for bandpass calibration. The initial system temperature correction was carried out using the MIR package\footnote{\href{https://www.cfa.harvard.edu/~cqi/mircook.html}{https://www.cfa.harvard.edu/$\sim$cqi/mircook.html}}, but all other calibration and mapping were done in CASA. The calibrated data were re-binned into $100\,{\rm km\,s^{-1}}$ channels, and the resulting naturally weighted map has a synthesized beam of $3^{\prime\prime}.44\times 3^{\prime\prime}.03$ at a position angle of $73.23^{\degree}$ and an average RMS noise of $3.8\,{\rm mJy\,beam^{-1}}$ in each $100\,{\rm km\,s^{-1}}$ channel. Analysis of the resulting data cube used the same custom set of IDL scripts as the PdBI observations.

\subsection{Karl G. Jansky Very Large Array}

To complement our GBT detection, the CO(1--0) line was observed at the VLA (proposal IDs 10B-230, 12A-331, and 13B-051). A single track was taken on 2010 November 6 with the array in the C configuration (using 19 antennas and a maximum baseline of $3.4\,{\rm km}$), and six additional tracks were taken in the D (2013 February 16 and March 29, 30, 31) and DnC (2013 May 12, 13) configurations (using 26 antennas and maximum baselines of $1.5\,{\rm km}$ and $2.1\,{\rm km}$, respectively). Zenith opacities were between $0.023$--$0.030\,{\rm nepers}$ for all observations. In 2010 we observed with the Wideband Interferometric Digital Architecture (WIDAR) correlator tuned to $30.8047\,{\rm GHz}$ in ``Open Shared Risk Observing 2 (OSRO2) 1 Subband/Dual polarization" mode, using the lowest available spectral resolution (256 channels at $500\,{\rm kHz}$ resolution). In 2013 we observed with the more-flexible WIDAR correlator using the lowest available spectral resolution for two polarization products (128 channels at $1\,{\rm kHz}$ resolution) and only three subbands (out of eight). The $384\,{\rm kHz}$ bandwidth used in the 8-bit B0/D0 baseband was centered on the (observed frame) line frequency of $30.8047\,{\rm GHz}$. A total of 10.8 hours was spent on source.

Flux calibration used the quasar 3C48. The adopted flux density of 3C48 is $0.93\,{\rm Jy}$ at $30.89\,{\rm GHz}$ assuming the CASA default ``Perley-Butler 2010" calibration standard \citep{perley2013}. We assume a 10\% flux calibration uncertainty. Phase and amplitude calibration relied on interleaved observations of the nearby quasar J0056+1625. Calibration and mapping were carried out in CASA. The integrated line map has a synthesized beam of $1^{\prime\prime}.08\times 1^{\prime\prime}.02$ at position angle $-65.16^{\degree}$ and RMS noise of $70.7\,{\rm \mu Jy\,beam^{-1}}$ per $50\,{\rm km\,s^{-1}}$ channel. Analysis of the resulting data cube used the same custom set of IDL scripts as the other observations. For the analysis of the spectral line profile, we excluded the higher-resolution and lower-S/N C configuration data. The synthesized beam of the combined D and DnC configuration data is $2^{\prime\prime}.59\times 2^{\prime\prime}.02$ at position angle $-76.50^{\degree}$; the RMS noise is $62.8\,{\rm \mu Jy\,beam^{-1}}$ per $50\,{\rm km\,s^{-1}}$ channel.

\section{Results}
\label{sec:results}

\begin{figure*}
\epsscale{.8}
\plotone{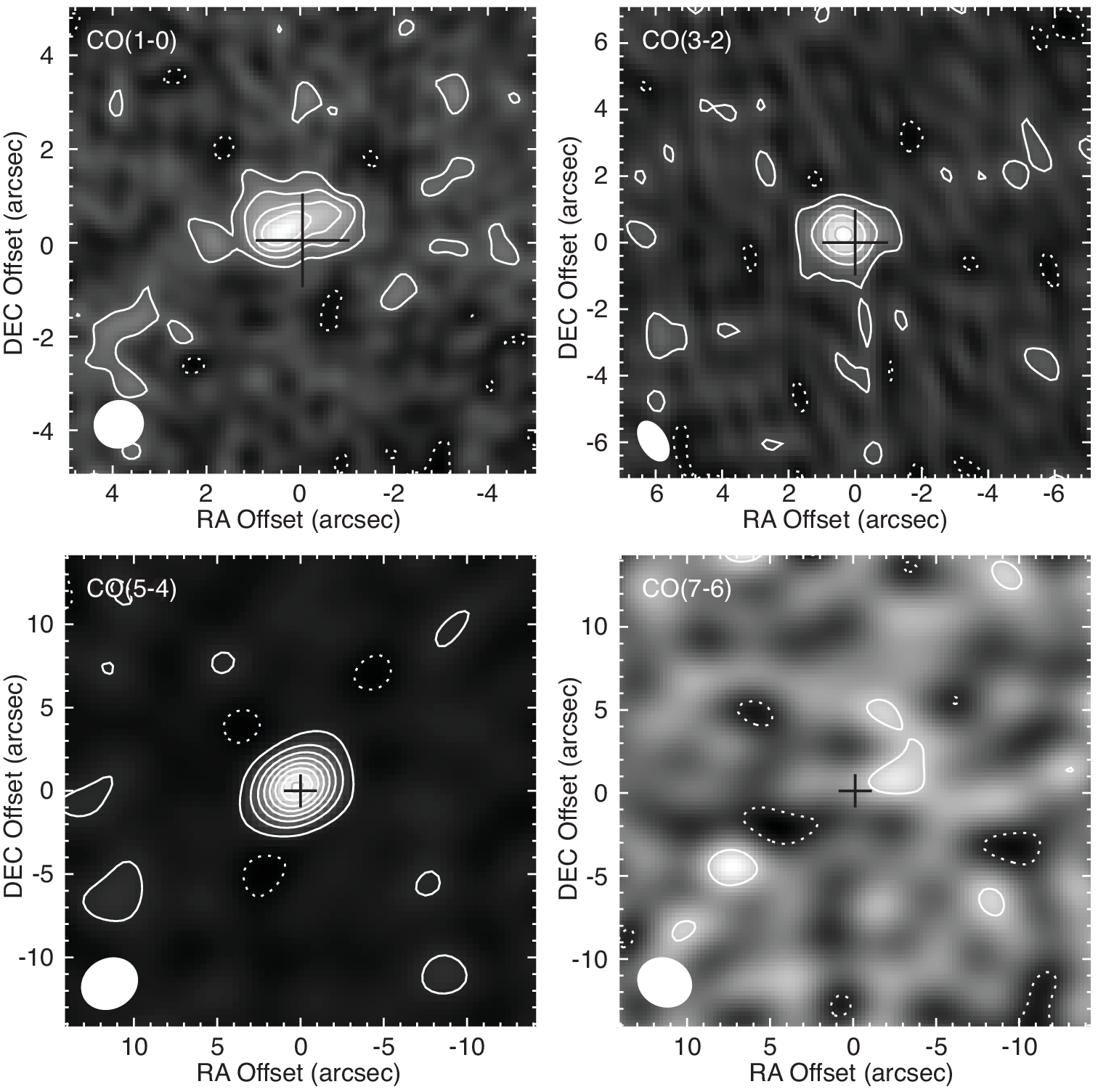}
\caption{\mbox{CO(1--0)} (top left), \mbox{CO(3--2)} (top right), \mbox{CO(5--4)} (bottom left), and \mbox{CO(7--6)} (bottom right) integrated line maps of J00266 centered on the \citet{frayer2000} continuum detection ($\alpha {\rm (J2000)} = 00^{\rm h}26^{\rm m}34.^{\rm s}10$ and $\delta {\rm (J2000)} = +17\degree08^{\prime}33.^{\prime\prime}7$). In order of increasing $J$, the contours are multiples of $\pm2\sigma$ ($\sigma=0.30\,{\rm mJy\, beam^{-1}}$), $\pm4\sigma$ (but starting at $\pm2\sigma$; $\sigma=1.6\,{\rm mJy\,beam^{-1}}$), $\pm4\sigma$ (but starting at $\pm2\sigma$; $\sigma=3.8\,{\rm mJy\,beam}^{-1}$), and $\pm2\sigma$ ($\sigma=1.5\,{\rm mJy\,beam}^{-1}$), where negative contours are dotted. Synthesized beams are shown in lower left corners. The center cross indicates the position and astrometric uncertainty of \citet{frayer2000} scaled up by a factor of two for visual clarity. Continuum subtraction has been performed for the \mbox{CO(7--6)} map; no other line has a significant continuum contribution. \label{fig:intmap}}
\end{figure*}

We have successfully detected the \mbox{CO(1--0)} line with both the GBT/Zpectrometer and the VLA, as well as the \mbox{CO(3--2)} and \mbox{CO(5--4)} lines at the PdBI. J00266 was undetected in \mbox{CO(7--6)} at the SMA. The position-switching partner for the GBT/Zpectrometer observations, SMM\,J00265+1710, was not detected in \mbox{CO(1--0)}. The integrated line maps for the interferometric observations of J00266 are shown in Figure~\ref{fig:intmap}, plotted relative to the position of the \citet{frayer2000} $1\,{\rm mm}$ continuum detection (J2000 \mbox{$00^{\rm h}\,26^{\rm m}\,34.10^{\rm s}\pm0.04^{\rm s}$} \mbox{$+17^{\degree}\,08^{\prime}\, 33.^{\prime\prime}7\pm0.^{\prime\prime}5$}). To determine a \mbox{CO(7--6)} upper limit, we integrate over the $-625\rightarrow275\,{\rm km\,s}^{-1}$ velocity range relative to $z_{\rm red}$ (see below) that best matches the FWZI of the lower-$J$ lines. For the VLA \mbox{CO(1--0)} and PdBI \mbox{CO(3--2)} maps, the line emission is spatially resolved. An elliptical Gaussian fit to the \mbox{CO(3--2)} $uv$ data gives a major axis of $1.^{\prime\prime}40\pm0.^{\prime\prime}11$ at position angle $-59\pm9$ degrees and a minor axis of $0.^{\prime\prime}91\pm0.^{\prime\prime}11$. Fits to the \mbox{CO(1--0)} $uv$ data do not converge, but image-plane fits give a major axis of $1.^{\prime\prime}97\pm0.^{\prime\prime}13$ at position angle $95\pm3$ degrees and a minor axis of $0.^{\prime\prime}62\pm0.^{\prime\prime}27$ when accounting for the beam size. The \mbox{CO(1--0)} emission is spatially extended relative to the \mbox{CO(3--2)} emission along its major axis. The larger \mbox{CO(1--0)} size is consistent with the relative sizes of regions producing \mbox{CO(3--2)} and \mbox{CO(1--0)} emission in other SMGs, and is evidence for multiple phases of molecular gas with different larger-scale distributions \citep{ivison2011}. The \mbox{CO(3--2)} data provide the strongest constraints on the source position; the improved position of J00266 is \mbox{$00^{\rm h}\,26^{\rm m}\,34.076^{\rm s}\pm0.005^{\rm s}$} \mbox{$+17^{\degree}\,08^{\prime}\, 33.^{\prime\prime}88\pm0.^{\prime\prime}07$}

\begin{figure}
\plotone{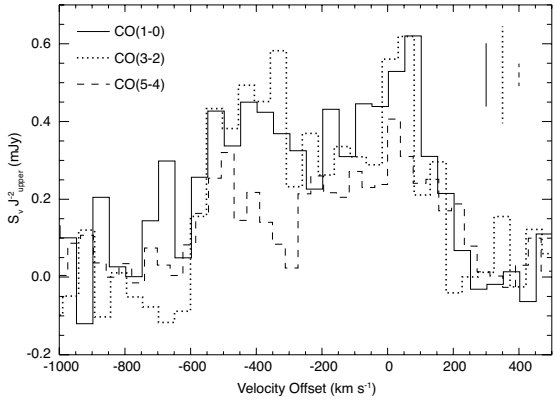}
\caption{The (VLA) \mbox{CO(1--0)} (solid), \mbox{CO(3--2)} (dotted), and \mbox{CO(5--4)} (dashed) spectral lines, plotted relative to the average velocity centroid of the red component at $z_{\rm red}=2.7420$ (divided by a factor of $J_{\rm upper}^2$ for clarity). The vertical lines in the top right corner show the average $1\sigma$ uncertainty for each channel. A second peak, blue-shifted from the \mbox{CO(1--0)} centroid by $\sim500\,{\rm km\,s^{-1}}$, is clearly visible in the mid-$J$ spectra.
\label{fig:spectra}}
\end{figure}

The VLA and PdBI profiles for J00266 (Figure~\ref{fig:spectra}) show clear double-peaked structures in all lines. We use the average redshift from the Gaussian fits to the stronger redder peak as the systemic velocity throughout this work (motivated by our interpretation that J00266 is composed of two merging galaxies; see Section~\ref{sec:structure}), obtaining $z_{\rm red}=2.7420\pm0.0002$; the parameters for the double-Gaussian line fits are given in Table\,\ref{tab:obssum}. The \mbox{CO(1--0)} observations made using the VLA and GBT/Zpectrometer have consistent total integrated line strengths, indicating that there is no spatially extended flux missing from the VLA data. While the GBT/Zpectrometer line profile does not yield well-determined uncertainties, the large FWHM and the velocity offset of the line centroid are consistent with detection of emission from both the red and blue components (two-component fits do not converge on either best-fit values or their uncertainties). The total \mbox{CO(1--0)} line flux obtained at the VLA is $0.31\pm0.03(\pm0.03)\,{\rm Jy\,km\,s^{-1}}$, where the uncertainty associated with the flux calibration is given in parentheses. The integrated line fluxes of the \mbox{CO(3--2)} and \mbox{CO(5--4)} lines are $2.62\pm0.25(\pm0.26)\,{\rm Jy\,km\,s^{-1}}$ and $4.59\pm0.27(\pm0.69)\,{\rm Jy\,km\,s^{-1}}$, respectively. The $3\sigma$ upper limit on the \mbox{CO(7--6)} line is $3.93\,{\rm Jy\, km\, s}^{-1}$ for a point-like source and $1000\,{\rm km\,s^{-1}}$ line width.

\begin{deluxetable*}{ccccccc}
\tablewidth{0pt}
\tablecaption{ SMM\,J00266+1708 ($z_{\rm red}=2.7420\pm0.0002$, $\mu=2.41\pm0.49$) \label{tab:obssum}}
\tablehead{ {Transition} & {Peak} & {$S\Delta\nu$} & {$L^\prime_{\rm line}\tablenotemark{a}$} & {Peak $S_\nu$\tablenotemark{b}} & {FWHM\tablenotemark{b}} & {Offset\tablenotemark{b,c}} \\
{(Telescope)} & {} & {(${\rm Jy\,km\,s^{-1}}$)} & {($10^{10}\,{\rm K\,km\,s^{-1}\,pc^2}$)} & {(mJy)} & {(${\rm km\,s^{-1}}$)} & {(${\rm km\,s}^{-1}$)} }
\startdata
{CO(1--0)}\tablenotemark{d} & {Total} & $0.33$ & $4.86$ & $0.51$ & $609$ & -204 \\
{(GBT/Zpec)} & {} & {} & {} & {} & {} & {} \\
\tableline
{CO(1--0)} & {Total} & $0.31\pm0.04$ & $4.51\pm1.12$ & ... & ... & ... \\
{(VLA)} & {Blue} & $0.11\pm0.02$ & $1.67\pm0.47$ & $0.43\pm0.09$ & $373\pm121$ & $-396\pm48$ \\
{} & {Red} & $0.19\pm0.03$ & $2.84\pm0.74$ & $0.54\pm0.10$ & $265\pm80$ & $22\pm35$ \\
\tableline
CO(3--2) & Total & $2.62\pm0.36$ & $4.19\pm1.03$ & ... & ... & ... \\
{(PdBI)} & Blue & $1.09\pm0.29$ & $1.75\pm0.58$ & $4.76\pm0.76$ & $288\pm70$ & $-381\pm27$ \\
{} & Red & $1.53\pm0.38$ & $2.44\pm0.78$ & $4.51\pm0.87$ & $241\pm66$ & $13\pm27$ \\
\tableline
CO(5--4) & Total & $4.59\pm0.74$ & $2.64\pm0.69$ & ... & ... & ... \\
{(PdBI)} & Blue & $1.37\pm0.27$ & $0.79\pm0.20$ & $6.64\pm0.98$ & $164\pm31$ & $-497\pm13$ \\
{} & Red & $3.20\pm0.52$ & $1.84\pm0.48$ & $7.59\pm0.67$ & $442\pm47$ & $-22\pm18$ \\
\tableline
CO(7--6)\tablenotemark{e} & Total & $<3.93$ & $<1.18$ & ... & ... & ... \\
{(SMA)} & Blue & $<2.61$ & $<0.79$ & ... & ... & ... \\
{} & Red & $<2.92$ & $<0.88$ & ... & ... & ... \\
\tableline
C\,{\sc i} $^3{\rm P}_2\rightarrow {^3}{\rm P}_1$ & Total & $4.24\pm2.47$ & $0.65\pm0.40$ & ... & ... & ... \\
{(SMA)} & Blue\tablenotemark{e} & $<2.37$ & $<0.36$ & ... & ... & ... \\
{} & Red & $2.44\pm1.85$ & $0.37\pm0.29$ & ... & ... & ...
\enddata
\tablenotetext{a}{Where line refers to either CO or C\,{\sc i}; magnification corrected.}
\tablenotetext{b}{From Gaussian fits to the line profile.}
\tablenotetext{c}{Relative to the average velocity centroid of the red component from the VLA and PdBI observations ($z_{\rm red}$).}
\tablenotetext{d}{Zpectrometer analysis did not converge on line uncertainties.}
\tablenotetext{e}{$3\sigma$ upper limits.}
\end{deluxetable*}

\begin{figure}
\plotone{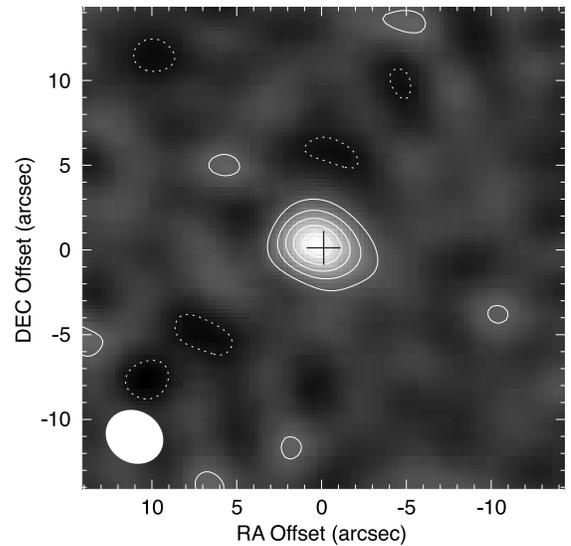}
\caption{The $1\,{\rm mm}$ continuum map centered on the \citet{frayer2000} continuum detection; $S_{1\,{\rm mm}}=5.34\pm 1.44\,{\rm mJy}$. Contours are multiples of $\pm2\sigma$ ($\sigma=0.41\,{\rm mJy\,beam}^{-1}$), where negative contours are dotted. The synthesized beam is shown at lower left. The center cross is as in Figure~\ref{fig:intmap}. \label{fig:continuum}}
\end{figure}

We do not detect significant continuum emission in the $1\,{\rm cm}$, $3\,{\rm mm}$, or $2\,{\rm mm}$ bands (with $3\sigma$ upper limits of $0.18\,{\rm mJy}$, $0.16\,{\rm mJy}$, and $0.45\,{\rm mJy}$, respectively, for a point-like source). The observational method of position switching used with the Zpectrometer does not allow for an unambiguous continuum detection; a difference spectrum whose continuum is consistent with zero implies only that J00266 has a continuum flux density nearly equal to that of SMM\,J00265+1710. We do, however, detect continuum emission in the $1\,{\rm mm}$ SMA observations (Figure~\ref{fig:continuum}), with $S_{1\,{\rm mm}}=5.34\pm0.97(\pm1.07)\,{\rm mJy}$. Our $1.4\,{\rm mm}$ continuum detection is consistent with $S_{\rm 1.3\,mm}=6.0\pm1.1\,{\rm mJy}$ from \citet{frayer2000}. In addition, the continuum fluxes measured in the upper and lower sidebands (separated by $12\,{\rm GHz}$ in their band centers) are consistent within their $1\sigma$ statistical uncertainties. Comparing our measured flux to the $850\,{\rm \mu m}$ SCUBA detection \citep{smail2002} yields a spectral index of $2.50\pm0.54$, consistent with the Rayleigh-Jeans tail of the dust spectral energy distribution. The position of our continuum detection is consistent with those of the CO line detections and the \citet{frayer2000} $1.3\,{\rm mm}$ detection.

\begin{figure*}
\epsscale{1.0}
\plotone{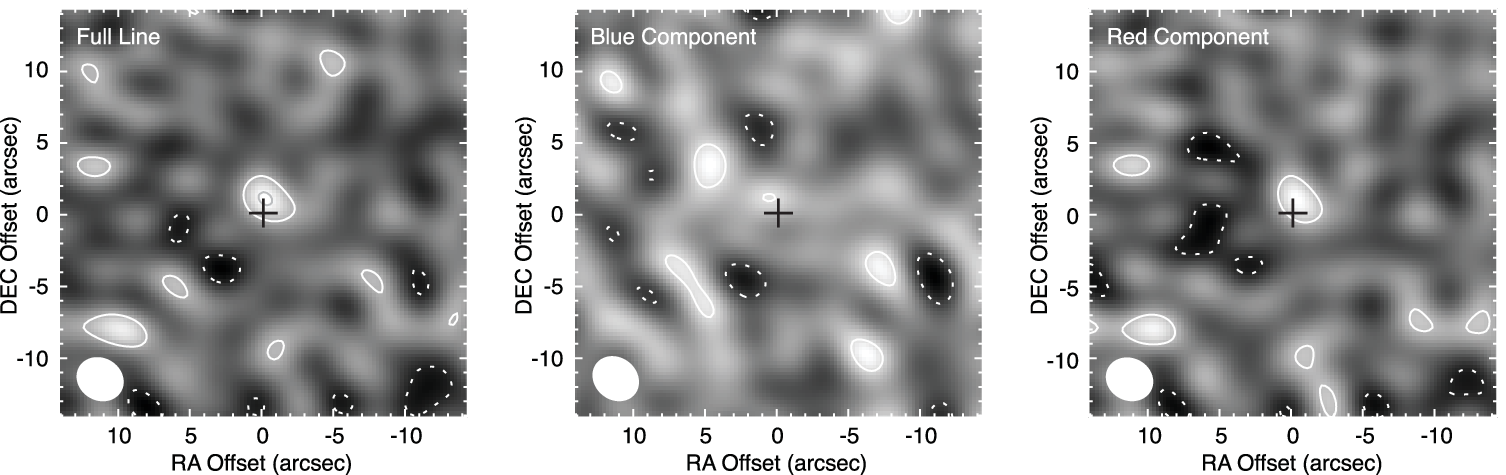}
\caption{Tentative detection of the C\,{\sc i} line for the full line (left), blue component (middle), and red component (right). Contours are multiples of $\pm2\sigma$ (the negative contours are dotted) where $\sigma=1.45\,{\rm mJy\,beam}^{-1}$, $2.25\,{\rm mJy\,beam}^{-1}$, and $1.81\,{\rm mJy\,beam}^{-1}$ from left to right. The synthesized beam is shown at lower left. The center cross is as in Figure~\ref{fig:intmap}. \label{fig:carbonmaps}}
\end{figure*}

While we do not detect any \mbox{CO(7--6)} emission from J00266, within the observed spectral bandpass we have marginally detected the C\,{\sc i} $^3{\rm P}_2\rightarrow {^3}{\rm P}_1$ fine structure line at $\gtrsim4\sigma$ at the observed peak (Figure~\ref{fig:carbonmaps}). The total line flux is $4.24\pm2.32(\pm0.85)\,{\rm Jy\, km\, s^{-1}}$. While the position of the C\,{\sc i} emission is consistent with those of our CO and $1\,{\rm mm}$ continuum detections, it is only marginally consistent with that of the \citet{frayer2000} continuum detection (although see Section~\ref{sec:structure}).

\section{Analysis}
\label{sec:analysis}

\subsection{Physical and dynamical structure}
\label{sec:structure}

\begin{figure}
\plotone{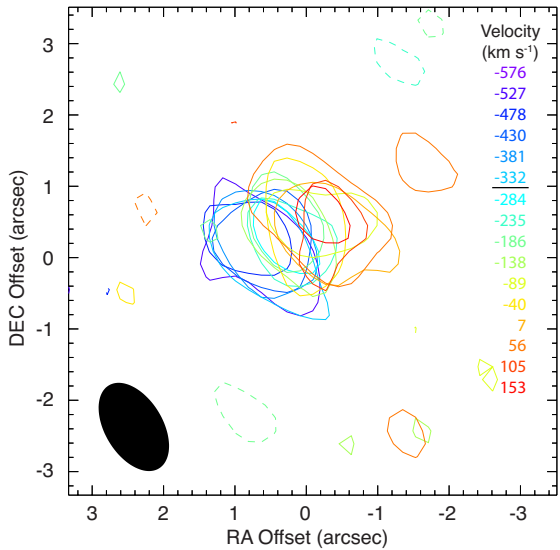}
\caption{Overlaid contours of the CO(3--2) channel maps, colorized by the channels' relative velocities; the black horizontal line marks the division between the two components. The synthesized beam is shown at lower left. Only the positive (solid) and negative (dashed) $3\sigma$ contours are shown ($1\sigma=3.6\,{\rm mJy\,beam}^{-1}$). \label{fig:renzogram}}
\end{figure}

The two peaks in the mid-$J$ CO spectra of J00266 suggest that the system contains either a rotating structure or two separate physical components. Several lines of evidence lead us to favor the latter scenario, with a division between the two components occurring at $\sim-300\,{\rm km\,s}^{-1}$ with respect to $z_{\rm red}=2.7420$. First, the higher S/N \mbox{CO(5--4)} spectrum shows a strong division and asymmetry between the two peaks, which is somewhat unlikely for a rotating disk. Second, we see different kinematics in the two components. Figure~\ref{fig:renzogram} shows the $\pm3\sigma$ contours of the \mbox{CO(3--2)} channel maps; the $-527\rightarrow-332\,{\rm km\,s}^{-1}$ channels of the blue peak are spatially coincident with each other and clearly separate from the redder channels, which show a notable velocity gradient. Examination of the peak flux positions in the individual \mbox{CO(3--2)} channel maps also reveals a clear progression in velocity along a position angle $\sim-70$ degrees for the channels redder than $-300\,{\rm km\,s}^{-1}$, while the peak flux positions in the remaining blue channels are scattered. Third, rescaling the \mbox{CO(1--0)} or \mbox{CO(3--2)} line profiles so that the peak flux of the red velocity component matches that of the \mbox{CO(5--4)} line does {\it not} result in similarly matching line strengths in the blue velocity component. The different mid-$J$ line ratios between the two peaks are indicative of different excitation conditions, which we investigate in Section~\ref{sec:excitation}. If the two CO peaks were produced by the same structure, the close physical association makes it unlikely that they would have very different excitation conditions (although it is possible for the components of late stage mergers to have similar excitation; e.\/g.\/, \citealt{weiss2005c}). We interpret the differing kinematics and CO line ratios are evidence that J00266 may be a merger between two galaxies with a relative velocity offset of $\sim500\,{\rm km\,s}^{-1}$ and close alignment along our line of sight.

Although the choice of division between the two components is motivated by both the \mbox{CO(5--4)} line shape and the centroid positions of the \mbox{CO(3--2)} channel maps, the merger interpretation could be weakened by alternative choices in the division between the two components. However, we also note that J00266 is an extremely compact and high surface brightness galaxy (see discussion below), which is expected in mergers where the molecular gas has been funneled into a central gas disk and has been observed in both local U/LIRGs \citep[e.\/g.\/,][]{downes1998,scoville2000,genzel2001} and other compact SMGs \citep[e.\/g.\/,][]{tacconi2006,tacconi2008,engel2010}. We therefore proceed under the assumption that J00266 is a merger between the two components defined below, but we also present global integrated measurements of J00266 for comparison.

For the rest of the analysis, we therefore consider the blue component (defined by the $-598\rightarrow-298\,{\rm km\,s}^{-1}$ velocity range for our \mbox{CO(1--0)} data, the $-600\rightarrow-308\,{\rm km\,s}^{-1}$ velocity range for our \mbox{CO(3--2)} data, and the $-624\rightarrow-312\,{\rm km\,s}^{-1}$ velocity range for our \mbox{CO(5--4)} data) and the red component (defined by the $-298\rightarrow202\,{\rm km\,s}^{-1}$ velocity range for our \mbox{CO(1--0)} data, the $-308\rightarrow178\,{\rm km\,s}^{-1}$ velocity range for our \mbox{CO(3--2)} data, and the $-273\rightarrow273\,{\rm km\,s}^{-1}$ velocity range for our \mbox{CO(5--4)} data) separately. We note that the double-Gaussian line profile fit given in Table\,\ref{tab:obssum} is not well-matched to our choice of velocity division for the two components. However, fixing the velocity offset of the blue component of the \mbox{CO(3--2)} line to match the \mbox{CO(5--4)} value results in a much closer match between the line FWHMs for (and therefore velocity division between) the two components, with only a slight change in the reduced-$\chi^2$ value of the two fits (0.66 without the constraint vs.~0.79 with the constraint).

\begin{figure*}
\epsscale{.65}
\plotone{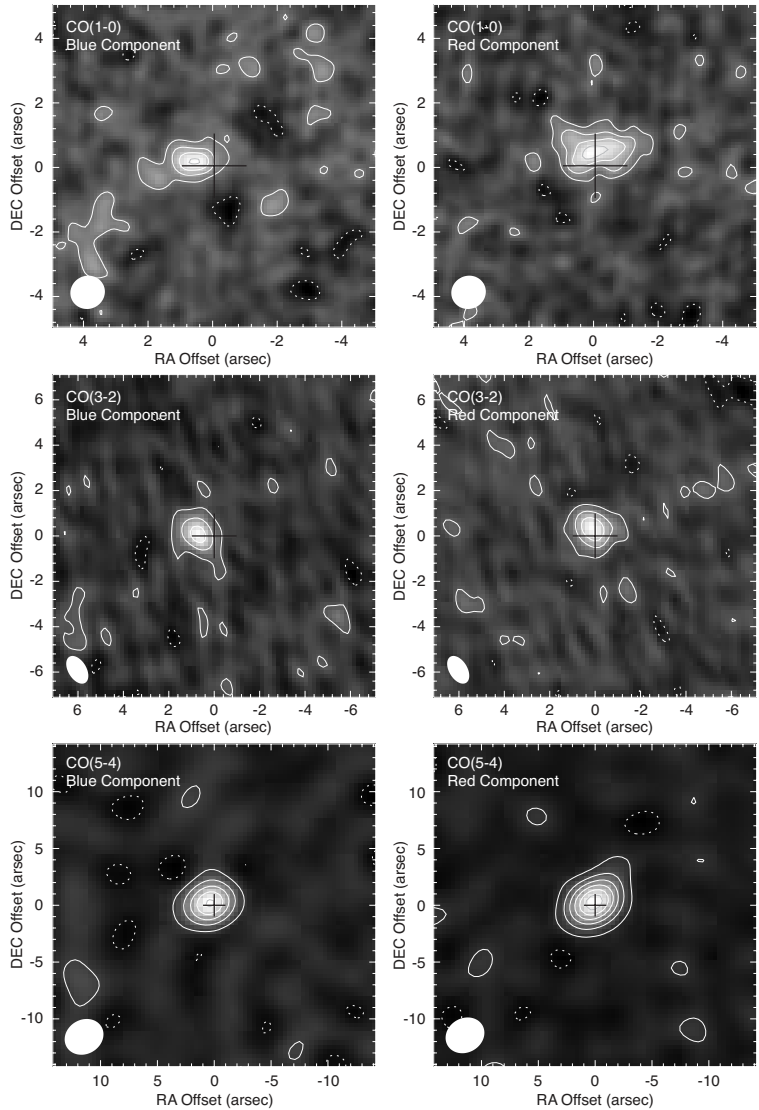}
\caption{The \mbox{CO(1--0)} (top row), \mbox{CO(3--2)} (middle row), and \mbox{CO(5--4)} (bottom row) integrated line maps of the blue component (left column) and red component (right column) centered on the \citet{frayer2000} continuum detection. In order of increasing $J$, the contours are multiples of $\pm2\sigma$ ($\sigma_{\rm blue}=0.19\,{\rm mJy\,beam^{-1}}$; $\sigma_{\rm red}=0.23\,{\rm mJy\,beam^{-1}}$), $\pm4\sigma$ (but starting at $\pm2\sigma$; $\sigma_{\rm blue}=0.93\,{\rm mJy\,beam^{-1}}$; $\sigma_{\rm red}=1.24\,{\rm mJy\,beam^{-1}}$), and $\pm4\sigma$ (but starting at $\pm2\sigma$; $\sigma_{\rm blue}=2.13\,{\rm mJy\,beam^{-1}}$; $\sigma_{\rm red}=3.09\,{\rm mJy\,beam^{-1}}$), where negative contours are dotted. The center cross is as in Figure~\ref{fig:intmap}; synthesized beams are shown in lower left corners. \label{fig:compmaps}}
\end{figure*}

The integrated line component maps are given in Figure~\ref{fig:compmaps} for the CO lines, and in Figure~\ref{fig:carbonmaps} for the C\,{\sc i} line. In order to obtain upper limits for our non-detections and line parameters for our weak detections, we consistently integrate over the $-275\rightarrow275\,{\rm km\, s^{-1}}$ velocity range for the red component and the $-625\rightarrow -275\,{\rm km\, s^{-1}}$ velocity range for the blue component. Our line parameters and upper limits for the SMA observations are insensitive to the exact separation of the two components, with reasonable variations in the velocity selection ($<100\,{\rm km\, s ^{-1}}$) yielding results that are consistent within the maps' $1\sigma$ statistical uncertainties. The component line measurements are summarized in Table\,\ref{tab:obssum}. The centroid positions for the red and blue components are consistent between the three detected CO lines. For the marginal detection of the C\,{\sc i} line, it appears that the bulk of the emission originates from the red component, supporting our merger interpretation for J00266 where the two components have different excitation conditions. The emission peak in the red component map is detected at $3.75\sigma$ and the integrated line flux is $2.44\pm1.78(\pm0.49)\,{\rm Jy\, km\, s^{-1}}$. The peak flux position is consistent with the red component maps of the CO lines.

\begin{figure*}
\epsscale{0.8}
\plotone{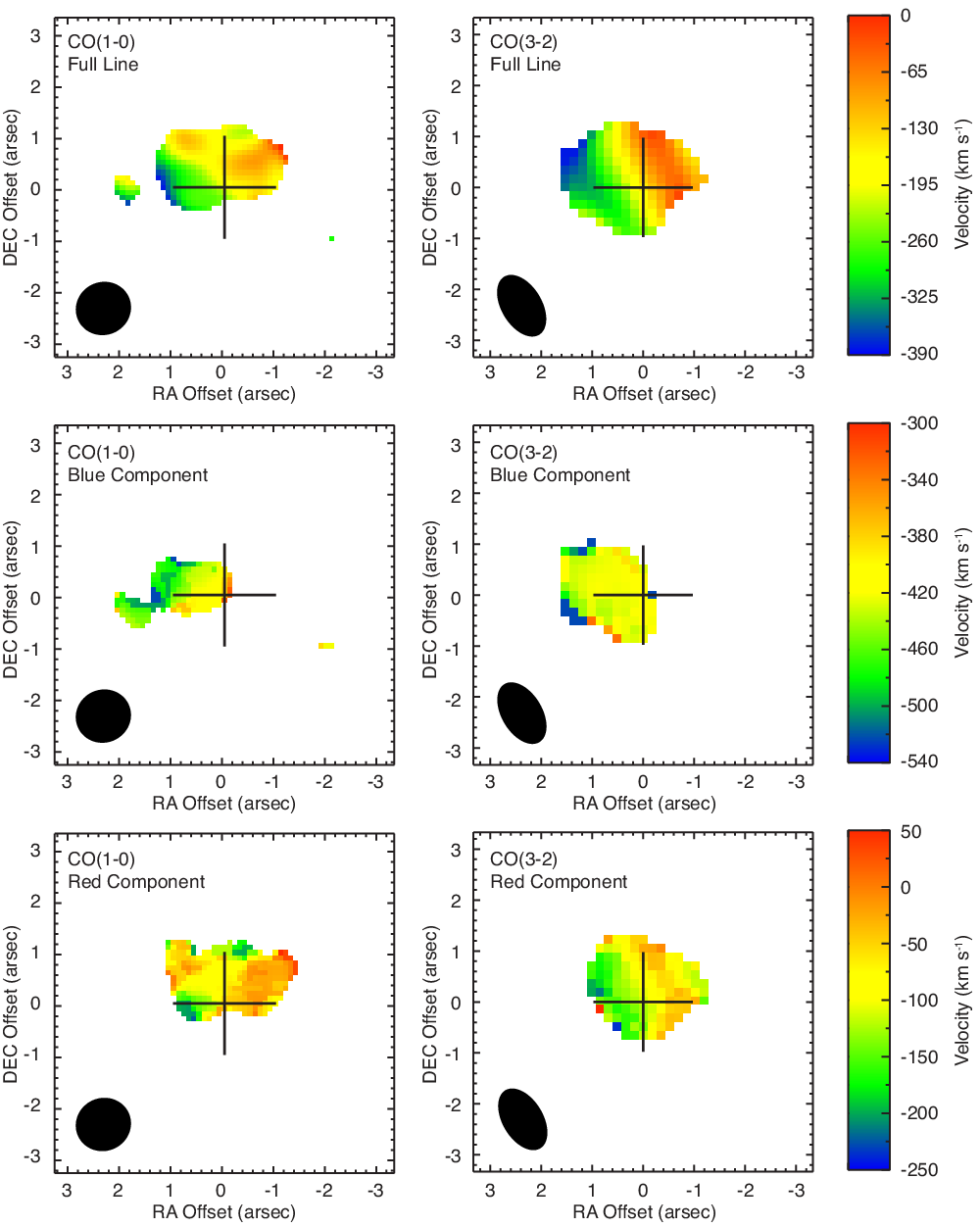}
\caption{The \mbox{CO(1--0)} (left) and \mbox{CO(3--2)} (right) first moment maps of the full line (top), blue component (middle), and red component (bottom). Positions are relative to the \citet{frayer2000} continuum detection (center cross). Pixels with values $<3\sigma$ in the corresponding integrated line map have been blanked. Synthesized beams are shown in lower left corners. These maps indicate that the blue component is dispersion-dominated, while the red component has a velocity gradient. \label{fig:firstmom}}
\end{figure*}

In Figure~\ref{fig:firstmom}, we show the first moment maps for our highest-resolution CO maps, of the \mbox{CO(1--0)} and \mbox{CO(3--2)} lines. The components' velocity structures are consistent for both emission lines. Second moment maps give us estimates for the turbulent velocity widths in both components of J00266 (although with the large beam size relative to the source size, we expect these to be affected by beam smearing). We find the velocity dispersion in the blue component to be $61\pm28\,{\rm km\,s^{-1}}$, consistent with the line FWHM. The red component has an average velocity dispersion of $101\pm37\,{\rm km\,s^{-1}}$, which is significantly less than the line FWHM. This distinction is consistent with a picture in which the blue component has dispersion-dominated kinematics while the red component has an additional velocity gradient.

$uv$-plane fits to the resolved \mbox{CO(3--2)} emission of the red and blue components show a centroid position offset of $0.^{\prime\prime}73\pm0.^{\prime\prime}06$, which is significant considering the $1.^{\prime\prime}36\times0.^{\prime\prime}81$ beam and $>10\sigma$ detections of the two components. To convert this separation to a physical scale, we must account for gravitational lensing. The best available mass modeling of the lensing cluster, Cl0024+1654 \citep{zitrin2009}, gives a magnification of $2.41\pm0.49$ and a shear angle of $-59.5^{\degree}\pm19.5^{\degree}$ \citep{zitrin2013}. The projected separation between the two components of J00266 is therefore $2.46\pm0.47\,{\rm kpc}$ at $z=2.742$, with an unknown separation along the line of sight. The blue component is best fit by a circular Gaussian with a FWHP of $0.^{\prime\prime}86\pm0.^{\prime\prime}11$ (major and minor axis radii of $2.79\pm0.34\,{\rm kpc}$ and $1.23\pm0.15\,{\rm kpc}$ when accounting for shear). The red component is best fit by an elliptical Gaussian of major and minor axis FWHPs of $1.^{\prime\prime}11\pm0.^{\prime\prime}12$ and $0.^{\prime\prime}85\pm0.^{\prime\prime}11$; the major axis is at a position angle of $-59\pm19$ degrees, which is closely aligned with the velocity gradient apparent in the first moment map (Figure~\ref{fig:firstmom}) as well as the lensing shear. The $uv$ fit result corresponds to a radius along the velocity gradient of $1.59\pm0.32\,{\rm kpc}$. $uv$-plane fits to the \mbox{CO(1--0)} emission from the two components do not converge. The image plane fitting also does not converge for the \mbox{CO(1--0)} blue component. For the red component, the best-fit model for the \mbox{CO(1--0)} emission is an elliptical Gaussian with major and minor axis FWHMs of $2.^{\prime\prime}02\pm0.^{\prime\prime}13\times0.^{\prime\prime}70\pm0.^{\prime\prime}25$ at a position angle of $86.0\pm3.1$ degrees. For the red component, the \mbox{CO(1--0)} emission appears spatially extended along the major axis relative to the \mbox{CO(3--2)} emission, which is evidence of multiple phases of molecular gas \citep[e.\/g.\/,][]{ivison2011}.

Using the \mbox{CO(3--2)}-measured source sizes and velocity structures observed in the two components, we infer dynamical masses of $(9.5\pm4.0)\times10^{9}\,M_\sun$ for the blue component (assuming a dispersion-dominated virialized sphere; $M_{\rm dyn}=1.2\times10^{10}\,M_\odot\,(\sigma/100\,{\rm km\, s^{-1}})^2(r/{\rm kpc})$; \citealt{pettini2002}) and $M_{\rm dyn}\sin^2(i)=(7.4\pm2.2)\times10^{10}\,M_\sun$ for the red component (assuming the velocity gradient of the red component is due to rotation;  $M_{\rm dyn}\sin^2(i)=2.325\times10^{9}\,M_\odot\,(\Delta v/100\,{\rm km\, s^{-1}})^2(r/{\rm kpc})$; \citealt{solomon2005}). Based on these estimates of the dynamical masses, we find that J00266 is a merger with a mass ratio of $(7.8\pm4.0)/\sin^2(i)$. Assuming an average inclination correction of $\langle\sin^2(i)\rangle=2/3$, the mass ratio becomes $11.7\pm6.0$. Alternatively, if we do not treat the two components of J00266 separately, the \mbox{CO(3--2)} emission in the $uv$-plane is best fit by an elliptical Gaussian with a major axis FWHP of $1.^{\prime\prime}40\pm0.^{\prime\prime}11$ at a position angle of $-59\pm9$ degrees and a minor axis FWHP of $0.^{\prime\prime}91\pm0.^{\prime\prime}11$. Accounting for shear, the radius along the velocity gradient is $2.05\pm0.39\,{\rm kpc}$. Using the variance-weighted average velocity separation between the two peaks as the line width, $M_{\rm dyn}\sin^2(i)=(1.06\pm0.21)\times10^{11}\,M_\sun$ if J00266 is treated as a single rotating disk.

Based on the lensing-corrected line luminosities given in Table~\ref{tab:obssum}, we estimate the gas masses for the two components of J00266. Using the VLA-measured \mbox{CO(1--0)} line fluxes, the molecular gas masses are $(2.27\pm0.59)\times10^{10}(\alpha_{\rm CO}/0.8)\,{\rm M_\sun}$ for the red component and $(1.34\pm0.38)\times10^{10}(\alpha_{\rm CO}/0.8)\,{\rm M_\sun}$ for the blue component, assuming a standard gas mass conversion factor of $\alpha_{\rm CO}=0.8\,M_\sun\,{\rm (K\,km\,s^{-1}\,pc^2)^{-1}}$ common to local U/LIRGs \citep{downes1998}. The red component's gas mass is smaller than its dynamical mass (regardless of the galaxy inclination), but the blue component's molecular gas and dynamical masses are nearly the same. While the gas mass fraction for the blue component is unphysical, more reasonable values are within the uncertainties so we do not find this tension significant. In addition, should the blue component's dynamics be dominated by rotation that is unresolved by our observations, the tension between the gas and dynamical masses would be resolved as well as the difference between the dynamical mass ratios and gas mass ratios between the two components. We note there is some evidence for variation in $\alpha_{\rm CO}$ within the population of SMGs \citep[e.\/g.\/,][]{tacconi2008,hodge2012,sharon2013}. Galactic values of the CO-to-${\rm H_2}$ conversion factor could make both components' gas masses larger than their dynamical masses (depending on the inclination correction for the red component's $M_{\rm dyn}$), and are therefore disfavored. The gas mass ratio of the two components is $(1.7\pm0.6)\times(\alpha_{\rm CO}/0.8)_{\rm red}/(\alpha_{\rm CO}/0.8)_{\rm blue}$. If we do not treat J00266 as two separate component, the gas mass fraction is $(0.34\pm0.11)sin^2(i)(\alpha_{\rm CO}/0.8)$; galactic values of $\alpha_{\rm CO}$ produce unphysical ratios between the gas mass and dynamical mass but more moderate assumptions on $\alpha_{\rm CO}$ are permitted depending on the inclination.

Our observations of the C\,{\sc i} line can also be used as alternative estimators of gas masses \citep[e.\/g.\/,][]{weiss2003}. Since we do not have measurements of both fine structure lines, we assume $T_{ex}=50\,{\rm K}$ for the blue component and $T_{ex}=20\,{\rm K}$ for the red component based on the LVG modeling presented in Section~\ref{sec:excitation}. These values yield a C\,{\sc i} mass of $M_{\rm C\,{\scriptsize\textsc i}}=(1.65\pm1.25)\times10^7\,M_\odot$ for the red component and a $3\sigma$ upper limit of $M_{\rm C\,{\scriptsize\textsc i}}<4.94\times10^6\,M_\odot$ for the blue component. Assuming the average C\,{\sc i}/CO abundance ratio for SMGs given in \citet{walter2011} (though corrected by a factor of 0.88 for the red component given our measured $r_{3,1}$), we estimate $M_{\rm H_2}=(3.71\pm3.40)\times10^{10}\,M_\odot$ for the red component and $M_{\rm H_2}<9.80\times10^{9}\,M_\odot$ for the blue component. The CO- and C\,{\sc i}-estimated molecular gas masses for the red component are consistent. For the blue component, there is some tension between the CO-determined gas mass and the C\,{\sc i}-determined $3\sigma$ upper limit, but given the uncertainties on the CO measurement, we do not find this discrepancy to be significant.

\subsection{Excitation conditions}
\label{sec:excitation}

\subsubsection{Methods}

We can compare CO excitation in J00266 to that in other SMGs by analyzing line ratios. For a specific pair of rotational transitions, $J\rightarrow J-1$ and $J^\prime\rightarrow J^\prime - 1$, we define the line ratio as

\begin{eqnarray}
r_{J,J^\prime}&=&\frac{\int T_R(J\rightarrow J-1)dv}{\int T_R(J^\prime\rightarrow J^\prime-1)dv} \nonumber \\
&=&\frac{\int S_\nu(J\rightarrow J-1)dv}{\int S_\nu(J^\prime\rightarrow J^\prime-1)dv}\left(\frac{\nu_{J^\prime\rightarrow J^\prime-1}}{\nu_{J\rightarrow J-1}}\right)^2 ,
\label{eq:lineratio}
\end{eqnarray}

\noindent
where the (beam-averaged) integrated line intensities are in brightness temperature units. Table\,\ref{tab:ratios} lists our measured values of $r_{J,J^\prime}$ for J00266, for both the full line and the two components separately (in the case of the \mbox{CO(7--6)} line, $3\sigma$ upper limits are given). All of our measured $r_{3,1}$ values, including for the two separate components, are larger than the average $r_{3,1}\sim0.6$ observed in other SMGs \citep[e.\/g.\/,][]{swinbank2010b, harris2010, ivison2011, danielson2011,bothwell2013} by at least $1\sigma$. For both components, $r_{3,1}$ is consistent with unity, similar to those of quasar host galaxies \citep{riechers2011f}, the SMG SMM\,J14011+0252 \citep{sharon2013}, and the potential ``wet-dry" merger SMM\,J04135+10277 \citep{riechers2013a}. Considering that the \mbox{CO(1--0)} emission is more extended than the \mbox{CO(3--2)} emission, a clear signature of multi-phase molecular gas, it is interesting that the extended cold gas reservoir does not result in a lower $r_{3,1}$ value for the red component (or for J00266 as a whole when treated as a single source). The \mbox{CO(5--4)}/\mbox{CO(3--2)} line ratio, $r_{5,3}$, indicates a possible ($<2\sigma$) excitation difference between the two components, with the blue component's $r_{5,3}$ lower than the population average \citep[$r_{5,3}=0.62$;][]{bothwell2013}.

\begin{deluxetable*}{ccccccc}
\tablewidth{0pt}
\tablecaption{ SMM\,J00266+1708 Line Ratios \label{tab:ratios}}
\tablehead{{Component} & {$r_{7,5}$\tablenotemark{a}} & {$r_{7,3}$\tablenotemark{a}} & {$r_{7,1}$\tablenotemark{a}} & {$r_{5,3}$} & {$r_{5,1}$} & {$r_{3,1}$}}
\startdata
Total & $<0.44$ & $<0.28$ & $<026$ & $0.63\pm0.07(\pm0.11)$ & $0.60\pm0.07(\pm0.11)$ & $0.95\pm0.13(\pm0.13)$\\
Blue & $<0.97$ & $<0.44$ & $<0.47$ & $0.45\pm0.12(\pm0.08)$ & $0.49\pm0.10(\pm0.09)$ & $1.07\pm0.32(\pm0.15)$\\
Red & $<0.47$ & $<0.35$ & $<0.31$ & $0.75\pm0.18(\pm0.14)$ & $0.66\pm0.10(\pm0.12)$ & $0.88\pm0.23(\pm0.12)$
\enddata
\tablecomments{Ratios that include the \mbox{CO(1--0)} line use the VLA-measured values. We give the statistical uncertainties, and in parentheses the flux calibration uncertainties.}
\tablenotetext{a}{$3\sigma$ upper limits.}
\end{deluxetable*}

In order to constrain the physical conditions of the molecular gas in J00266, we have compared our measured line ratios to the results of a radiative transfer model using the large velocity gradient (LVG) approximation. The LVG modeling follows the methods detailed in \citet{ward2002}. The model was adapted to include the cosmic microwave background (CMB) radiation, which can affect CO rotational level populations at high redshift where $T_{\rm CMB}$ and $T_{\rm ex}$ (for low-$J$ lines) are comparable. Collision rates between CO and both ortho- and para-${\rm H}_2$ were taken from \citet{yang2010}, and collision rates between CO and He were taken from \citet{cecchi2002}. These collision rates were interpolated onto the temperature spacing we probe in the LVG model using a quadratic spline. The ortho-to-para-${\rm H_2}$ ratio is determined as a function of temperature \citep{takahashi2001}, which asymptotes to the common assumption of a constant ratio equal to three for high temperatures ($\gtrsim100\,{\rm K}$; differences in the CO level populations resulting from temperature-dependent and constant values of the ortho-to-para ratio are negligible). We assume a spherical cloud geometry with the escape probability given by \citet{goldreich1974}.

The output of the LVG code is the calculated line luminosity per unit density, $\Lambda_{J,J-1}=h\nu_{J,J-1} A_{J,J-1} n_{\rm H_2} x_{\rm CO} \chi_J \beta_{J,J-1}$ for a specific $J\rightarrow J-1$ transition ($\chi_J$, the fraction of CO in the $J$th rotational state, and the escape probability, $\beta_{J,J-1}$, are the chief unknowns determined by the LVG code). The line luminosity is a function of three parameters that detail the physical conditions of the molecular gas: kinetic temperature, ${\rm H_2}$ density, and CO column density per unit line width ($T_{\rm kin}$, $n_{\rm H_2}$, and $N_{\rm CO}/\Delta v$). We generate a library of line luminosities for the three input parameters, which are sampled at $\Delta T_{\rm kin}=2\,{\rm K}$ for $0\,{\rm K}\leq T_{\rm kin}\leq 200\,{\rm K}$, $\Delta\log{(n_{\rm H_2}/ {\rm cm}^{-3})}=0.1$ for $0\leq \log{(n_{\rm H_2}/ {\rm cm}^{-3})}\leq 7$, and $\Delta\log{((N_{\rm CO}/\Delta v)/ {\rm cm^{-2}\, km^{-1}\, s})}=0.25$ for $15\leq \log{((N_{\rm CO}/\Delta v)/{\rm cm^{-2}\, km^{-1}\, s})}\leq 22$. By taking the ratio of the line luminosities for two transitions using the same input parameter triplet ($T_{\rm kin}$, $n_{\rm H_2}$, and $N_{\rm CO}/\Delta v$), we have a predicted line ratio as produced by a single-phase molecular ISM to compare to our measurements. By evaluating the model line ratios, we avoid uncertainty due to choice of $x_{\rm CO}$, the relative abundance of CO to ${\rm H_2}$. Given observed parameters like source size and velocity structure and a best-fit model, we can determine the value of $x_{\rm CO}$.

Since recent work indicates that the line ratios of SMGs are frequently inconsistent with single-phase models of the molecular ISM \citep[][]{swinbank2010b, harris2010, ivison2011, danielson2011,bothwell2013}, we also consider two-phase models. In order to determine the predicted line ratios for the two-phase LVG model, we iteratively determine the line intensity for every possible pair of input parameter triplets for a range of filling factors. The two-phase line ratio for given CO transitions $J\rightarrow J-1$ and $J^\prime\rightarrow J^\prime-1$ is given by

\begin{widetext}
\begin{equation}
r_{J,J^\prime}=\frac{f_{\rm c} \Lambda_J (T_{\rm c}, n_{\rm c}, N_{\rm c}/\Delta v) + f_{\rm w} \Lambda_J(T_{\rm w}, n_{\rm w}, N_{\rm w}/\Delta v)}{f_{\rm c} \Lambda_{J^\prime} (T_{\rm c}, n_{\rm c,} N_{\rm c}/\Delta v) + f_{\rm w} \Lambda_{J^\prime} (T_{\rm w}, n_{\rm w}, N_{\rm w}/\Delta v)} \left( \frac{\nu_{J^\prime}}{\nu_J}\right)^3 ,
\label{eq:modellineratio}
\end{equation}
\end{widetext}

\noindent
where the ``${\rm c}$" and ``${\rm w}$" subscripts denote the cold and warm phases, we have dropped the redundant ``kin", ${\rm H_2}$, CO, $J-1$, and $J^\prime -1$ subscripts, and the additional frequency term is used to put the predicted line ratio in brightness temperature units. Since $r_{J,J^\prime}$ depends only on the ratio of the filling factors, $f_{\rm c}/f_{\rm w}$, rather than on both parameters separately (the filling factors individually lie between zero and one, which means $f_{\rm c}/f_{\rm w}$ can have any value between zero and infinity), we can eliminate an excess parameter by probing only the filling factor ratio. This ratio was sampled every $\Delta\log{(f_{\rm c}/f_{\rm w})}=0.2$ for $0\leq \log{(f_{\rm c}/f_{\rm w})}\leq 5$.

Since both the single-phase and two-phase LVG models have numbers of free parameters comparable to or greater than the number of measurements for J00266, and these measurements have significant uncertainties, we use a Bayesian analysis to determine the likelihood distributions for the various model parameters. This approach \citep[detailed in][]{ward2003} is preferred to a $\chi^2$ minimization technique since it more accurately reflects the uncertainties in the measurements (although we report both the minimum $\chi^2$ and maximum probability results for the single-phase analysis for comparison). Priors have been carefully chosen to (1) reflect our lack of pre-existing knowledge about the physical conditions in the molecular gas, (2) rule out scenarios that are obviously physically unrealistic, and (3) compensate for biases towards high temperature/density models that result from the large logarithmically sampled parameter space. We adopt uniform priors in the logarithm of kinetic temperature, ${\rm H_2}$ density, CO column density per velocity gradient, and (in the case of the two-phase model) filling factor ratio. While this choice effectively weights the likelihood distribution towards colder-phase solutions, substantial fractions of SMGs' gas masses are likely in the cold phase, making its characterization of primary interest. For systems well-described by a single-phase molecular ISM, $\chi^2$ minimization and Bayesian techniques favor similar models. The best-fit models for J00266 are relatively insensitive to the choice between logarithmic and uniform priors, meaning that the logarithmic prior is not significantly biasing our results. Additional priors are used to rule out models with kinetic temperatures less than the CMB temperature and regions of the two-phase parameter space where the cool phase is warmer than the warm phase. 

In addition to the basic priors described above, which are in place for all applications of our Bayesian approach, in cases where the LVG model produces extremely degenerate results, it is worth considering additional (physically motivated) priors to limit the available parameter space. First, in a multi-phase ISM, optically thick emission should dominate the observed light, overwhelming any optically thin emission. We therefore consider models that only produce optically thick CO emission for the observed rotational transitions (although for the single-phase analysis we compare optically thin and thick best-fit models). Second, detectability constrains the CO brightness temperature to be greater than the CMB temperature (at the source redshift). This prior requires knowledge of the source size, which can be uncertain if unresolved structure exists in the source, and assumes that the emission in different CO lines is coming from the same physical regions (the latter assumption is implicit for all single-phase LVG models, as well as for some fraction of the emission in multi-phase models). In practice, even with the most severe assumptions about source size, this prior eliminates $<1\%$ of parameter space and duplicates other cuts such as the requirement of $\tau>1$. Third (and also dependent on source size), the CO column length should be less than the diameter of the object. Given an estimate of the source size and line width, cuts on the CO column can remove the highest values of $N_{\rm CO}/\Delta v$. A fourth potential requirement is that the molecular gas mass implied by the model be less than the dynamical mass ($n_{\rm H_2}\leq 3M_{\rm dyn}/(m_{\rm H_2} 4 \pi r^3)$). Fifth, in cases where we expect that most of the gas mass is in dense, star-forming, virialized clouds, we can introduce a prior based on the degree of virialization,

\begin{equation}
K_{\rm vir}=\frac{\Delta v/\Delta r_{\rm LVG}}{\Delta v/\Delta r_{\rm vir}} \sim 1.54\frac{\Delta v}{\Delta r}\left[ \frac{\langle n_{\rm H_2}\rangle}{10^3\,{\rm cm^{-3}}}\right]^{-1/2}
\label{eq:alpha}
\end{equation}

\noindent
\citep[e.\/g.\/,][]{goldsmith2001,greve2009}, where $K_{\rm vir}\sim1$ for virialized clouds. For an observed velocity gradient, any constraints on $K_{\rm vir}$ translate to constraints on $n_{\rm H_2}$; very large densities that produce $K_{\rm vir}\ll1$ can be eliminated, since velocity gradients smaller than that of virialized gas (for a specific density) are dynamically unobtainable. In principle, since the dust and molecular gas are mixed, as a final constraint the gas kinetic temperature can be set equal to the dust temperature. However, a dust temperature determined from far-IR photometry lacks the necessary spatial or velocity information required to properly attribute temperatures to the two nearly coincidental components of J00266. Similarly, a single infrared-estimated dust temperature would probe the dust commingled with all phases of the molecular gas, making the dust temperature an ambiguous constraint in the case of multiple gas phases.

\subsubsection{Blue component}

\begin{figure}
\epsscale{1.0}
\plotone{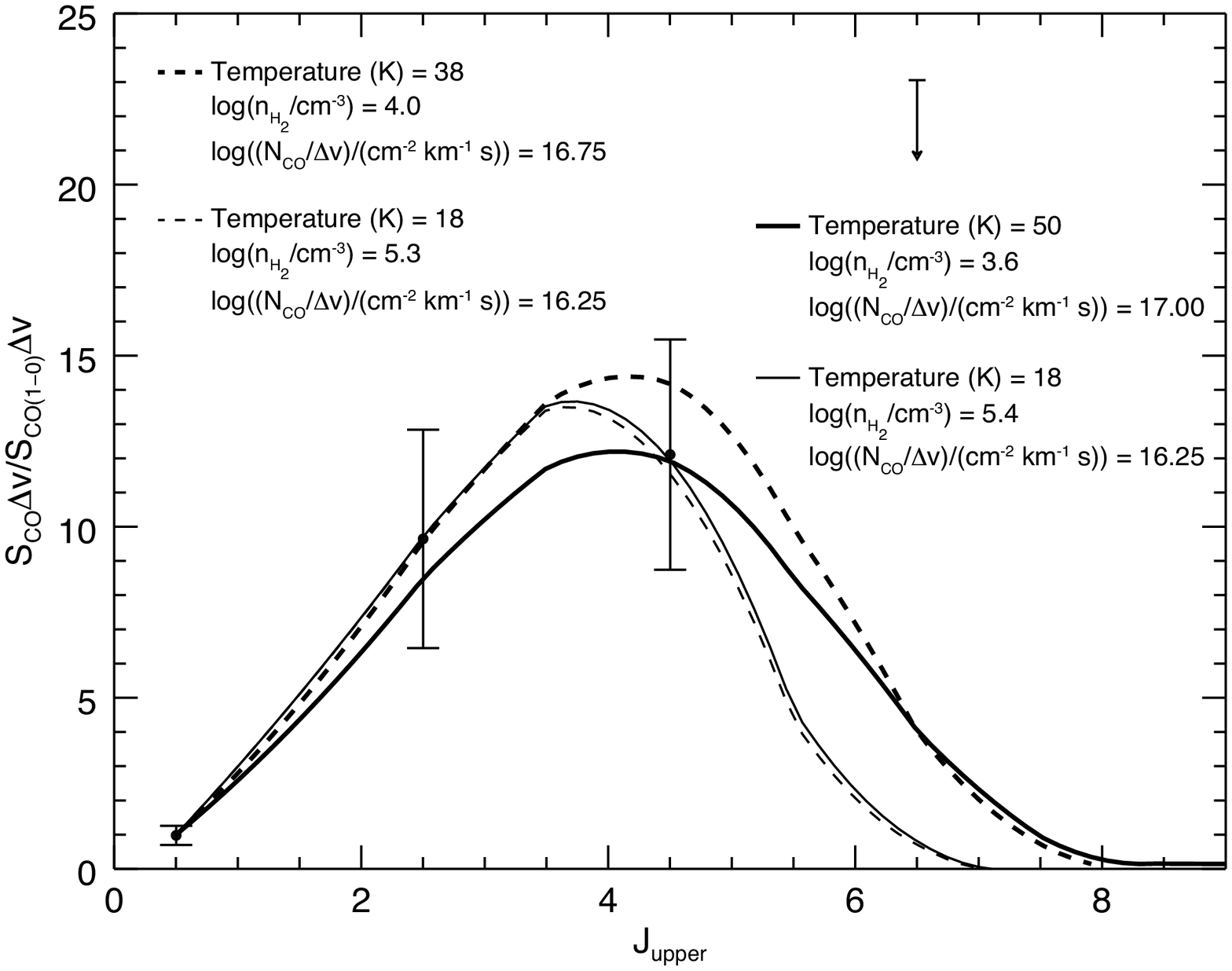}
\caption{CO SLEDs for the best-fitting models (lines) and the measured line ratios and $1\sigma$ errors (points) for the blue component of J00266. The $3\sigma$ upper limit on the \mbox{CO(7--6)} line is given by the arrow. The solid lines are for the highest-probability models from the Bayesian analysis, and the dashed lines are for models that minimize $\chi^2$. Thick lines are for the models where the full LVG parameter space was considered (with the exception of $T<T_{\rm CMB}$), and thin lines are for models that require all of the line emission to be optically thick. \label{fig:bluesled}}
\end{figure}

The line ratios for the the blue component of J00266 can be reproduced by single-phase LVG models. Based on our Bayesian analysis, the best-fit model parameters are $T_{\rm kin}=18\,{\rm K}$, $n_{\rm H_2}=10^{5.4}\,{\rm cm^{-3}}$, and $N_{\rm CO}/\Delta v=10^{16.25}\,{\rm cm^{-2}\, km^{-1}\, s}$; the best-fit CO SLED is shown in Figure~\ref{fig:bluesled}. However, a wide range of parameters can reproduce the line ratios within the measured errors. To illustrate this degeneracy, we show three additional CO SLEDs in Figure~\ref{fig:bluesled}: the minimum $\chi^2$ model ($T_{\rm kin}=18\,{\rm K}$, $n_{\rm H_2}=10^{5.3}\,{\rm cm^{-3}}$, $N_{\rm CO}/\Delta v=10^{16.25}\,{\rm cm^{-2}\, km^{-1}\, s}$), the highest probability model in which all line emission is optically thick ($T_{\rm kin}=50\,{\rm K}$, $n_{\rm H_2}=10^{3.6}\,{\rm cm^{-3}}$, $N_{\rm CO}/\Delta v=10^{17.00}\,{\rm cm^{-2}\, km^{-1}\, s}$), and the minimum $\chi^2$ model in which all line emission is optically thick ($T_{\rm kin}=38\,{\rm K}$, $n_{\rm H_2}=10^{4.0}\,{\rm cm^{-3}}$, $N_{\rm CO}/\Delta v=10^{16.75}\,{\rm cm^{-2}\, km^{-1}\, s}$).

\begin{figure*}
\epsscale{1.0}
\plotone{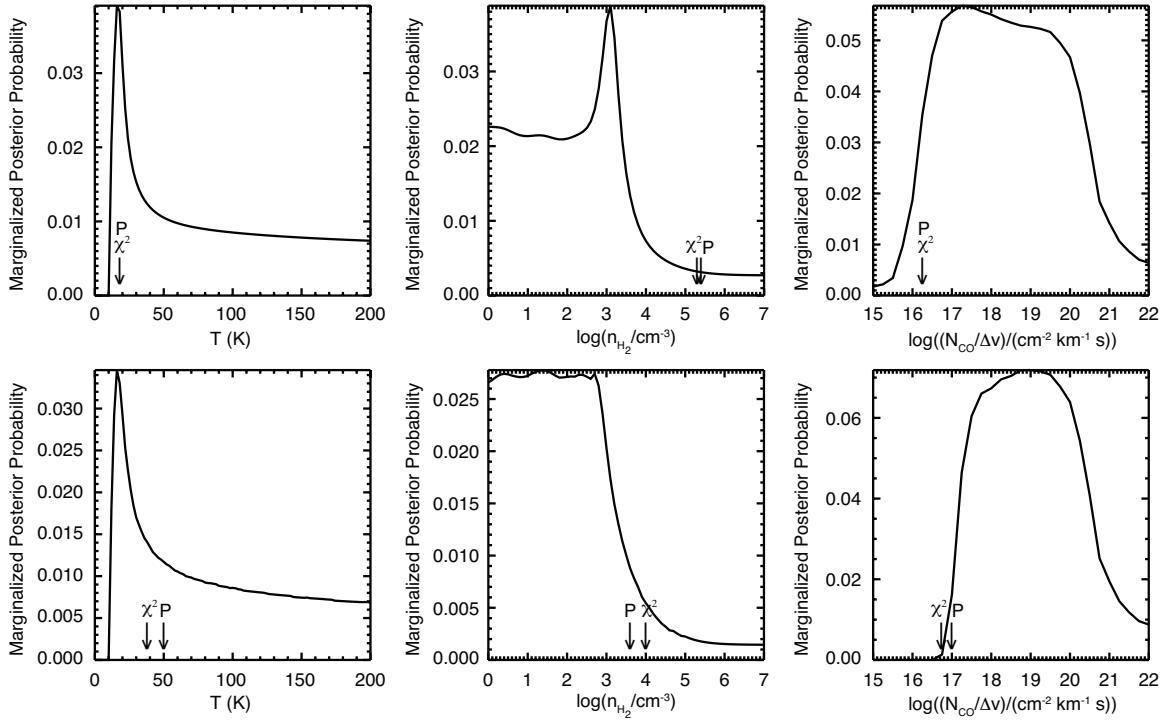}
\caption{The marginalized likelihood distributions for the single-phase model for the blue component of J00266, where the entire parameter space is available (top row) and where models with optically thin line emission have been removed (bottom row). The exact parameters of the highest-probability and minimum-$\chi^2$ models are indicated with arrows (labeled with ``P" and ``$\chi^2$", respectively). As discussed in Section~4.2.2, the marginalized probability distributions may not peak at the locations of the maximum probability (or minimum $\chi^2$) model parameters due to the shapes of the best-fit model parameter regions (see Figure~\ref{fig:bluecontours1} and Figure~\ref{fig:bluecontours2}). \label{fig:bluemarginalized}}
\end{figure*}

\begin{figure*}
\epsscale{1.0}
\plotone{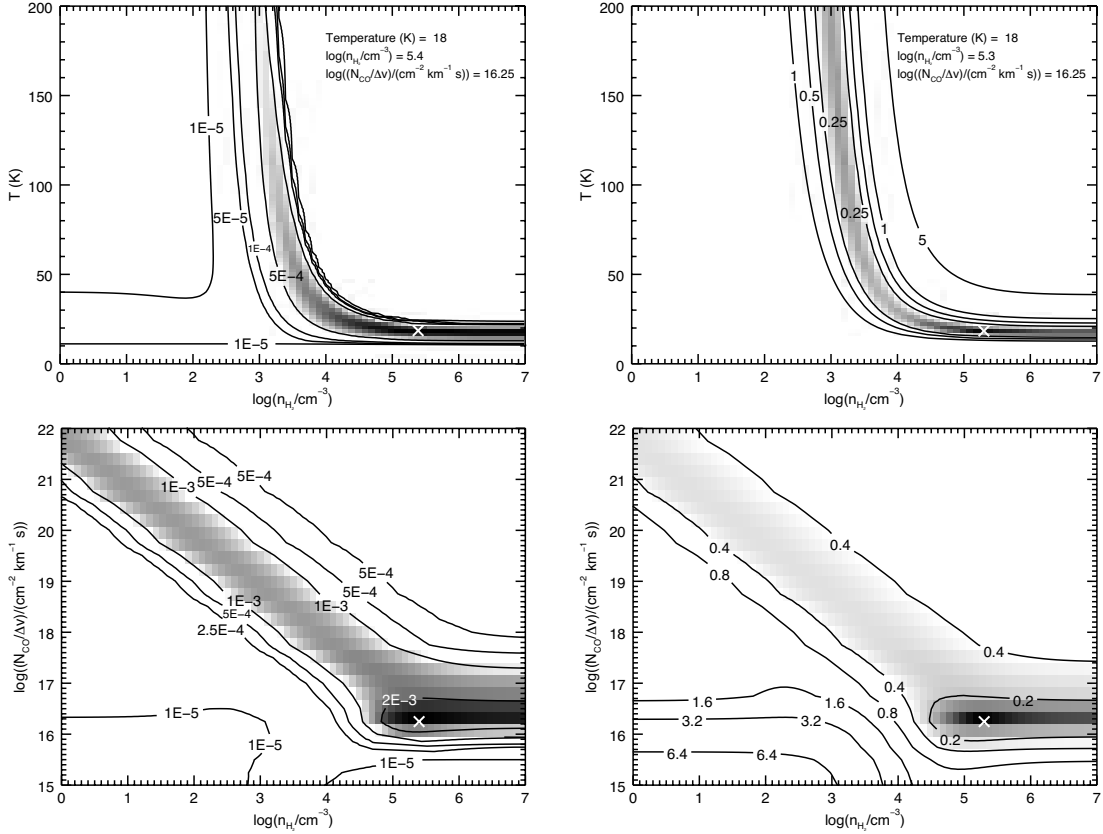}
\caption{Probability contours (left column) and $\chi^2$ contours (right column) for two slices through the best-fit models for the blue component, fixing the best-fit $N_{\rm CO}/\Delta v$ value (upper row) and the best-fit temperature (lower row). The best-fit model values are marked with white crosses, and the values are given in the top panel for each. The shading for the lowest contours is logarithmic to illustrate the shape of the parameter space where contours become unclear. \label{fig:bluecontours1}}
\end{figure*}

\begin{figure*}
\epsscale{1.0}
\plotone{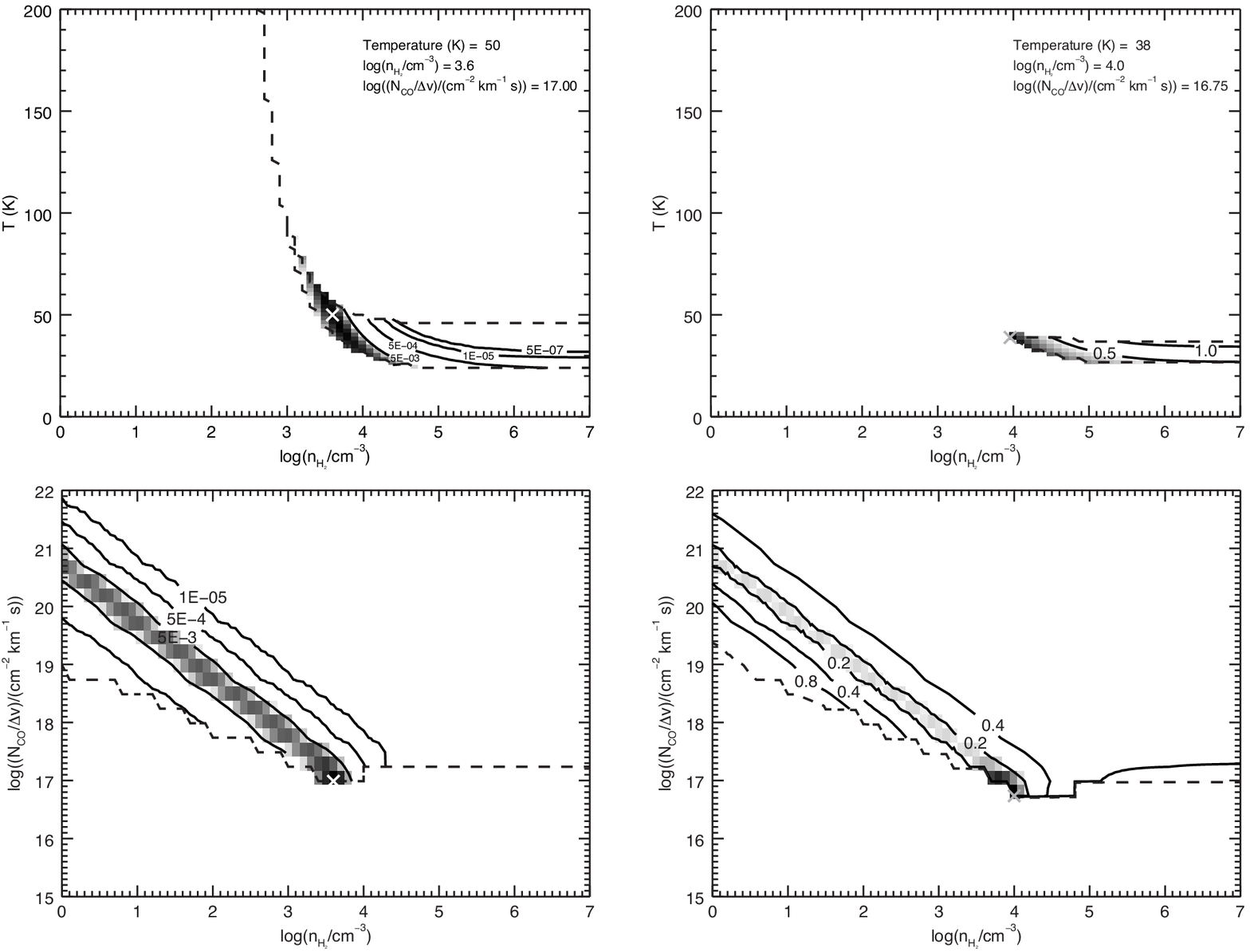}
\caption{Probability contours (left column) and $\chi^2$ contours (right column) for two slices through the best-fit models for the blue component, through the best-fit $N_{\rm CO}/\Delta v$ value (upper row) and best-fit temperature (lower row), after optically thin models have been removed. The best-fit model values are marked with white crosses, and the values are given in the top panel for each. The shading for the lowest contours is logarithmic to illustrate the shape of the parameter space where contours become unclear. Cuts imposed by the optical depth requirements are marked with dashed lines. \label{fig:bluecontours2}}
\end{figure*}

The posterior probability distributions for these parameters are given in Figure~\ref{fig:bluemarginalized}. The marginalized probability distributions generally do not peak at the locations of the maximum probability (or minimum $\chi^2$) model parameters. This apparent inconsistency is due to the significant degeneracies between models that adequately reproduce the observed line ratios (illustrated in Figure~\ref{fig:bluecontours1} and Figure~\ref{fig:bluecontours2}), in particular the ``L"-shaped degenerate parameter space along $T_{\rm kin}$ and $n_{\rm H_2}$ seen in the top panels of Figure~\ref{fig:bluecontours1}. Since the number of collisions that occurs is proportional to both CO density and ${\rm H_2}$ density, the additional degeneracy between $n_{\rm H_2}$ and $N_{\rm CO}/\Delta v$ is illustrated in the bottom panels of Figure~\ref{fig:bluecontours1}. However, at very low $N_{\rm CO}/\Delta v$, the emission becomes optically thin (Figure~\ref{fig:bluecontours2}), and only increased ${\rm H_2}$ density will keep the CO in excited states. The Bayesian and $\chi^2$ analyses both prefer optically thick line emission ($1<\tau \lesssim5$) for the three \emph{detected} CO lines, even without enforcing the $\tau>1$ prior. Requiring the undetected \mbox{CO(7--6)} line also to be optically thick imposes the cuts in parameter space seen in Figure~\ref{fig:bluecontours2}. In this case, the best-fit models have $\tau\sim1-2$ for the \mbox{CO(7--6)} line, and moderately optically thick emission ($1<\tau<15$) for the lower-$J$ lines.

Given the degeneracies in LVG model parameters that can reproduce the blue component's line ratios, the best-fit models should be interpreted as rough estimates of the range of gas conditions that dominate the CO emission. Based on the likelihood distributions and degeneracies exhibited over the full three-dimensional parameter space, the blue component's temperature and density are poorly constrained, with $T_{\rm kin}=15$--$50\,{\rm K}$ and $n_{\rm H_2}\gtrsim10^{3.5}\,{\rm cm^{-3}}$, but we see a reasonably consistent $N_{\rm CO}/\Delta v=10^{16.5\pm0.5}\,{\rm cm^{-2}\, km^{-1}\, s}$. The temperature range is consistent with the $39\pm1\,{\rm K}$ dust temperature calculated in \citet{magnelli2012}. Barring additional line measurements to constrain acceptable LVG models, we consider additional priors. Requiring that the ${\rm H_2}$ densities produce gas masses lower than the dynamical mass does not produce useful cuts on the ${\rm H_2}$ density, allowing all densities probed by our LVG models. If the molecular gas in the blue component is virialized, then its cloud densities should produce $0.5<K_{\rm vir}<2$. This prior limits the molecular gas density to be between $10^{2.3}$--$10^{4.5}\,{\rm cm^{-3}}$, consistent with the lower-density and higher-temperature best-fit models favored by the requirement that the \mbox{CO(7--6)} emission be optically thick.

\subsubsection{Red component}
\label{sec:red}

\begin{figure}
\epsscale{1.0}
\plotone{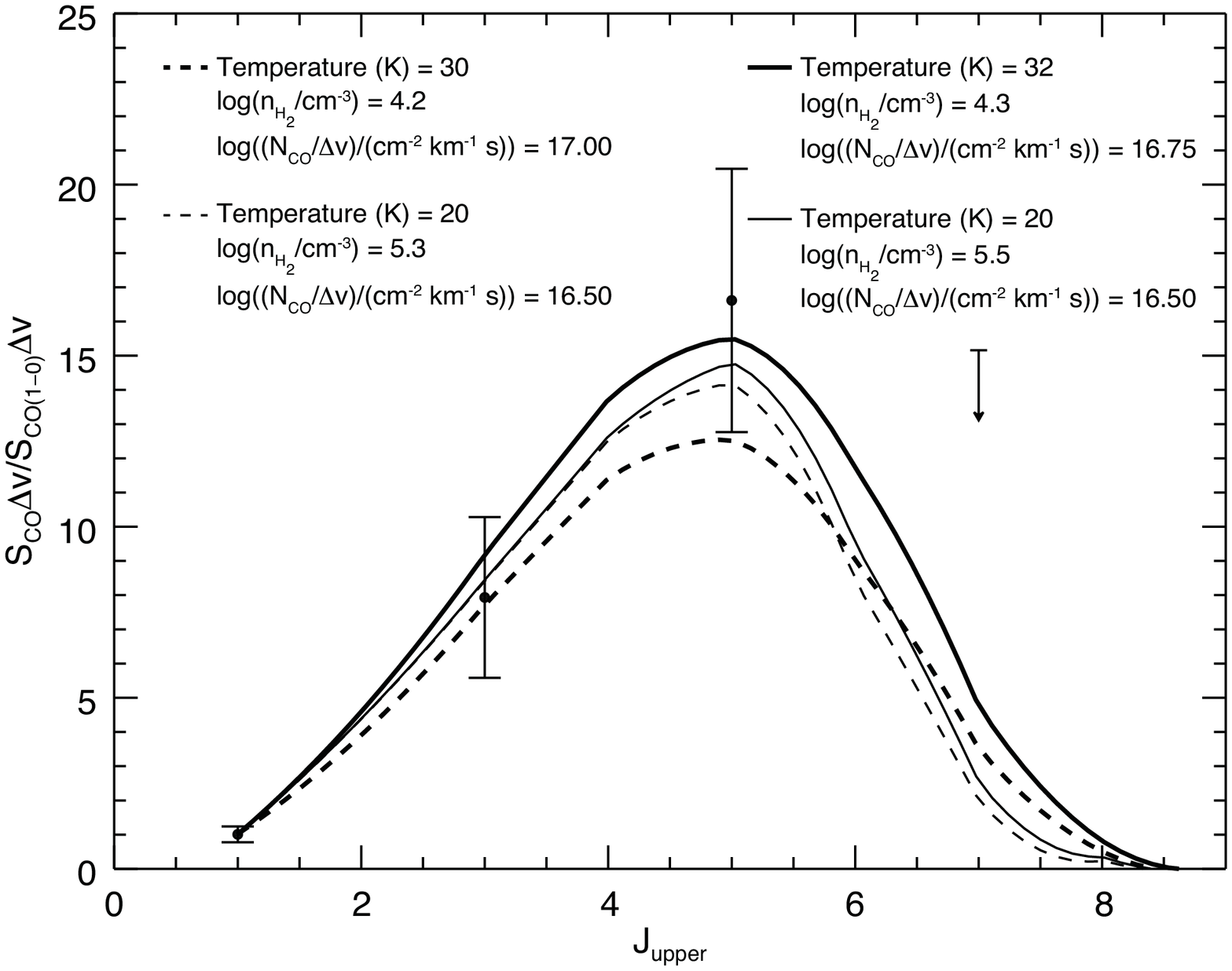}
\caption{CO SLEDs for the best fitting single-phase models (lines) and the measured line ratios and $1\sigma$ errors (points) for the red component of J00266. The $3\sigma$ upper limit for the CO(7--6) line is given by the arrow. The solid lines are for the highest-probability models from the Bayesian analysis, and the dashed lines are for models that minimize $\chi^2$. Thick lines are for the models where the full LVG parameter space was considered (with the exception of $T<T_{\rm CMB}$), and thin lines are for models that require all of the line emission to be optically thick. \label{fig:redsled}}
\end{figure}

\begin{figure*}
\epsscale{1.0}
\plotone{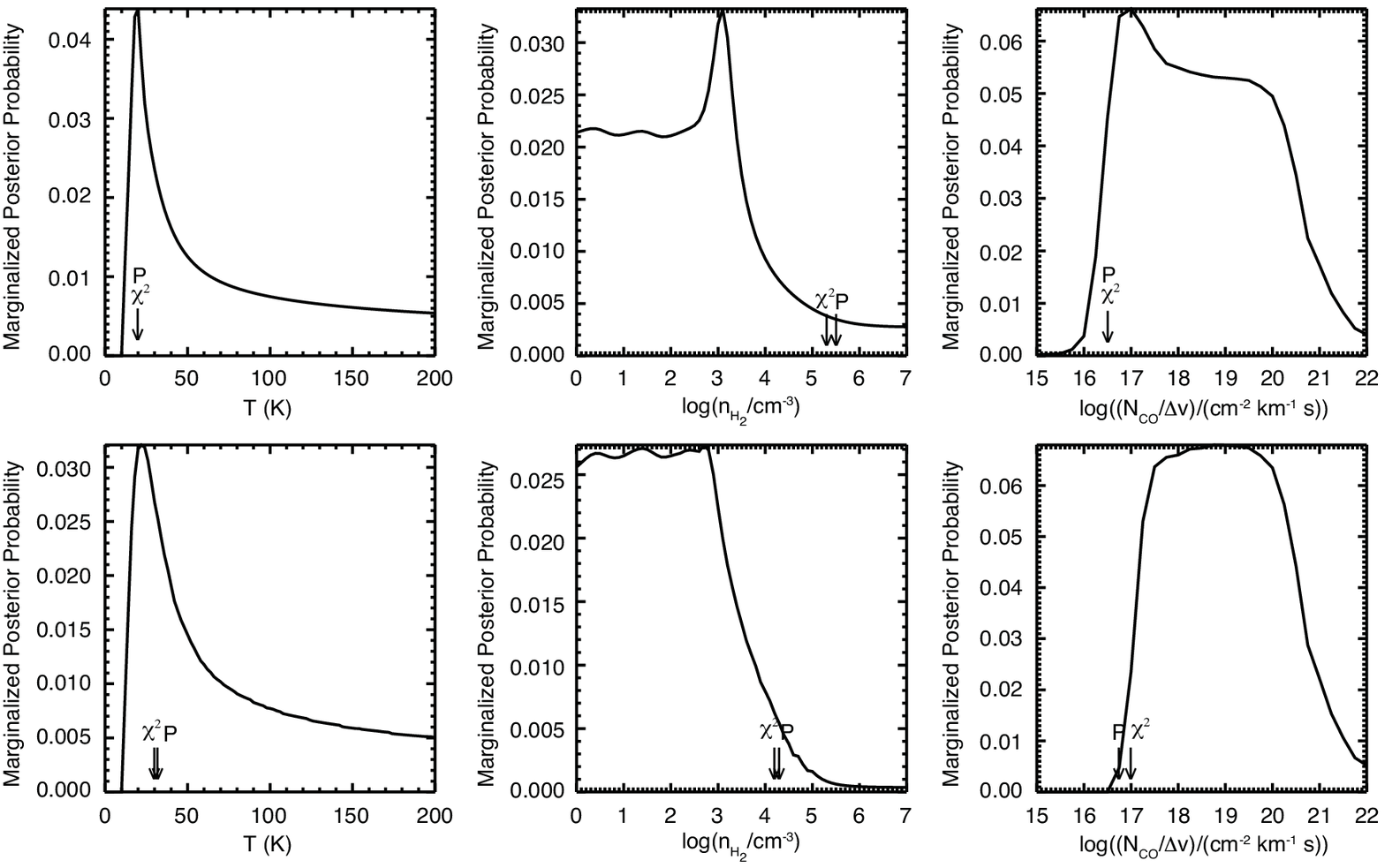}
\caption{The marginalized likelihood distributions for the single-phase model for the red component of J00266, where the entire parameter space is available (top row) and where models with optically thin line emission have been removed (bottom row). The exact parameters of the highest-probability and minimum-$\chi^2$ models are indicated with arrows (labeled with ``P" and ``$\chi^2$", respectively). \label{fig:redmarginalized}}
\end{figure*}

\begin{figure*}
\epsscale{1.0}
\plotone{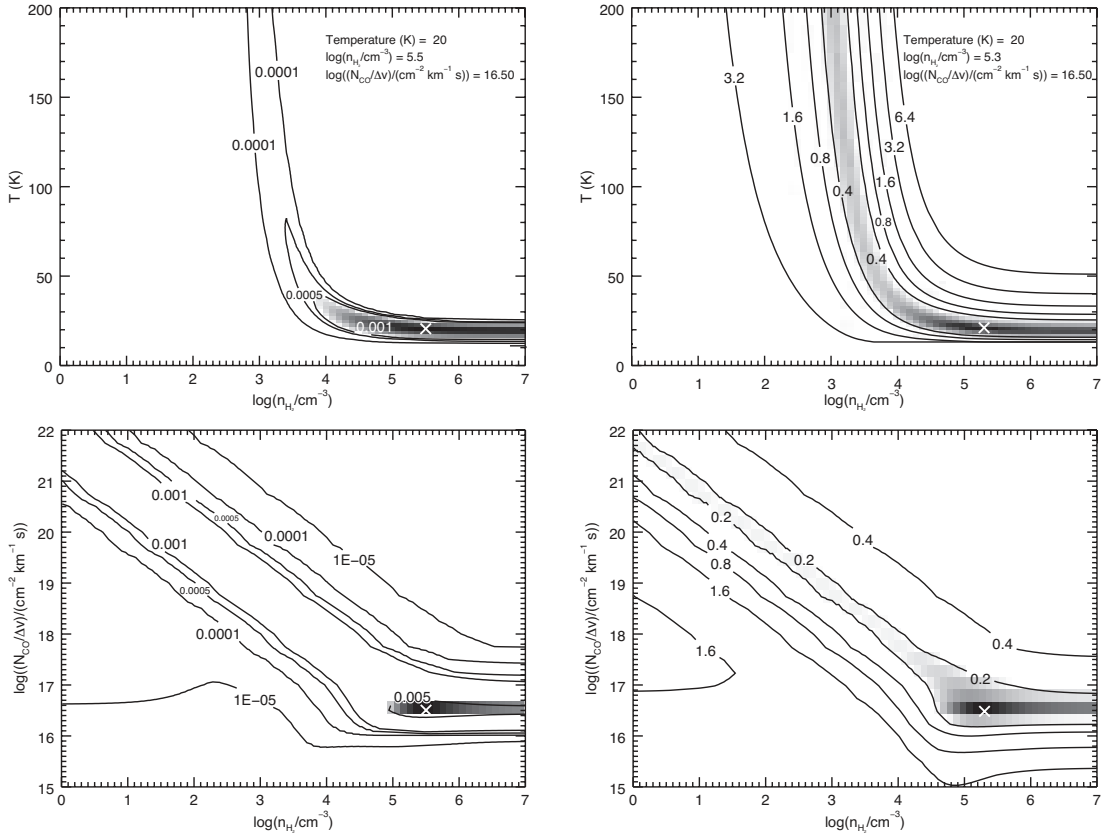}
\caption{Probability contours (left column) and $\chi^2$ contours (right column) for two slices through the best-fit single-phase models for the red component, fixing the best-fit $N_{\rm CO}/\Delta v$ value (upper row) and the best-fit temperature (lower row). Other notations are as in Figure~\ref{fig:bluecontours1}. \label{fig:redcontours1}}
\end{figure*}

\begin{figure*}
\epsscale{1.0}
\plotone{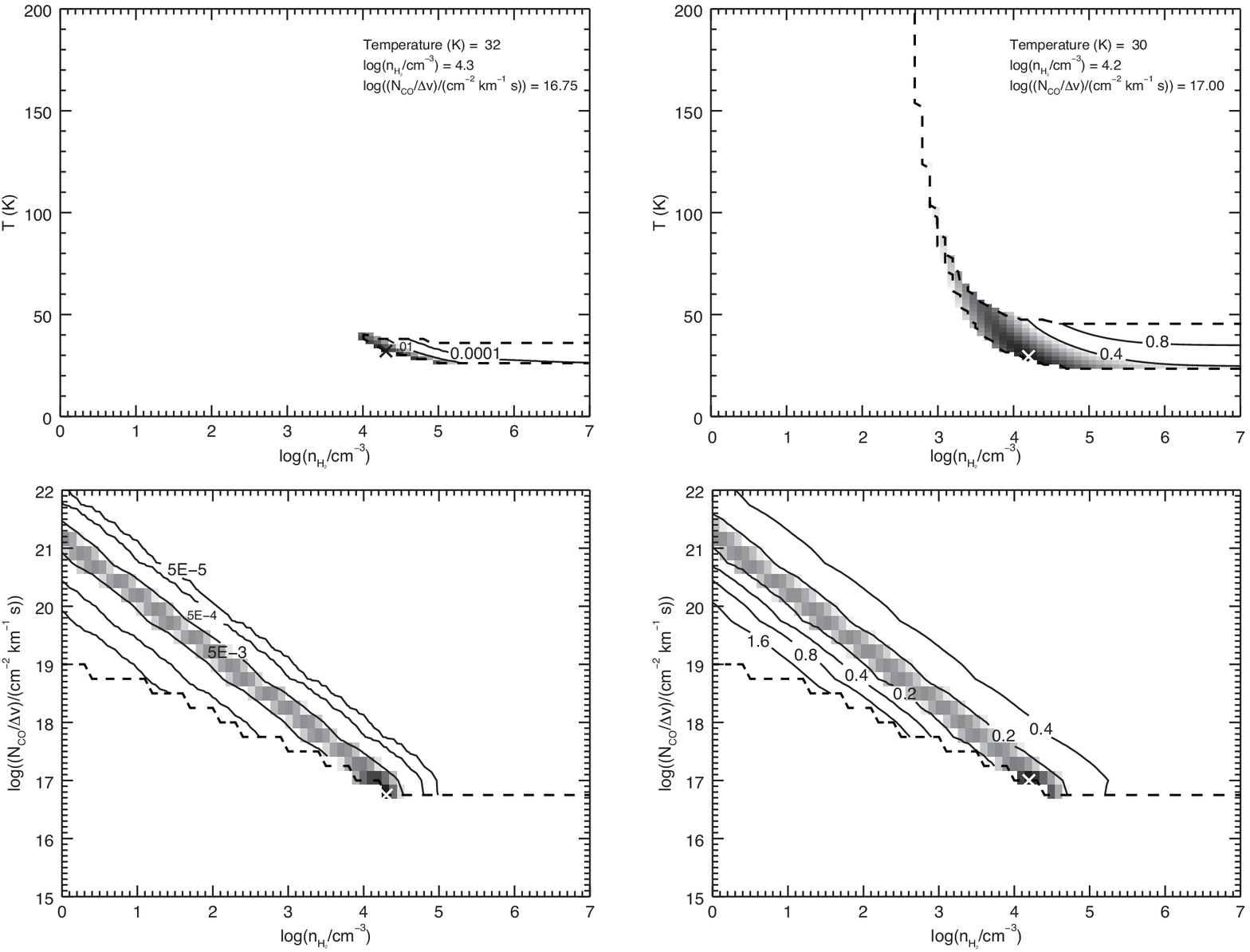}
\caption{Probability contours (left column) and $\chi^2$ contours (right column) for two slices through the best-fit models for the red component, through the best-fit $N_{\rm CO}/\Delta v$ value (upper row) and best-fit temperature (lower row), after optically thin models have been removed. Other notations are as in Figure~\ref{fig:bluecontours2}. \label{fig:redcontours2}}
\end{figure*}

The red component line ratios can also be reproduced by single-phase LVG models. The CO SLED for the minimum $\chi^2$ and highest probability models favor very similar parameters and are shown in Figure~\ref{fig:redsled}; the two models have common values of $T_{\rm kin}=20\,{\rm K}$ and $N_{\rm CO}/\Delta v=10^{16.50}\,{\rm cm^{-2}\, km^{-1}\, s}$, but differ slightly with $n_{\rm H_2}=10^{5.5}\,{\rm cm^{-3}}$ for the maximum probability model and $n_{\rm H_2}=10^{5.3}\,{\rm cm^{-3}}$ for the minimum $\chi^2$ model. As for the blue component, these models have $\tau>1$ for the emission lines with $J_{\rm upper}\leq5$, but optically thin emission for the \mbox{CO(7--6)} line ($\tau_{1-0}=1.5$, $\tau_{3-2}=7.8$, $\tau_{5-4}=2.7$, and $\tau_{7-6}<0.2$). If we restrict the possible parameter space to only models that produce optically thick line emission for all transitions, including \mbox{CO(7--6)}, the models favor higher temperatures and lower densities ($T_{\rm kin}=32\,{\rm K}$, $n_{\rm H_2}=10^{4.3}\,{\rm cm^{-3}}$, and $N_{\rm CO}/\Delta v=10^{16.75}\,{\rm cm^{-2}\, km^{-1}\, s}$ for the maximum probability model and $T_{\rm kin}=30\,{\rm K}$, $n_{\rm H_2}=10^{4.2}\,{\rm cm^{-3}}$, and $N_{\rm CO}/\Delta v=10^{17.0}\,{\rm cm^{-2}\, km^{-1}\, s}$ for the minimum $\chi^2$ model). For these models, the optical depth of the \mbox{CO(3--2)} and \mbox{CO(5--4)} lines increase by a factor of $\sim2$, while the $\tau_{1-0}$ increases for the minimum $\chi^2$ model to $2.5$ and stays the same for the maximum probability model. The marginalized probability distributions and contour plots for the three-dimensional probability distributions are shown in Figures~\ref{fig:redmarginalized}--\ref{fig:redcontours2} and are generally similar to those for the blue component. Based on the likelihood distributions and degeneracies exhibited over the full three-dimensional parameter space, the red component's density is poorly constrained, with $n_{\rm H_2}\gtrsim10^{4.0}\,{\rm cm^{-3}}$, but we see a reasonably consistent temperature and density, $T_{\rm kin}=20-35\,{\rm K}$ and  $N_{\rm CO}/\Delta v=10^{16.75\pm0.25}\,{\rm cm^{-2}\, km^{-1}\, s}$.

All of these models underproduce the \mbox{CO(5--4)} emission relative to the other lines. Forcing the models to better match the \mbox{CO(5--4)} emission (by removing the \mbox{CO(3--2)} line from the analysis) favors models where all emission lines are optically thin (with $\tau<0.5$) and dramatically over predicts the relative amount of \mbox{CO(3--2)} emission (by a factor of $>5$). Forcing these fits to use parameters that produce only optically thick emission recovers parameters similar to those of the CO SLEDs in Figure~\ref{fig:redsled}.

Based on the evidence for a spatially extended cold gas reservoir (as measured by the \mbox{CO(1--0)} and \mbox{CO(3--2)} source sizes along the major axis of the red component), we also consider a two-phase LVG model to try to better reproduce the red component's CO line ratios. Given the small number of line measurements relative to the number of parameters in the two-phase LVG analysis, and the number of degeneracies among model parameters, we can only constrain likely ranges of physical conditions as opposed to determining a single best-fit model. For the red component of J00266, we have also made use of several priors to restrict the allowed parameter space. For these models, in addition to the standard logarithmic prior for each parameter and the requirement $T_{\rm kin}>T_{\rm CMB}=10.2\,{\rm K}$, we also require the emission to be optically thick for transitions with $J_{\rm upper}\leq7$, the line brightness temperatures to be greater than $T_{\rm CMB}$, the CO column length to be less than the source diameter, and $K_{\rm vir}>0.01$. For the cut on allowed column length, we conservatively use the major axis from the red component \mbox{CO(3--2)} detection and do not take into account the filling fraction for each component. This prior also requires assuming a value of the CO-to-${\rm H_2}$ abundance, for which we use $x_{\rm CO}=10^{-4}$ as an upper limit. Using the measured FWHM of the red component, we find $N_{\rm CO}/\Delta v/n_{\rm H_2}<10^{15.5}\,{\rm cm\,km^{-1} s}$, which does not eliminate any single value of the ${\rm H_2}$ density or $N_{\rm CO}/\Delta v$ within the ranges probed, but does restrict combinations of those parameters. Similarly, we also choose a very conservative cut on the degree of virialization, $K_{\rm vir}>0.01$ (since a diffuse phase does not have to be virialized, and stricter limits on $K_{\rm vir}$ tend to limit ${\rm H_2}$ densities to unrealistic values), which eliminates $n_{\rm H_2}>10^{5.05}\,{\rm cm^{-3}}$. Since the requirement that defines the two phases, $T_{\rm c}<T_{\rm w}$, allows for \emph{nearly} identical temperatures and the same ranges in density and $N_{\rm CO}/\Delta v$, there are significant degeneracies between the preferred parameters of the two phases, leading to favored models effectively consistent with a single phase. Motivated by the arrangement proposed to explain the common $r_{3,1}\sim0.6$ seen in other SMGs---small clouds of high density/excitation molecular gas embedded within a larger volume filled with cold/low-density molecular gas \citep{harris2010,ivison2011}---we introduce an additional prior requiring that the cold-phase ${\rm H_2}$ density must be less than the warm-phase density. 

\begin{figure*}
\epsscale{1.0}
\plotone{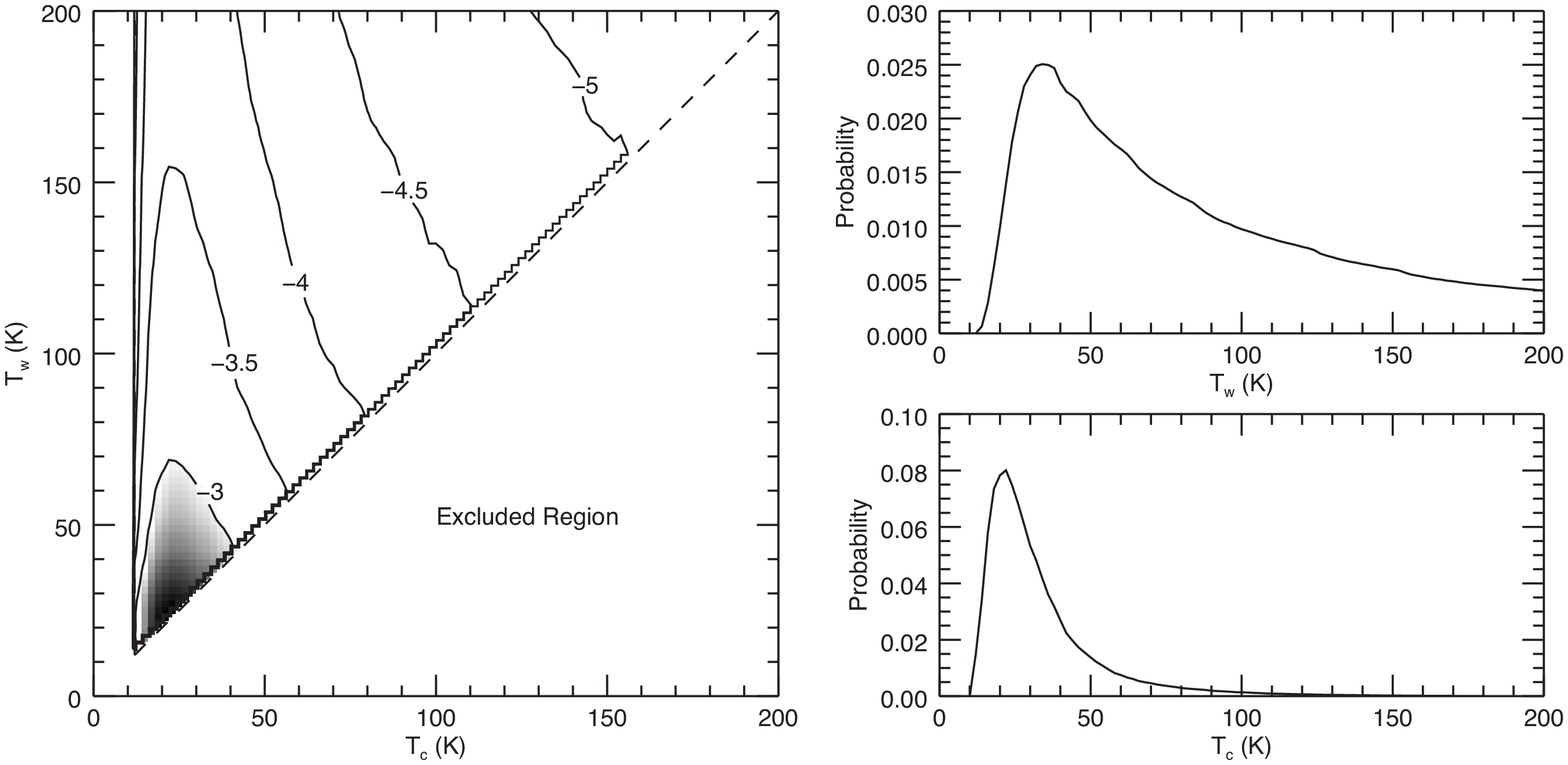}
\caption{Marginalized posterior probability distributions for the temperatures of the cold and warm-phase molecular gas in a two-phase model of the red component. The left panel show the marginalized distribution in the two-temperature plane with logarithmic shading/contours; the lower right half is eliminated by the requirement that $T_{\rm c}<T_{\rm w}$. The slight periodic jumps in the 1D probability distribution are an artifact of the interpolation of the collision coefficients between temperatures. \label{fig:lvg2_tvst}}
\end{figure*}

\begin{figure*}
\epsscale{1.0}
\plotone{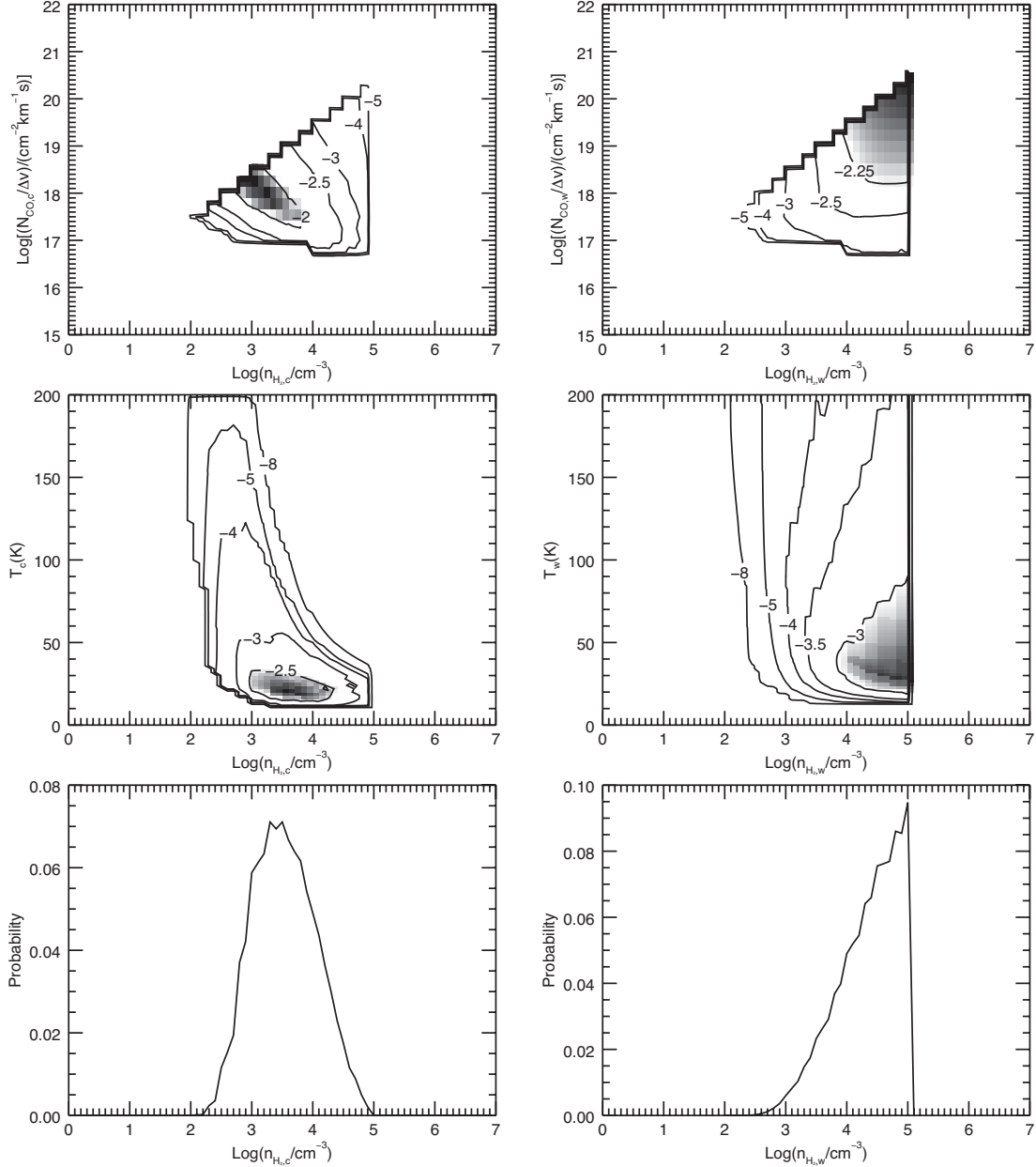}
\caption{Marginalized posterior probability distributions for the densities of the cold phase (left column) and warm phase (right column) molecular gas in a two-phase model of the red component. The marginalized probability distributions of the $N_{\rm CO}/\Delta v$-density plane (top row) and temperature-density plane (middle row) have logarithmic shading and contours. The probability distributions in the bottom row have been marginalized over all other parameters. \label{fig:lvg2_tvsd}}
\end{figure*}

\begin{figure*}
\epsscale{1.0}
\plotone{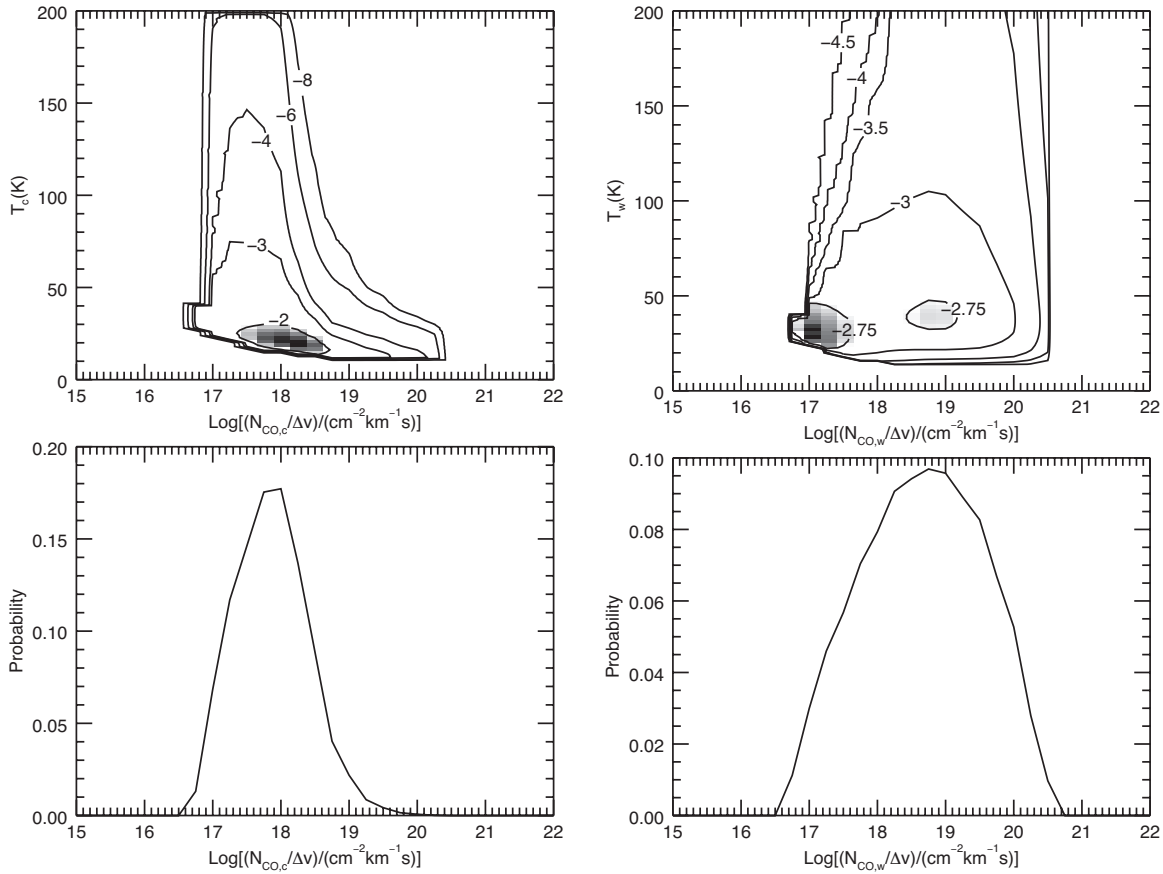}
\caption{Marginalized posterior probability distributions for the cold-phase $N_{\rm CO}/\Delta v$ (left column) and warm-phase $N_{\rm CO}/\Delta v$ (right column) in a two-phase model of the red component. The marginalized probability distributions of the temperature-$N_{\rm CO}/\Delta v$ plane (upper row) have logarithmic shading and contours. The probability distributions in the lower row have been marginalized over all other parameters. \label{fig:lvg2_tvsnco}}
\end{figure*}

\begin{figure*}
\plotone{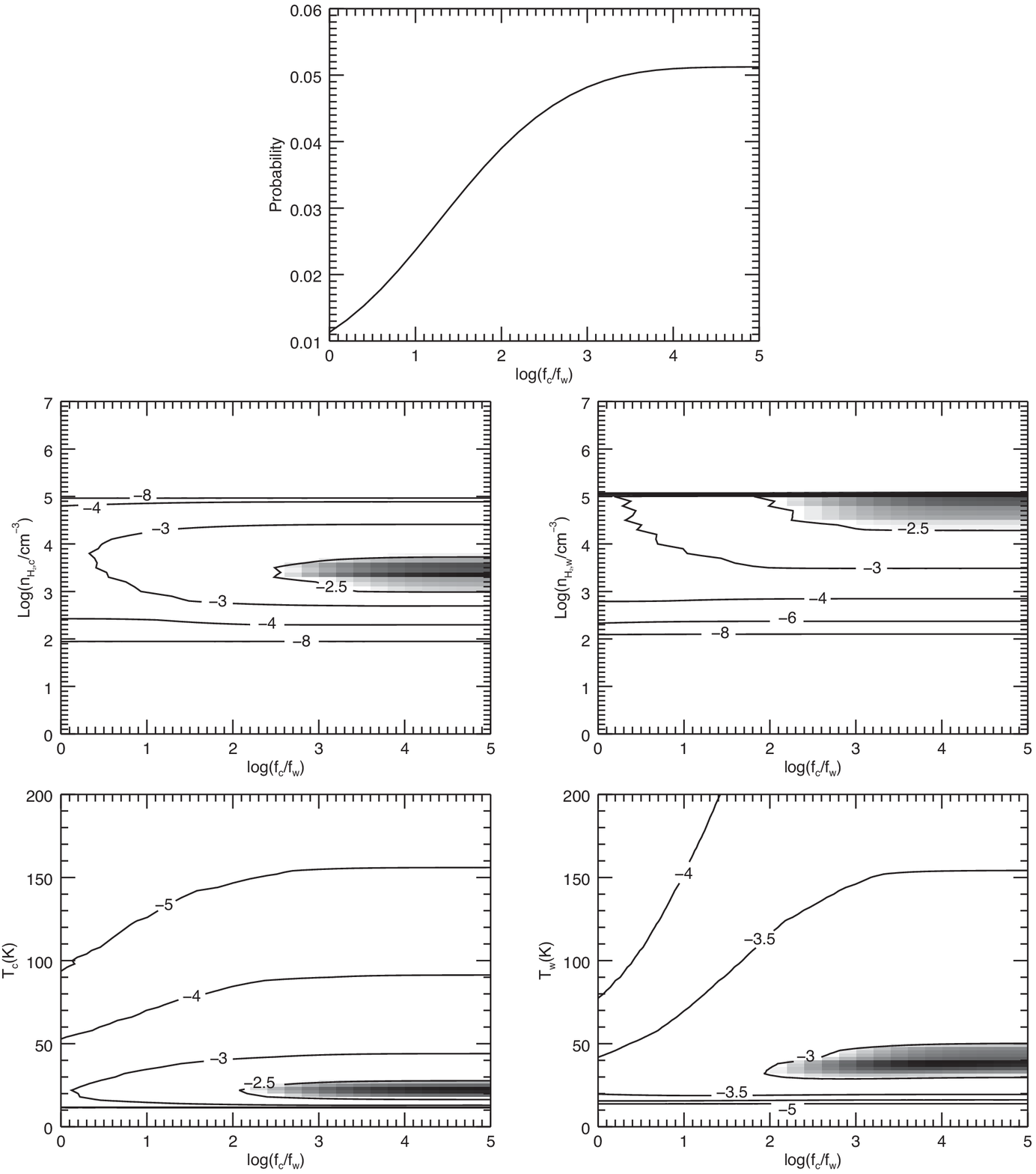}
\caption{Marginalized posterior probability distributions for the ratio of the cold and warm-phase filling fractions in a two-phase model of the red component. The marginalized probability distributions of the temperature-fraction plane (bottom row) and density-fraction plane (middle row) have logarithmic shading and contours.\label{fig:lvg2_frac}}
\end{figure*}

The results of the two-phase analysis are shown in Figures\,\ref{fig:lvg2_tvst}--\ref{fig:lvg2_frac}. Due to the large parameter space probed and the methods used, we are limited to analyzing marginalized distributions of LVG model parameters. The marginalized posterior probability distributions for the temperatures of the cold and warm-phase molecular gas are shown in Figure~\ref{fig:lvg2_tvst}. Both models are strongly peaked at the coldest temperatures, with the peak of the marginalized distribution of the cold phase at $T_{\rm c}=22\,{\rm K}$ and the peak for the warm phase at $T_{\rm w}=34\,{\rm K}$. Since the peak probability of the marginalized warm-phase temperature distribution is lower than the probability at the same temperature for the cold phase, it is clear that the warm-phase temperature is poorly constrained. It is more likely that the cold phase has $20\,{\rm K}<T_{\rm c}<40\,{\rm K}$ (consistent with $T_{\rm dust}=39\pm1\,{\rm K}$; \citealp{magnelli2012}) and the warm phase has some higher temperature. The cold and warm-phase marginalized distributions for $n_{\rm H_2}$, the temperature vs.~density plane, and the $N_{\rm CO}/\Delta v$ vs.~density plane are shown in Figure~\ref{fig:lvg2_tvsd}. The marginalized posterior probabilities for the densities of the two phases appear choppy due to the coarse sampling of $N_{\rm CO}/\Delta v$; marginalizing over $N_{\rm CO}/\Delta v$ sums distributions in the temperature-density plane that have the characteristic ``L"-shape seen in the single-phase models (Figure~\ref{fig:redcontours2}). Since the priors limit the ${\rm H_2}$ density range to $\sim10^2$--$10^5\,{\rm cm^{-3}}$, it is unsurprising that the additional prior of $n_{\rm H_2,\,c}<n_{\rm H_2,\,w}$ effectively divides this parameter range in half, with $n_{\rm H_2,\,c}\approx10^3$--$10^4\,{\rm cm^{-3}}$ and $n_{\rm H_2,\,w}\approx10^4$--$10^5\,{\rm cm^{-3}}$. For $N_{\rm CO}/\Delta v$ (Figure~\ref{fig:lvg2_tvsnco}), the cold-phase probability peaks at $\sim10^{18}\,{\rm cm^{-2}\,km^{-1}\,s}$ and the warm-phase probability peaks at densities a factor of $\sim10$ higher, although with a broader probability distribution. The LVG model can also constrain the ratio of the filling fractions (Figure~\ref{fig:lvg2_frac}). However, the degeneracies cause the probability distribution for the ratio of filling fractions to asymptote towards what is effectively a singe-phase model, with $f_{\rm c}/f_{\rm w}\gtrsim1000$. In summary, the two-phase LVG models do not provide any additional constraints on the gas physical conditions, and the red component's CO line ratios are adequately reproduced by single-phase models.

\subsection{CO to ${\rm H_2}$ abundance}

Given the most likely parameter ranges for the molecular gas in J00266 implied by the LVG analysis, we can determine the CO abundance factor relative to ${\rm H_2}$, $x_{\rm CO}=(N_{\rm CO}/\Delta v)\times(\Delta v/l)/n_{\rm H_2}$, where $l$ and $\Delta v$ are the globally determined average column length through the source and line width as estimated from our observations. For the two-phase model, either we can treat each phase separately, thus allowing for variations in the CO abundance, or we can create a weighted average using the filling factors:

\begin{equation}
x_{\rm CO}=\frac{f_{\rm c}N_{\rm c}/\Delta v + f_{\rm w}N_{\rm w}/\Delta v}{f_{\rm c}n_{\rm c} + f_{\rm w}n_{\rm w}}\left(\frac{\Delta v}{l}\right) .
\label{eq:xco2comp}
\end{equation}

For the blue component, which is consistent with a single-phase model, the range of best-fit models (maximum probability or minimum $\chi^2$, with or without the optical depth requirement) produce $x_{\rm CO}\sim10^{-9}$--$10^{-7}$. The single-phase red component models produce the same range of values. For both components, the models with the optical depth requirements tend to produce CO abundances at the upper end of this range. These abundances are nevertheless extremely low; typical assumptions elsewhere in the literature are $x_{\rm CO}\sim10^{-4}$. It is likely that the {\it global} velocity gradients of the two components do not accurately describe the kinematic structures within the system. The global velocity gradient may fail to describe the gas motions if the two components of J00266 possess smaller clouds that are not resolved by our observations. Unresolved velocity structure as implied by the unphysical results of analyzing LVG models with global velocity gradients has also been seen in another SMG, SMM\,J14011+0252 \citep{sharon2013}, which suggests that the failure of global velocity gradients to accurately describe gas physical conditions may be common among SMGs. For the two-phase LVG analysis of the red component, it is difficult to estimate the CO abundance due to the broad range of parameters allowed by the LVG analysis. Using the mean values of the ranges given in Section~\ref{sec:red}, and the minimum value of $f_{\rm c}/f_{\rm w}\sim1000$, we estimate $x_{\rm CO}\sim4\times10^{-4}$--$10^{-5}$. While this value is in line with canonical assumptions of the abundance, we do not ascribe much significance to this result (nor to its difference from the single-phase value) since parameter estimates from the marginalized distributions of our large and degenerate parameter space may not represent the true best-fit model (an effect which can even be seen in the marginalized distributions of the single-phase models; Figures~\ref{fig:bluemarginalized} and \ref{fig:redmarginalized}).

In some published LVG analyses, instead of using $N_{\rm CO}/\Delta v$ to quantify the crucial ``large velocity gradient" that simplifies analysis of optically thick emission, authors use $x_{\rm CO}/(\Delta v/\Delta r)$. The CO abundance per unit velocity gradient (which we effectively calculate above if we leave off the potentially problematic factor of the measured velocity gradient) is often assumed to be $10^{-5}\,{\rm (km\,s^{-1})^{-1}\,pc}$ \citep[e.\/g.\/,][]{riechers2006,weiss2007,aravena2010a}, since models that vary $x_{\rm CO}/(\Delta v/\Delta r)$ by an order of magnitude do not seem to produce very different line ratios \citep[e.\/g.\/,][]{weiss2005b,iono2006}. While this assumed value is within the ranges possible for the red and blue components, we note that fixing $x_{\rm CO}/(\Delta v/\Delta r)$ (or the alternative parameterization, $N_{\rm CO}/\Delta v$) and examining marginalized probability distributions ignores substantial degeneracies between the physical parameters used in the LVG model and can produce artificially narrow ranges in best-fit results. We note that for our best-fit single phase models, $x_{\rm CO}/(\Delta v/\Delta r)\sim10^{-5}$ is only achieved for optically thick models; the best-fit optically thin models have $x_{\rm CO}/(\Delta v/\Delta r)\sim10^{-7}$. Higher-resolution observations are necessary in order to accurately determine local velocity gradients in SMGs, which will allow LVG models to capture more realistic values and uncertainties for the physical conditions of the molecular gas.

The best fit models from the LVG analysis can also be used to constrain the CO-to-${\rm H_2}$ conversion factor, $\alpha_{\rm CO}$, using the model given in \cite{dickman1986}. For this model, $\alpha_{\rm CO}\propto n^{1/2}T^{-1}$ assuming that the molecular clouds are virialized and the CO emission is optically thick; this model produces near-unifrom values of $\alpha_{\rm CO}$ for a wide range of measured molecular cloud temperatures and densities in the local universe \citep[e.\/g.\/,][]{young1991}. Among the single-phase models considered, only those where we enforced $\tau_{7-6}>1$ (which also favored models with larger optical depths in the lower-$J$ CO lines) result in values of $\alpha_{\rm CO}$ close to the observed range of values ($\alpha_{\rm CO}\sim1$--$5$, depending on the assumed cloud geometry and density profile). While the large uncertainties in the LVG model parameters do no allow us to constrain the CO-to-${\rm H_2}$ conversion factor in J00266, it is reassuring that our models using the optical depth priors produce physically realistic (i.\/e.\/, virialized) molecular clouds as probed by $\alpha_{\rm CO}$. For the red component two-phase LVG models, the model parameter space has already been restricted to physically realistic molecular clouds by the virialization prior, which again yields $\alpha_{\rm CO}\sim1$--$5$.

\subsection{Star formation properties}

As most characterizations of J00266 made at other wavelengths have not resolved the source (spatially or spectrally), it is difficult to derive separate parameters for star formation in its two components. If we assume FIR luminosity traces star formation (determined from the total IR luminosity in \citealt{magnelli2012} scaled to match our magnification and redshift), and that the star formation rate is proportional to the gas mass with a power law index of $n\sim1$--$1.5$, we can use the ratio of the \mbox{CO(1--0)} line fluxes to partition the global star formation rate (SFR) between the two components. Assuming the \citet{kennicutt1998} conversion factor (corrected to the initial mass function adopted by \citet{calzetti2007}), and our new magnification factor, the SFRs of the blue and red components are $200$ and $340\,M_\sun\,{\rm yr^{-1}}$, respectively (for $n=1$), or $170$ and $370\,M_\sun\,{\rm yr^{-1}}$ for $n=1.5$. Using the source sizes and gas masses of the two components, we can place the two components of J00266 on the Schmidt-Kennicutt relation \citep{schmidt1959, kennicutt1998} using average values of the SFR surface density and molecular gas surface density. For the both components we assume the de-lensed source sizes given by the Gaussian fits to the \mbox{CO(3--2)} emission. The blue component average SFR surface density is $5.3\,M_\sun\,{\rm yr^{-1}\,kpc^{-2}}$ ($n=1$) or $4.5\,M_\sun\,{\rm yr^{-1}\,kpc^{-2}}$ ($n=1.5$), and the gas mass surface density is $10^{2.5}(\alpha_{\rm CO}/0.8)\,M_\sun\,{\rm pc^{-2}}$. The red component SFR surface density is approximately a factor of two larger: $12$--$13\,M_\sun\,{\rm yr^{-1}\,kpc^{-2}}$ for $n=1.0$--$1.5$. The average gas mass surface density for the red component is $10^{2.9}(\alpha_{\rm CO}/0.8)\,M_\sun\,{\rm pc^{-2}}$. Both the blue and red components of J00266 fall among the starburst galaxies of \citet{kennicutt1998}, regardless of our choice of star-formation law index. The components also remain among the starburst galaxies for choices of $\alpha_{\rm CO}$ within the ranges typically assumed for star forming galaxies ($\alpha_{\rm CO}=0.8$--$4.6\,M_\sun\,{\rm (K\,km\,s^{-1}\,pc^2)^{-1}}$). The apparent division between the ``starburst sequence" and the ``disk galaxy sequence" in the Schmidt-Kennicutt relation \citep[e.\/g.\/,][]{daddi2010,genzel2010} has been argued to be an artificial product of the bi-modal choice in $\alpha_{\rm CO}$ for those two populations \citep[e.\/g.\/,][]{narayanan2012}. However, the position of the two components of J00266 among the starburst sequence of galaxies appears to be robust to reasonable choices in conversion factor. J00266 also remains within the scatter of the starburst sequence for $\alpha_{\rm CO}=0.8$--$4.6\,M_\sun\,{\rm (K\,km\,s^{-1}\,pc^2)^{-1}}$ if we do not partition the SFR between components and instead treat J00266 as a single galaxy; in this case, the SFR surface density is $14\,M_\sun\,{\rm yr^{-1}\,kpc^{-2}}$ and gas mass surface density is $925(\alpha_{\rm CO}/0.8)\,M_\sun\,{\rm pc^{-2}}$.

\section{Discussion}
\label{sec:discuss}

Interpretation of the kinematics in low-resolution CO observations has made answering the question of SMGs' structure (i.\/e.\/, merger-driven starbursts or star-forming disks) difficult; velocity gradients and double-peaked line profiles could arise from either ordered rotation or a merger. \citet{hayward2013} propose that submillimeter-bright galaxies are a heterogeneous sample containing disks, galaxy pairs (coincidently caught within the large beams of most single-dish submillimeter telescopes), and mergers \citep[see also][]{wang2011,barger2012,hodge2013}. They find that the relative contribution of each population to the number counts varies with (sub)millimeter flux, with mergers and galaxy pairs dominating at the bright end. J00266, a merging system with $S_{850\,{\rm \mu m}}=7.7\pm1.0\,{\rm mJy}$ (magnification-corrected), is in line with these predictions. However, the SMG GN20 has a significantly larger submillimeter flux \citep[$20.3\pm2.1\,{\rm mJy}$;][]{pope2006}, well into the regime where mergers are expected to dominate the population, and is a massive ($M_{\rm H_2}=1.3\pm0.4\times10^{11}(\alpha_{\rm CO}/0.8)\,M_\sun$) $z=4.05$ disk galaxy \citep{hodge2012}. The contrasting structures of J00266 and GN20 highlight the heterogeneity of the SMG population, but muddy evolutionary explanations of large FIR luminosities.

J00266 also illustrates the importance of obtaining kinematic information for addressing the multiplicity of SMGs. \citet{hodge2013} undertook a high-resolution continuum followup of the LABOCA ECDFS Submillimeter Survey \citep[LESS;][]{weiss2009} sources with the Atacama Large Millimeter/submillimeter Array (ALMA), and found that $\sim35$--$45\%$ of detected LESS sources were resolved into multiple SMGs (where the LESS beam FWHM is $19.^{\prime\prime}2$ and the median ALMA beam FWHM is $\sim1.^{\prime\prime}5$; see also \citealt{barger2012}). However, at the resolution of the ALMA LESS observations, J00266 would \emph{not} have been detected as a multiple-component system. Assuming that the distribution of the submillimeter continuum emission matches that of the CO lines, the spatially extended red component of J00266 overlaps the blue component in the plane of the sky, and \emph{only} the additional velocity information provided by spectral line observations allows us to distinguish between the two galaxies. While many of the SMGs discovered in the SLS, including J00266, have extensive multi-wavelength followup \citep[e.\/g.\/,][]{smail1998,smail1999,ivison2000,smail2002}, significantly fewer have the high-resolution spatial \emph{and} spectroscopic observations necessary to address multiplicity. Given one example (J00266) in the 15 SLS galaxies, we can naively estimate that a minimum of $\gtrsim7\%$ of SMGs are closely aligned galaxy pairs that require spectroscopic information to identify the two components (in \emph{addition} to sources that are multiples in higher resolution continuum observations). In some instances it may be possible to use a combination of lower-resolution CO mapping that is sufficient to detect (but not resolve) velocity gradients and higher-resolution continuum imaging to determine if a SMG is a closely aligned galaxy pair/merger (e.\/g.\/, SMM\,J04331+0210; \citealt{aguirre2013}). Therefore, while higher-resolution continuum observations with ALMA and other interferometers certainly help address source confusion in submillimeter single-dish surveys, spectral line observations are still necessary to fully characterize galaxy pairs that are very nearly aligned with our line of sight and/or closely interacting.

In the growing sample of well-studied SMGs, close examination of merging SMGs \citep[e.\/g.\/][]{tacconi2008,engel2010,riechers2011c,riechers2013a} may help us understand the origin of the average population characteristics and possibly the evolutionary connection between SMGs and other high-redshift galaxies. \citet{riechers2011c} report \mbox{CO(1--0)} observations of a merging SMG (SMM\,J123707+6214). As for J00266, the two components of SMM\,J123707+6214 differ in terms of line width and velocity structure. However, the line ratios of the two components are similar, with \mbox{CO(3--2)}/\mbox{CO(1--0)} line ratios significantly below the SMG average ($r_{3,1}=0.39\pm0.09$ and $0.37\pm0.10$, more in line with $z\sim1.5$ BzK galaxies; e.\/g.\/, \citealt{dannerbauer2009,aravena2010a}). \citet{riechers2013a} reports a candidate ``wet-dry" merger between an SMG and a quasar, where the molecular gas emission from the quasar host galaxy has not been detected and the galaxy pair is widely separated compared to the component separation in J00266 ($41.5\,{\rm kpc}$ in projection and $700\,{\rm km\,s^{-1}}$ in velocity). However, the ``wet" SMG has $r_{3,1}=0.82\pm0.15$, a line ratio similar to the red component of J00266. Of the small number of SLS SMGs that have high-resolution multi-$J$ CO line observations sufficient to address both merger state and excitation (J00266, SMM\,J02399-0136, and SMM\,J14011+0252; \citealt{frayer1999,downes2003,genzel2003,ivison2010a,sharon2013}) two thirds are identified as mergers and the three systems span $r_{3,1}\sim0.5$--$1$.

The potential correspondence of different kinematics with different ISM conditions in J00266 is suggestive; disk galaxies at other redshifts seem to have a multi-phase molecular ISM, while quasar host galaxies might be expected to have funneled all their gas into a compact central region (making velocity gradients difficult to resolve) that is well-described by a single phase. The \citet{valiante2007} {\it Spitzer}/IRS spectrum of J00266 contains signatures of both AGN (mid-IR continuum) and star formation (PAH emission), as could arise if J00266 comprises a merger between two galaxies of different evolutionary states. However, the low spectral resolution of the {\it Spitzer}/IRS spectrum in \citet{valiante2007} does not allow us to use the velocity offset in the two components to distinguish between their contributions to the $7.7\,{\rm \mu m}$ PAH feature. A potential connection between gas excitation and dynamics (or AGN presence) is weakened by observations of dusty star-forming galaxies at both low and high redshifts. At low redshift, galaxies with ordered motion and $r_{3,1}\sim1$ that lack an AGN have been observed among local IR-bright galaxies \citep[e.\/g.\/,][]{mauersberger1999,yao2003}, making a strict gas excitation-dynamics (or excitation-AGN) connection unlikely even though galaxies selected for U/LIRG luminosities at low- and high-redshift may represent different populations \citep[like major mergers and ``normal" disk galaxies; e.\/g.\/,][]{genzel2010,saintonge2013}. At high redshift, SMM\,J14011+0252 appears to be an exception to many SMG trends, including a potential excitation-dynamics connection, since it exhibits a velocity gradient (albeit a weak one), $r_{3,1}\sim1$, and no AGN \citep{sharon2013}.

J00266 is also unique in that it has extremely high obscuration; it is fainter in $K$ band than all other IRAM CO-detected SMGs in the sample of \citet{frayer2004}. High obscuration is consistent with simulations that indicate SMGs are especially dust-bright during major mergers \citep[if viewed in the correct orientation; e.\/g.\/,][]{narayanan2010}. High obscuration caused by dust screen geometry is also consistent with its more moderate FIR luminosity \citep{magnelli2012} compared to other SMGs. High dust obscuration caused by dust screen geometry supports our argument that J00266 may be a merger (regardless of it being two components or a single compact source) rather than J00266 belonging to the population of $z\sim2$ dusty but UV-detected disk galaxies which are significantly less dust-obscured \citep[e.\/g.\/,][]{genzel2010,saintonge2013,tacconi2013}. It may be that the relatively high-excitation CO properties of J00266 ($r_{3,1}\sim0.8$--$1.0$) only appear unique (relative to the average $r_{3,1}\sim0.6$ observed in SMGs) because it belongs to a class of heavily obscured objects that would not have been detected in CO studies that rely on optical counterpart identification \citep[e.\/g.\/,][]{chapman2005}, and that have thus biased our conclusions about the SMG population. Testing this hypothesis requires followup CO observations of submillimeter-bright objects that have not relied on radio and optical counterparts for their redshift measurements \citep[e.\/g.\/,][]{harris2012,vieira2013}, or more deliberate followup of SMGs for which optical spectroscopy fails to measure the redshift \citep[e.\/g.\/,][]{chapman2005}.

\section{Conclusions}
\label{sec:summary}

We present CO observations of the SMG SMM\,J00266+1708 and provide the first spectroscopic measurement of its redshift ($z=2.742$). We find that J00266 has two distinct components separated by $\sim 500\,{\rm km\,s^{-1}}$ and closely aligned along our line of sight. The two components of J00266 appear to have different kinematics, with the blue-shifted component (relative to the line center) having dispersion-dominated kinematics and the red-shifted component showing a velocity gradient. The larger spatial extent of the \mbox{CO(1--0)} emission relative to the \mbox{CO(3--2)} emission along the red-shifted component's major axis (parallel to the velocity gradient) indicates the presence of multiple phases of molecular gas. Based on the components' differing kinematics and line ratios, we conclude that the SMG J00266 is likely undergoing a merger with a mass ratio of $(7.8\pm4.0)/{\sin}^2(i)$ for the two components.

We have performed LVG analyses of the CO line ratios in order to constrain the physical conditions of the two components. Based on the likelihood distributions and degeneracies exhibited over the full three-dimensional parameter space, we find that the line ratios of the blue component are consistent with a single-phase molecular ISM with $T_{\rm kin}=15-50\,{\rm K}$, $n_{\rm H_2}\gtrsim10^{3.5}\,{\rm cm^{-3}}$, and $N_{\rm CO}/\Delta v=10^{16.5\pm0.5}\,{\rm cm^{-2}\, km^{-1}\, s}$. While the red component is also consistent with a single phase ISM ($T_{\rm kin}=20-35\,{\rm K}$, $n_{\rm H_2}\gtrsim10^{4.0}\,{\rm cm^{-3}}$, and $N_{\rm CO}/\Delta v=10^{16.75\pm0.25}\,{\rm cm^{-2}\, km^{-1}\, s}$), we also attempt a multi-phase LVG analysis as motivated by its spatially extended \mbox{CO(1--0)} emission relative to its \mbox{CO(3--2)} emission and the multi-phase ISM seen in other SMGs. The limited number of line measurements (three plus an additional upper limit) relative to the number of free parameters (seven) leads to results that are highly degenerate. We constrain the cold phase temperature to be between $20$--$40\,{\rm K}$ and the warm phase to be at some higher temperature. We find that the ${\rm H_2}$ density of the cold phase is $n_{\rm H_2,\,c}\approx10^3$--$10^4\,{\rm cm^{-3}}$ and the warm phase has $n_{\rm H_2,\,w}\approx10^4$--$10^5\,{\rm cm^{-3}}$. We also find $N_{\rm CO,\,c}/\Delta v\sim10^{18}\,{\rm cm^{-2}\,km^{-1}\,s}$, while the warm phase probability peaks at columns a factor of $\sim10$ greater. The ratio of the cold-to-warm phase filling factors is constrained to be $f_{\rm c}/f_{\rm w}\gtrsim1000$ (consistent with the fact that the red component line ratios can be reproduced by single-phase LVG models). While the results of the LVG modeling do not put strong constraints on the physical conditions of the molecular gas in J00266, these results are useful as a compact form of quantifying the CO SLED shape (and uncertainty) which facilitates comparisons to other galaxies.

CO excitation is also a useful way to distinguish physical structures that are (nearly) unresolved. In this case, the different excitation supports the argument that J00266 is a merger, helping to provide new confirmation that at least some SMGs are ongoing mergers. The fact that the two components of J00266 would not be identified in high-resolution continuum studies of SMGs also illustrates the value of additional velocity information provided by spectral line studies, important for identifying mergers or separating SMG pairs that overlap in the plane of the sky. Our observations of J00266 are suggestive of an early stage merger between a dispersion-dominated quasar host galaxy and a rotating disk galaxy with a multiphase ISM. However, the diversity of high-redshift galaxies with characteristics similar to those of J00266 (even among the small number with comparably detailed observations) significantly weakens any conclusions about connections between gas excitation, kinematics, and merging state.

\acknowledgments{We thank an anonymous referee for helpful comments. This research has been supported by NSF grants AST-0708653 and AST-0955810. CES held an American Fellowship from AAUW for part of this work. The National Radio Astronomy Observatory is a facility of the National Science Foundation operated under cooperative agreement by Associated Universities, Inc. The Submillimeter Array is a joint project between the Smithsonian Astrophysical Observatory and the Academia Sinica Institute of Astronomy and Astrophysics and is funded by the Smithsonian Institution and the Academia Sinica.}

{\it Facilities:} \facility{GBT (Zpectrometer)}, \facility{IRAM:Interferometer}, \facility{SMA}, \facility{VLA}


\begin{thebibliography}{123}
\expandafter\ifx\csname natexlab\endcsname\relax\def\natexlab#1{#1}\fi

\bibitem[{{Aguirre} {et~al.}(2013){Aguirre}, {Baker}, {Menanteau}, {Lutz}, \&
  {Tacconi}}]{aguirre2013}
{Aguirre}, P., {Baker}, A.~J., {Menanteau}, F., {Lutz}, D., \& {Tacconi}, L.~J.
  2013, \apj, 768, 164

\bibitem[{{Alexander} {et~al.}(2003){Alexander}, {Bauer}, {Brandt},
  {Hornschemeier}, {Vignali}, {Garmire}, {Schneider}, {Chartas}, \&
  {Gallagher}}]{alexander2003}
{Alexander}, D.~M., {Bauer}, F.~E., {Brandt}, W.~N., {et~al.} 2003, \aj, 125,
  383

\bibitem[{{Aravena} {et~al.}(2010){Aravena}, {Carilli}, {Daddi}, {Wagg},
  {Walter}, {Riechers}, {Dannerbauer}, {Morrison}, {Stern}, \&
  {Krips}}]{aravena2010a}
{Aravena}, M., {Carilli}, C., {Daddi}, E., {et~al.} 2010, \apj, 718, 177

\bibitem[{{Barger} {et~al.}(1998){Barger}, {Cowie}, {Sanders}, {Fulton},
  {Taniguchi}, {Sato}, {Kawara}, \& {Okuda}}]{barger1998}
{Barger}, A.~J., {Cowie}, L.~L., {Sanders}, D.~B., {et~al.} 1998, \nat, 394,
  248

\bibitem[{{Barger} {et~al.}(2012){Barger}, {Wang}, {Cowie}, {Owen}, {Chen}, \&
  {Williams}}]{barger2012}
{Barger}, A.~J., {Wang}, W.-H., {Cowie}, L.~L., {et~al.} 2012, \apj, 761, 89

\bibitem[{{Berta} {et~al.}(2010){Berta}, {Magnelli}, {Lutz}, {Altieri},
  {Aussel}, {Andreani}, {Bauer}, {Bongiovanni}, {Cava}, {Cepa}, {Cimatti},
  {Daddi}, {Dominguez}, {Elbaz}, {Feuchtgruber}, {F{\"o}rster Schreiber},
  {Genzel}, {Gruppioni}, {Katterloher}, {Magdis}, {Maiolino}, {Nordon},
  {P{\'e}rez Garc{\'{\i}}a}, {Poglitsch}, {Popesso}, {Pozzi}, {Riguccini},
  {Rodighiero}, {Saintonge}, {Santini}, {Sanchez-Portal}, {Shao}, {Sturm},
  {Tacconi}, {Valtchanov}, {Wetzstein}, \& {Wieprecht}}]{berta2010}
{Berta}, S., {Magnelli}, B., {Lutz}, D., {et~al.} 2010, \aap, 518, L30

\bibitem[{{Berta} {et~al.}(2011){Berta}, {Magnelli}, {Nordon}, {Lutz}, {Wuyts},
  {Altieri}, {Andreani}, {Aussel}, {Casta{\~n}eda}, {Cepa}, {Cimatti}, {Daddi},
  {Elbaz}, {F{\"o}rster Schreiber}, {Genzel}, {Le Floc'h}, {Maiolino},
  {P{\'e}rez-Fournon}, {Poglitsch}, {Popesso}, {Pozzi}, {Riguccini},
  {Rodighiero}, {Sanchez-Portal}, {Sturm}, {Tacconi}, \&
  {Valtchanov}}]{berta2011}
{Berta}, S., {Magnelli}, B., {Nordon}, R., {et~al.} 2011, \aap, 532, A49

\bibitem[{{B{\'e}thermin} {et~al.}(2012){B{\'e}thermin}, {Le Floc'h}, {Ilbert},
  {Conley}, {Lagache}, {Amblard}, {Arumugam}, {Aussel}, {Berta}, {Bock},
  {Boselli}, {Buat}, {Casey}, {Castro-Rodr{\'{\i}}guez}, {Cava}, {Clements},
  {Cooray}, {Dowell}, {Eales}, {Farrah}, {Franceschini}, {Glenn}, {Griffin},
  {Hatziminaoglou}, {Heinis}, {Ibar}, {Ivison}, {Kartaltepe}, {Levenson},
  {Magdis}, {Marchetti}, {Marsden}, {Nguyen}, {O'Halloran}, {Oliver}, {Omont},
  {Page}, {Panuzzo}, {Papageorgiou}, {Pearson}, {P{\'e}rez-Fournon}, {Pohlen},
  {Rigopoulou}, {Roseboom}, {Rowan-Robinson}, {Salvato}, {Schulz}, {Scott},
  {Seymour}, {Shupe}, {Smith}, {Symeonidis}, {Trichas}, {Tugwell}, {Vaccari},
  {Valtchanov}, {Vieira}, {Viero}, {Wang}, {Xu}, \& {Zemcov}}]{bethermin2012}
{B{\'e}thermin}, M., {Le Floc'h}, E., {Ilbert}, O., {et~al.} 2012, \aap, 542,
  A58

\bibitem[{{Blain} {et~al.}(2004){Blain}, {Chapman}, {Smail}, \&
  {Ivison}}]{blain2004}
{Blain}, A.~W., {Chapman}, S.~C., {Smail}, I., \& {Ivison}, R. 2004, \apj, 611,
  725

\bibitem[{{Bothwell} {et~al.}(2013){Bothwell}, {Smail}, {Chapman}, {Genzel},
  {Ivison}, {Tacconi}, {Alaghband-Zadeh}, {Bertoldi}, {Blain}, {Casey}, {Cox},
  {Greve}, {Lutz}, {Neri}, {Omont}, \& {Swinbank}}]{bothwell2013}
{Bothwell}, M.~S., {Smail}, I., {Chapman}, S.~C., {et~al.} 2013, \mnras, 429,
  3047

\bibitem[{{Calzetti} {et~al.}(2007){Calzetti}, {Kennicutt}, {Engelbracht},
  {Leitherer}, {Draine}, {Kewley}, {Moustakas}, {Sosey}, {Dale}, {Gordon},
  {Helou}, {Hollenbach}, {Armus}, {Bendo}, {Bot}, {Buckalew}, {Jarrett}, {Li},
  {Meyer}, {Murphy}, {Prescott}, {Regan}, {Rieke}, {Roussel}, {Sheth}, {Smith},
  {Thornley}, \& {Walter}}]{calzetti2007}
{Calzetti}, D., {Kennicutt}, R.~C., {Engelbracht}, C.~W., {et~al.} 2007, \apj,
  666, 870

\bibitem[{{Carilli} {et~al.}(2002){Carilli}, {Cox}, {Bertoldi}, {Menten},
  {Omont}, {Djorgovski}, {Petric}, {Beelen}, {Isaak}, \&
  {McMahon}}]{carilli2002}
{Carilli}, C.~L., {Cox}, P., {Bertoldi}, F., {et~al.} 2002, \apj, 575, 145

\bibitem[{{Carilli} {et~al.}(2010){Carilli}, {Daddi}, {Riechers}, {Walter},
  {Weiss}, {Dannerbauer}, {Morrison}, {Wagg}, {Dav{\'e}}, {Elbaz}, {Stern},
  {Dickinson}, {Krips}, \& {Aravena}}]{carilli2010}
{Carilli}, C.~L., {Daddi}, E., {Riechers}, D., {et~al.} 2010, \apj, 714, 1407

\bibitem[{{Cecchi-Pestellini} {et~al.}(2002){Cecchi-Pestellini}, {Bodo},
  {Balakrishnan}, \& {Dalgarno}}]{cecchi2002}
{Cecchi-Pestellini}, C., {Bodo}, E., {Balakrishnan}, N., \& {Dalgarno}, A.
  2002, \apj, 571, 1015

\bibitem[{{Chapman} {et~al.}(2005){Chapman}, {Blain}, {Smail}, \&
  {Ivison}}]{chapman2005}
{Chapman}, S.~C., {Blain}, A.~W., {Smail}, I., \& {Ivison}, R.~J. 2005, \apj,
  622, 772

\bibitem[{{Conselice} {et~al.}(2003){Conselice}, {Chapman}, \&
  {Windhorst}}]{conselice2003}
{Conselice}, C.~J., {Chapman}, S.~C., \& {Windhorst}, R.~A. 2003, \apjl, 596,
  L5

\bibitem[{{Coppin} {et~al.}(2010){Coppin}, {Pope}, {Men{\'e}ndez-Delmestre},
  {Alexander}, {Dunlop}, {Egami}, {Gabor}, {Ibar}, {Ivison}, {Austermann},
  {Blain}, {Chapman}, {Clements}, {Dunne}, {Dye}, {Farrah}, {Hughes},
  {Mortier}, {Page}, {Rowan-Robinson}, {Scott}, {Simpson}, {Smail}, {Swinbank},
  {Vaccari}, \& {Yun}}]{coppin2010}
{Coppin}, K., {Pope}, A., {Men{\'e}ndez-Delmestre}, K., {et~al.} 2010, \apj,
  713, 503

\bibitem[{{Daddi} {et~al.}(2010){Daddi}, {Elbaz}, {Walter}, {Bournaud},
  {Salmi}, {Carilli}, {Dannerbauer}, {Dickinson}, {Monaco}, \&
  {Riechers}}]{daddi2010}
{Daddi}, E., {Elbaz}, D., {Walter}, F., {et~al.} 2010, \apjl, 714, L118

\bibitem[{{Danielson} {et~al.}(2011){Danielson}, {Swinbank}, {Smail}, {Cox},
  {Edge}, {Weiss}, {Harris}, {Baker}, {De Breuck}, {Geach}, {Ivison}, {Krips},
  {Lundgren}, {Longmore}, {Neri}, \& {Flaquer}}]{danielson2011}
{Danielson}, A.~L.~R., {Swinbank}, A.~M., {Smail}, I., {et~al.} 2011, \mnras,
  410, 1687

\bibitem[{{Dannerbauer} {et~al.}(2009){Dannerbauer}, {Daddi}, {Riechers},
  {Walter}, {Carilli}, {Dickinson}, {Elbaz}, \& {Morrison}}]{dannerbauer2009}
{Dannerbauer}, H., {Daddi}, E., {Riechers}, D.~A., {et~al.} 2009, \apjl, 698,
  L178

\bibitem[{{Dav{\'e}} {et~al.}(2010){Dav{\'e}}, {Finlator}, {Oppenheimer},
  {Fardal}, {Katz}, {Kere{\v s}}, \& {Weinberg}}]{dave2010}
{Dav{\'e}}, R., {Finlator}, K., {Oppenheimer}, B.~D., {et~al.} 2010, \mnras,
  404, 1355

\bibitem[{{Devlin} {et~al.}(2009){Devlin}, {Ade}, {Aretxaga}, {Bock}, {Chapin},
  {Griffin}, {Gundersen}, {Halpern}, {Hargrave}, {Hughes}, {Klein}, {Marsden},
  {Martin}, {Mauskopf}, {Moncelsi}, {Netterfield}, {Ngo}, {Olmi}, {Pascale},
  {Patanchon}, {Rex}, {Scott}, {Semisch}, {Thomas}, {Truch}, {Tucker},
  {Tucker}, {Viero}, \& {Wiebe}}]{devlin2009}
{Devlin}, M.~J., {Ade}, P.~A.~R., {Aretxaga}, I., {et~al.} 2009, \nat, 458, 737

\bibitem[{{Dickman} {et~al.}(1986){Dickman}, {Snell}, \&
  {Schloerb}}]{dickman1986}
{Dickman}, R.~L., {Snell}, R.~L., \& {Schloerb}, F.~P. 1986, \apj, 309, 326

\bibitem[{{Downes} \& {Solomon}(1998)}]{downes1998}
{Downes}, D., \& {Solomon}, P.~M. 1998, \apj, 507, 615

\bibitem[{{Downes} \& {Solomon}(2003)}]{downes2003}
{Downes}, D., \& {Solomon}, P.~M. 2003, \apj, 582, 37

\bibitem[{{Engel} {et~al.}(2010){Engel}, {Tacconi}, {Davies}, {Neri}, {Smail},
  {Chapman}, {Genzel}, {Cox}, {Greve}, {Ivison}, {Blain}, {Bertoldi}, \&
  {Omont}}]{engel2010}
{Engel}, H., {Tacconi}, L.~J., {Davies}, R.~I., {et~al.} 2010, \apj, 724, 233

\bibitem[{{Fixsen} {et~al.}(1998){Fixsen}, {Dwek}, {Mather}, {Bennett}, \&
  {Shafer}}]{fixsen1998}
{Fixsen}, D.~J., {Dwek}, E., {Mather}, J.~C., {Bennett}, C.~L., \& {Shafer},
  R.~A. 1998, \apj, 508, 123

\bibitem[{{Frayer} {et~al.}(2004){Frayer}, {Reddy}, {Armus}, {Blain},
  {Scoville}, \& {Smail}}]{frayer2004}
{Frayer}, D.~T., {Reddy}, N.~A., {Armus}, L., {et~al.} 2004, \aj, 127, 728

\bibitem[{{Frayer} {et~al.}(2000){Frayer}, {Smail}, {Ivison}, \&
  {Scoville}}]{frayer2000}
{Frayer}, D.~T., {Smail}, I., {Ivison}, R.~J., \& {Scoville}, N.~Z. 2000, \aj,
  120, 1668

\bibitem[{{Frayer} {et~al.}(1999){Frayer}, {Ivison}, {Scoville}, {Evans},
  {Yun}, {Smail}, {Barger}, {Blain}, \& {Kneib}}]{frayer1999}
{Frayer}, D.~T., {Ivison}, R.~J., {Scoville}, N.~Z., {et~al.} 1999, \apjl, 514,
  L13

\bibitem[{{Frayer} {et~al.}(2011){Frayer}, {Harris}, {Baker}, {Ivison},
  {Smail}, {Negrello}, {Maddalena}, {Aretxaga}, {Baes}, {Birkinshaw},
  {Bonfield}, {Burgarella}, {Buttiglione}, {Cava}, {Clements}, {Cooray},
  {Dannerbauer}, {Dariush}, {De Zotti}, {Dunlop}, {Dunne}, {Dye}, {Eales},
  {Fritz}, {Gonzalez-Nuevo}, {Herranz}, {Hopwood}, {Hughes}, {Ibar}, {Jarvis},
  {Lagache}, {Leeuw}, {Lopez-Caniego}, {Maddox}, {Micha{\l}owski}, {Omont},
  {Pohlen}, {Rigby}, {Rodighiero}, {Scott}, {Serjeant}, {Smith}, {Swinbank},
  {Temi}, {Thompson}, {Valtchanov}, {van der Werf}, \& {Verma}}]{frayer2011}
{Frayer}, D.~T., {Harris}, A.~I., {Baker}, A.~J., {et~al.} 2011, \apjl, 726,
  L22

\bibitem[{{Genzel} {et~al.}(2003){Genzel}, {Baker}, {Tacconi}, {Lutz}, {Cox},
  {Guilloteau}, \& {Omont}}]{genzel2003}
{Genzel}, R., {Baker}, A.~J., {Tacconi}, L.~J., {et~al.} 2003, \apj, 584, 633

\bibitem[{{Genzel} {et~al.}(2001){Genzel}, {Tacconi}, {Rigopoulou}, {Lutz}, \&
  {Tecza}}]{genzel2001}
{Genzel}, R., {Tacconi}, L.~J., {Rigopoulou}, D., {Lutz}, D., \& {Tecza}, M.
  2001, \apj, 563, 527

\bibitem[{{Genzel} {et~al.}(2010){Genzel}, {Tacconi}, {Gracia-Carpio},
  {Sternberg}, {Cooper}, {Shapiro}, {Bolatto}, {Bouch{\'e}}, {Bournaud},
  {Burkert}, {Combes}, {Comerford}, {Cox}, {Davis}, {Schreiber},
  {Garcia-Burillo}, {Lutz}, {Naab}, {Neri}, {Omont}, {Shapley}, \&
  {Weiner}}]{genzel2010}
{Genzel}, R., {Tacconi}, L.~J., {Gracia-Carpio}, J., {et~al.} 2010, \mnras,
  407, 2091

\bibitem[{{Goldreich} \& {Kwan}(1974)}]{goldreich1974}
{Goldreich}, P., \& {Kwan}, J. 1974, \apj, 189, 441

\bibitem[{{Goldsmith}(2001)}]{goldsmith2001}
{Goldsmith}, P.~F. 2001, \apj, 557, 736

\bibitem[{{Granato} {et~al.}(2001){Granato}, {Silva}, {Monaco}, {Panuzzo},
  {Salucci}, {De Zotti}, \& {Danese}}]{granato2001}
{Granato}, G.~L., {Silva}, L., {Monaco}, P., {et~al.} 2001, \mnras, 324, 757

\bibitem[{{Greve} {et~al.}(2003){Greve}, {Ivison}, \&
  {Papadopoulos}}]{greve2003}
{Greve}, T.~R., {Ivison}, R.~J., \& {Papadopoulos}, P.~P. 2003, \apj, 599, 839

\bibitem[{{Greve} {et~al.}(2004){Greve}, {Ivison}, \&
  {Papadopoulos}}]{greve2004}
{Greve}, T.~R., {Ivison}, R.~J., \& {Papadopoulos}, P.~P. 2004, \aap, 419, 99

\bibitem[{{Greve} {et~al.}(2009){Greve}, {Papadopoulos}, {Gao}, \&
  {Radford}}]{greve2009}
{Greve}, T.~R., {Papadopoulos}, P.~P., {Gao}, Y., \& {Radford}, S.~J.~E. 2009,
  \apj, 692, 1432

\bibitem[{{Greve} {et~al.}(2005){Greve}, {Bertoldi}, {Smail}, {Neri},
  {Chapman}, {Blain}, {Ivison}, {Genzel}, {Omont}, {Cox}, {Tacconi}, \&
  {Kneib}}]{greve2005}
{Greve}, T.~R., {Bertoldi}, F., {Smail}, I., {et~al.} 2005, \mnras, 359, 1165

\bibitem[{{Guesten} {et~al.}(1993){Guesten}, {Serabyn}, {Kasemann},
  {Schinckel}, {Schneider}, {Schulz}, \& {Young}}]{guesten1993}
{Guesten}, R., {Serabyn}, E., {Kasemann}, C., {et~al.} 1993, \apj, 402, 537

\bibitem[{{Guilloteau} \& {Lucas}(2000)}]{guilloteau2000}
{Guilloteau}, S., \& {Lucas}, R. 2000, in Astronomical Society of the Pacific
  Conference Series, Vol. 217, Imaging at Radio through Submillimeter
  Wavelengths, ed. {J.~G.~Mangum \& S.~J.~E.~Radford}, 299

\bibitem[{{Guilloteau} {et~al.}(1992){Guilloteau}, {Delannoy}, {Downes},
  {Greve}, {Guelin}, {Lucas}, {Morris}, {Radford}, {Wink}, {Cernicharo},
  {Forveille}, {Garcia-Burillo}, {Neri}, {Blondel}, {Perrigourad}, {Plathner},
  \& {Torres}}]{guilloteau1992}
{Guilloteau}, S., {Delannoy}, J., {Downes}, D., {et~al.} 1992, \aap, 262, 624

\bibitem[{{Hainline} {et~al.}(2006){Hainline}, {Blain}, {Greve}, {Chapman},
  {Smail}, \& {Ivison}}]{hainline2006}
{Hainline}, L.~J., {Blain}, A.~W., {Greve}, T.~R., {et~al.} 2006, \apj, 650,
  614

\bibitem[{{Hainline} {et~al.}(2011){Hainline}, {Blain}, {Smail}, {Alexander},
  {Armus}, {Chapman}, \& {Ivison}}]{hainline2011}
{Hainline}, L.~J., {Blain}, A.~W., {Smail}, I., {et~al.} 2011, \apj, 740, 96

\bibitem[{{Harris} {et~al.}(2010){Harris}, {Baker}, {Zonak}, {Sharon},
  {Genzel}, {Rauch}, {Watts}, \& {Creager}}]{harris2010}
{Harris}, A.~I., {Baker}, A.~J., {Zonak}, S.~G., {et~al.} 2010, {\apj}, 723,
  1130

\bibitem[{{Harris} {et~al.}(2007){Harris}, {Baker}, {Jewell}, {Rauch}, {Zonak},
  {O'Neil}, {Shelton}, {Norrod}, {Ray}, \& {Watts}}]{harris2007}
{Harris}, A.~I., {Baker}, A.~J., {Jewell}, P.~R., {et~al.} 2007, in
  Astronomical Society of the Pacific Conference Series, Vol. 375, From
  Z-Machines to ALMA: (Sub)Millimeter Spectroscopy of Galaxies, ed.
  {A.~J.~Baker, J.~Glenn, A.~I.~Harris, J.~G.~Mangum, \& M.~S.~Yun }, 82

\bibitem[{{Harris} {et~al.}(2012){Harris}, {Baker}, {Frayer}, {Smail},
  {Swinbank}, {Riechers}, {van der Werf}, {Auld}, {Baes}, {Bussmann},
  {Buttiglione}, {Cava}, {Clements}, {Cooray}, {Dannerbauer}, {Dariush}, {De
  Zotti}, {Dunne}, {Dye}, {Eales}, {Fritz}, {Gonz{\'a}lez-Nuevo}, {Hopwood},
  {Ibar}, {Ivison}, {Jarvis}, {Maddox}, {Negrello}, {Rigby}, {Smith}, {Temi},
  \& {Wardlow}}]{harris2012}
{Harris}, A.~I., {Baker}, A.~J., {Frayer}, D.~T., {et~al.} 2012, \apj, 752, 152

\bibitem[{{Hayward} {et~al.}(2011){Hayward}, {Kere{\v s}}, {Jonsson},
  {Narayanan}, {Cox}, \& {Hernquist}}]{hayward2011}
{Hayward}, C.~C., {Kere{\v s}}, D., {Jonsson}, P., {et~al.} 2011, \apj, 743,
  159

\bibitem[{{Hayward} {et~al.}(2013){Hayward}, {Narayanan}, {Kere{\v s}},
  {Jonsson}, {Hopkins}, {Cox}, \& {Hernquist}}]{hayward2013}
{Hayward}, C.~C., {Narayanan}, D., {Kere{\v s}}, D., {et~al.} 2013, \mnras,
  428, 2529

\bibitem[{{Hodge} {et~al.}(2012){Hodge}, {Carilli}, {Walter}, {de Blok},
  {Riechers}, {Daddi}, \& {Lentati}}]{hodge2012}
{Hodge}, J.~A., {Carilli}, C.~L., {Walter}, F., {et~al.} 2012, \apj, 760, 11

\bibitem[{{Hodge} {et~al.}(2013){Hodge}, {Karim}, {Smail}, {Swinbank},
  {Walter}, {Biggs}, {Ivison}, {Weiss}, {Alexander}, {Bertoldi}, {Brandt},
  {Chapman}, {Coppin}, {Cox}, {Danielson}, {Dannerbauer}, {De Breuck},
  {Decarli}, {Edge}, {Greve}, {Knudsen}, {Menten}, {Rix}, {Schinnerer},
  {Simpson}, {Wardlow}, \& {van der Werf}}]{hodge2013}
{Hodge}, J.~A., {Karim}, A., {Smail}, I., {et~al.} 2013, \apj, 768, 91

\bibitem[{{Hughes} {et~al.}(1998){Hughes}, {Serjeant}, {Dunlop},
  {Rowan-Robinson}, {Blain}, {Mann}, {Ivison}, {Peacock}, {Efstathiou}, {Gear},
  {Oliver}, {Lawrence}, {Longair}, {Goldschmidt}, \& {Jenness}}]{hughes1998}
{Hughes}, D.~H., {Serjeant}, S., {Dunlop}, J., {et~al.} 1998, \nat, 394, 241

\bibitem[{{Iono} {et~al.}(2006){Iono}, {Tamura}, {Nakanishi}, {Kawabe},
  {Kohno}, {Okuda}, {Yamada}, {Hatsukade}, \& {Sameshima}}]{iono2006}
{Iono}, D., {Tamura}, Y., {Nakanishi}, K., {et~al.} 2006, \pasj, 58, 957

\bibitem[{{Iono} {et~al.}(2009){Iono}, {Wilson}, {Yun}, {Baker}, {Petitpas},
  {Peck}, {Krips}, {Cox}, {Matsushita}, {Mihos}, \& {Pihlstrom}}]{iono2009}
{Iono}, D., {Wilson}, C.~D., {Yun}, M.~S., {et~al.} 2009, \apj, 695, 1537

\bibitem[{{Ivison} {et~al.}(2011){Ivison}, {Papadopoulos}, {Smail}, {Greve},
  {Thomson}, {Xilouris}, \& {Chapman}}]{ivison2011}
{Ivison}, R.~J., {Papadopoulos}, P.~P., {Smail}, I., {et~al.} 2011, \mnras,
  412, 1913

\bibitem[{{Ivison} {et~al.}(2000){Ivison}, {Smail}, {Barger}, {Kneib}, {Blain},
  {Owen}, {Kerr}, \& {Cowie}}]{ivison2000}
{Ivison}, R.~J., {Smail}, I., {Barger}, A.~J., {et~al.} 2000, \mnras, 315, 209

\bibitem[{{Ivison} {et~al.}(2010){Ivison}, {Smail}, {Papadopoulos}, {Wold},
  {Richard}, {Swinbank}, {Kneib}, \& {Owen}}]{ivison2010a}
{Ivison}, R.~J., {Smail}, I., {Papadopoulos}, P.~P., {et~al.} 2010, \mnras,
  404, 198

\bibitem[{{Kennicutt}(1998)}]{kennicutt1998}
{Kennicutt}, Jr., R.~C. 1998, \apj, 498, 541

\bibitem[{{Klamer} {et~al.}(2005){Klamer}, {Ekers}, {Sadler}, {Weiss},
  {Hunstead}, \& {De Breuck}}]{klamer2005}
{Klamer}, I.~J., {Ekers}, R.~D., {Sadler}, E.~M., {et~al.} 2005, \apjl, 621, L1

\bibitem[{{Komatsu} {et~al.}(2011){Komatsu}, {Smith}, {Dunkley}, {Bennett},
  {Gold}, {Hinshaw}, {Jarosik}, {Larson}, {Nolta}, {Page}, {Spergel},
  {Halpern}, {Hill}, {Kogut}, {Limon}, {Meyer}, {Odegard}, {Tucker}, {Weiland},
  {Wollack}, \& {Wright}}]{komatsu2011}
{Komatsu}, E., {Smith}, K.~M., {Dunkley}, J., {et~al.} 2011, \apjs, 192, 18

\bibitem[{{Kormendy} \& {Sanders}(1992)}]{kormendy1992}
{Kormendy}, J., \& {Sanders}, D.~B. 1992, \apjl, 390, L53

\bibitem[{{Lindner} {et~al.}(2012){Lindner}, {Baker}, {Beelen}, {Owen}, \&
  {Polletta}}]{lindner2012}
{Lindner}, R.~R., {Baker}, A.~J., {Beelen}, A., {Owen}, F.~N., \& {Polletta},
  M. 2012, \apj, 757, 3

\bibitem[{{Lutz} {et~al.}(2005){Lutz}, {Valiante}, {Sturm}, {Genzel},
  {Tacconi}, {Lehnert}, {Sternberg}, \& {Baker}}]{lutz2005}
{Lutz}, D., {Valiante}, E., {Sturm}, E., {et~al.} 2005, \apjl, 625, L83

\bibitem[{{Magnelli} {et~al.}(2012){Magnelli}, {Lutz}, {Santini}, {Saintonge},
  {Berta}, {Albrecht}, {Altieri}, {Andreani}, {Aussel}, {Bertoldi},
  {B{\'e}thermin}, {Bongiovanni}, {Capak}, {Chapman}, {Cepa}, {Cimatti},
  {Cooray}, {Daddi}, {Danielson}, {Dannerbauer}, {Dunlop}, {Elbaz}, {Farrah},
  {F{\"o}rster Schreiber}, {Genzel}, {Hwang}, {Ibar}, {Ivison}, {Le Floc'h},
  {Magdis}, {Maiolino}, {Nordon}, {Oliver}, {P{\'e}rez Garc{\'{\i}}a},
  {Poglitsch}, {Popesso}, {Pozzi}, {Riguccini}, {Rodighiero}, {Rosario},
  {Roseboom}, {Salvato}, {Sanchez-Portal}, {Scott}, {Smail}, {Sturm},
  {Swinbank}, {Tacconi}, {Valtchanov}, {Wang}, \& {Wuyts}}]{magnelli2012}
{Magnelli}, B., {Lutz}, D., {Santini}, P., {et~al.} 2012, \aap, 539, A155

\bibitem[{{Mao} {et~al.}(2000){Mao}, {Henkel}, {Schulz}, {Zielinsky},
  {Mauersberger}, {St{\"o}rzer}, {Wilson}, \& {Gensheimer}}]{mao2000}
{Mao}, R.~Q., {Henkel}, C., {Schulz}, A., {et~al.} 2000, \aap, 358, 433

\bibitem[{{Mauersberger} {et~al.}(1999){Mauersberger}, {Henkel}, {Walsh}, \&
  {Schulz}}]{mauersberger1999}
{Mauersberger}, R., {Henkel}, C., {Walsh}, W., \& {Schulz}, A. 1999, \aap, 341,
  256

\bibitem[{{Men{\'e}ndez-Delmestre} {et~al.}(2009){Men{\'e}ndez-Delmestre},
  {Blain}, {Smail}, {Alexander}, {Chapman}, {Armus}, {Frayer}, {Ivison}, \&
  {Teplitz}}]{menendez2009}
{Men{\'e}ndez-Delmestre}, K., {Blain}, A.~W., {Smail}, I., {et~al.} 2009, \apj,
  699, 667

\bibitem[{{Narayanan} {et~al.}(2010){Narayanan}, {Hayward}, {Cox}, {Hernquist},
  {Jonsson}, {Younger}, \& {Groves}}]{narayanan2010}
{Narayanan}, D., {Hayward}, C.~C., {Cox}, T.~J., {et~al.} 2010, \mnras, 401,
  1613

\bibitem[{{Narayanan} {et~al.}(2012){Narayanan}, {Krumholz}, {Ostriker}, \&
  {Hernquist}}]{narayanan2012}
{Narayanan}, D., {Krumholz}, M.~R., {Ostriker}, E.~C., \& {Hernquist}, L. 2012,
  \mnras, 421, 3127

\bibitem[{{Neri} {et~al.}(2003){Neri}, {Genzel}, {Ivison}, {Bertoldi}, {Blain},
  {Chapman}, {Cox}, {Greve}, {Omont}, \& {Frayer}}]{neri2003}
{Neri}, R., {Genzel}, R., {Ivison}, R.~J., {et~al.} 2003, \apjl, 597, L113

\bibitem[{{Papadopoulos} {et~al.}(2001){Papadopoulos}, {Ivison}, {Carilli}, \&
  {Lewis}}]{papadopoulos2001}
{Papadopoulos}, P., {Ivison}, R., {Carilli}, C., \& {Lewis}, G. 2001, \nat,
  409, 58

\bibitem[{{Perley} \& {Butler}(2013)}]{perley2013}
{Perley}, R.~A., \& {Butler}, B.~J. 2013, \apjs, 204, 19

\bibitem[{{Pettini} {et~al.}(2002){Pettini}, {Rix}, {Steidel}, {Adelberger},
  {Hunt}, \& {Shapley}}]{pettini2002}
{Pettini}, M., {Rix}, S.~A., {Steidel}, C.~C., {et~al.} 2002, \apj, 569, 742

\bibitem[{{Pope} {et~al.}(2006){Pope}, {Scott}, {Dickinson}, {Chary},
  {Morrison}, {Borys}, {Sajina}, {Alexander}, {Daddi}, {Frayer}, {MacDonald},
  \& {Stern}}]{pope2006}
{Pope}, A., {Scott}, D., {Dickinson}, M., {et~al.} 2006, \mnras, 370, 1185

\bibitem[{{Pope} {et~al.}(2008){Pope}, {Chary}, {Alexander}, {Armus},
  {Dickinson}, {Elbaz}, {Frayer}, {Scott}, \& {Teplitz}}]{pope2008}
{Pope}, A., {Chary}, R.-R., {Alexander}, D.~M., {et~al.} 2008, \apj, 675, 1171

\bibitem[{{Puget} {et~al.}(1996){Puget}, {Abergel}, {Bernard}, {Boulanger},
  {Burton}, {Desert}, \& {Hartmann}}]{puget1996}
{Puget}, J., {Abergel}, A., {Bernard}, J., {et~al.} 1996, \aap, 308, L5

\bibitem[{{Riechers}(2013)}]{riechers2013a}
{Riechers}, D.~A. 2013, \apjl, 765, L31

\bibitem[{{Riechers} {et~al.}(2006){Riechers}, {Walter}, {Carilli}, {Knudsen},
  {Lo}, {Benford}, {Staguhn}, {Hunter}, {Bertoldi}, {Henkel}, {Menten},
  {Weiss}, {Yun}, \& {Scoville}}]{riechers2006}
{Riechers}, D.~A., {Walter}, F., {Carilli}, C.~L., {et~al.} 2006, \apj, 650,
  604

\bibitem[{{Riechers} {et~al.}(2011{\natexlab{a}}){Riechers}, {Carilli},
  {Maddalena}, {Hodge}, {Harris}, {Baker}, {Walter}, {Wagg}, {Vanden Bout},
  {Wei{\ss}}, \& {Sharon}}]{riechers2011f}
{Riechers}, D.~A., {Carilli}, C.~L., {Maddalena}, R.~J., {et~al.}
  2011{\natexlab{a}}, \apjl, 739, L32

\bibitem[{{Riechers} {et~al.}(2011{\natexlab{b}}){Riechers}, {Carilli},
  {Walter}, {Weiss}, {Wagg}, {Bertoldi}, {Downes}, {Henkel}, \&
  {Hodge}}]{riechers2011c}
{Riechers}, D.~A., {Carilli}, L.~C., {Walter}, F., {et~al.} 2011{\natexlab{b}},
  \apjl, 733, L11

\bibitem[{{Saintonge} {et~al.}(2013){Saintonge}, {Lutz}, {Genzel}, {Magnelli},
  {Nordon}, {Tacconi}, {Baker}, {Bandara}, {Berta}, {F{\"o}rster Schreiber},
  {Poglitsch}, {Sturm}, {Wuyts}, \& {Wuyts}}]{saintonge2013}
{Saintonge}, A., {Lutz}, D., {Genzel}, R., {et~al.} 2013, \apj, 778, 2

\bibitem[{{Sanders} \& {Mirabel}(1996)}]{sanders1996}
{Sanders}, D.~B., \& {Mirabel}, I.~F. 1996, \araa, 34, 749

\bibitem[{{Schmidt}(1959)}]{schmidt1959}
{Schmidt}, M. 1959, \apj, 129, 243

\bibitem[{{Scoville} {et~al.}(2000){Scoville}, {Evans}, {Thompson}, {Rieke},
  {Hines}, {Low}, {Dinshaw}, {Surace}, \& {Armus}}]{scoville2000}
{Scoville}, N.~Z., {Evans}, A.~S., {Thompson}, R., {et~al.} 2000, \aj, 119, 991

\bibitem[{{Sharon} {et~al.}(2013){Sharon}, {Baker}, {Harris}, \&
  {Thomson}}]{sharon2013}
{Sharon}, C.~E., {Baker}, A.~J., {Harris}, A.~I., \& {Thomson}, A.~P. 2013,
  \apj, 765, 6

\bibitem[{{Smail} {et~al.}(1997){Smail}, {Ivison}, \& {Blain}}]{smail1997}
{Smail}, I., {Ivison}, R.~J., \& {Blain}, A.~W. 1997, \apjl, 490, L5

\bibitem[{{Smail} {et~al.}(1998){Smail}, {Ivison}, {Blain}, \&
  {Kneib}}]{smail1998}
{Smail}, I., {Ivison}, R.~J., {Blain}, A.~W., \& {Kneib}, J. 1998, \apjl, 507,
  L21

\bibitem[{{Smail} {et~al.}(2002){Smail}, {Ivison}, {Blain}, \&
  {Kneib}}]{smail2002}
{Smail}, I., {Ivison}, R.~J., {Blain}, A.~W., \& {Kneib}, J. 2002, \mnras, 331, 495

\bibitem[{{Smail} {et~al.}(1999){Smail}, {Ivison}, {Kneib}, {Cowie}, {Blain},
  {Barger}, {Owen}, \& {Morrison}}]{smail1999}
{Smail}, I., {Ivison}, R.~J., {Kneib}, J.-P., {et~al.} 1999, \mnras, 308, 1061

\bibitem[{{Solomon} \& {Vanden Bout}(2005)}]{solomon2005}
{Solomon}, P.~M., \& {Vanden Bout}, P.~A. 2005, \araa, 43, 677

\bibitem[{{Somerville} {et~al.}(2008){Somerville}, {Hopkins}, {Cox},
  {Robertson}, \& {Hernquist}}]{somerville2008}
{Somerville}, R.~S., {Hopkins}, P.~F., {Cox}, T.~J., {Robertson}, B.~E., \&
  {Hernquist}, L. 2008, \mnras, 391, 481

\bibitem[{{Swinbank} {et~al.}(2008){Swinbank}, {Lacey}, {Smail}, {Baugh},
  {Frenk}, {Blain}, {Chapman}, {Coppin}, {Ivison}, {Gonzalez}, \&
  {Hainline}}]{swinbank2008}
{Swinbank}, A.~M., {Lacey}, C.~G., {Smail}, I., {et~al.} 2008, \mnras, 391, 420

\bibitem[{{Swinbank} {et~al.}(2010){Swinbank}, {Smail}, {Chapman}, {Borys},
  {Alexander}, {Blain}, {Conselice}, {Hainline}, \& {Ivison}}]{swinbank2010b}
{Swinbank}, A.~M., {Smail}, I., {Chapman}, S.~C., {et~al.} 2010, \mnras, 405,
  234

\bibitem[{{Swinbank} {et~al.}(2011){Swinbank}, {Papadopoulos}, {Cox}, {Krips},
  {Ivison}, {Smail}, {Thomson}, {Neri}, {Richard}, \& {Ebeling}}]{swinbank2011}
{Swinbank}, A.~M., {Papadopoulos}, P.~P., {Cox}, P., {et~al.} 2011, \apj, 742,
  11

\bibitem[{{Tacconi} {et~al.}(2006){Tacconi}, {Neri}, {Chapman}, {Genzel},
  {Smail}, {Ivison}, {Bertoldi}, {Blain}, {Cox}, {Greve}, \&
  {Omont}}]{tacconi2006}
{Tacconi}, L.~J., {Neri}, R., {Chapman}, S.~C., {et~al.} 2006, \apj, 640, 228

\bibitem[{{Tacconi} {et~al.}(2008){Tacconi}, {Genzel}, {Smail}, {Neri},
  {Chapman}, {Ivison}, {Blain}, {Cox}, {Omont}, {Bertoldi}, {Greve},
  {F{\"o}rster Schreiber}, {Genel}, {Lutz}, {Swinbank}, {Shapley}, {Erb},
  {Cimatti}, {Daddi}, \& {Baker}}]{tacconi2008}
{Tacconi}, L.~J., {Genzel}, R., {Smail}, I., {et~al.} 2008, \apj, 680, 246

\bibitem[{{Tacconi} {et~al.}(2013){Tacconi}, {Neri}, {Genzel}, {Combes},
  {Bolatto}, {Cooper}, {Wuyts}, {Bournaud}, {Burkert}, {Comerford}, {Cox},
  {Davis}, {F{\"o}rster Schreiber}, {Garc{\'{\i}}a-Burillo}, {Gracia-Carpio},
  {Lutz}, {Naab}, {Newman}, {Omont}, {Saintonge}, {Shapiro Griffin}, {Shapley},
  {Sternberg}, \& {Weiner}}]{tacconi2013}
{Tacconi}, L.~J., {Neri}, R., {Genzel}, R., {et~al.} 2013, \apj, 768, 74

\bibitem[{{Takahashi}(2001)}]{takahashi2001}
{Takahashi}, J. 2001, \apj, 561, 254

\bibitem[{{Tecza} {et~al.}(2004){Tecza}, {Baker}, {Davies}, {Genzel},
  {Lehnert}, {Eisenhauer}, {Lutz}, {Nesvadba}, {Seitz}, {Tacconi}, {Thatte},
  {Abuter}, \& {Bender}}]{tecza2004}
{Tecza}, M., {Baker}, A.~J., {Davies}, R.~I., {et~al.} 2004, \apjl, 605, L109

\bibitem[{{The Astronomical Almanac}(2011)}]{astroalmanac}
{The Astronomical Almanac} 2011, {U.S. Naval Observatory and Rutherford
  Appleton Laboratory}, {(Washington, DC: U.S. Govt. Printing Office)}

\bibitem[{{Thomas} {et~al.}(2005){Thomas}, {Maraston}, {Bender}, \& {Mendes de
  Oliveira}}]{thomas2005}
{Thomas}, D., {Maraston}, C., {Bender}, R., \& {Mendes de Oliveira}, C. 2005,
  \apj, 621, 673

\bibitem[{{Thomson} {et~al.}(2012){Thomson}, {Ivison}, {Smail}, {Swinbank},
  {Weiss}, {Kneib}, {Papadopoulos}, {Baker}, {Sharon}, \& {van
  Moorsel}}]{thomson2012}
{Thomson}, A.~P., {Ivison}, R.~J., {Smail}, I., {et~al.} 2012, \mnras, 425,
  2203

\bibitem[{{Valiante} {et~al.}(2007){Valiante}, {Lutz}, {Sturm}, {Genzel},
  {Tacconi}, {Lehnert}, \& {Baker}}]{valiante2007}
{Valiante}, E., {Lutz}, D., {Sturm}, E., {et~al.} 2007, \apj, 660, 1060

\bibitem[{{Vieira} {et~al.}(2013){Vieira}, {Marrone}, {Chapman}, {De Breuck},
  {Hezaveh}, {Weiss}, {Aguirre}, {Aird}, {Aravena}, {Ashby}, {Bayliss},
  {Benson}, {Biggs}, {Bleem}, {Bock}, {Bothwell}, {Bradford}, {Brodwin},
  {Carlstrom}, {Chang}, {Crawford}, {Crites}, {de Haan}, {Dobbs}, {Fomalont},
  {Fassnacht}, {George}, {Gladders}, {Gonzalez}, {Greve}, {Gullberg},
  {Halverson}, {High}, {Holder}, {Holzapfel}, {Hoover}, {Hrubes}, {Hunter},
  {Keisler}, {Lee}, {Leitch}, {Lueker}, {Luong-Van}, {Malkan}, {McIntyre},
  {McMahon}, {Mehl}, {Menten}, {Meyer}, {Mocanu}, {Murphy}, {Natoli}, {Padin},
  {Plagge}, {Reichardt}, {Rest}, {Ruel}, {Ruhl}, {Sharon}, {Schaffer}, {Shaw},
  {Shirokoff}, {Spilker}, {Stalder}, {Staniszewski1}, {Stark}, {Story},
  {Vanderlinde}, {Welikala}, \& {Williamson}}]{vieira2013}
{Vieira}, J.~D., {Marrone}, D.~P., {Chapman}, S.~C., {et~al.} 2013, Nature,
  495, 344

\bibitem[{{Walter} {et~al.}(2011){Walter}, {Wei{\ss}}, {Downes}, {Decarli}, \&
  {Henkel}}]{walter2011}
{Walter}, F., {Wei{\ss}}, A., {Downes}, D., {Decarli}, R., \& {Henkel}, C.
  2011, \apj, 730, 18

\bibitem[{{Wang} {et~al.}(2011){Wang}, {Cowie}, {Barger}, \&
  {Williams}}]{wang2011}
{Wang}, W.-H., {Cowie}, L.~L., {Barger}, A.~J., \& {Williams}, J.~P. 2011,
  \apjl, 726, L18

\bibitem[{{Ward}(2002)}]{ward2002}
{Ward}, J.~S. 2002, PhD thesis, California Institute of Technology

\bibitem[{{Ward} {et~al.}(2003){Ward}, {Zmuidzinas}, {Harris}, \&
  {Isaak}}]{ward2003}
{Ward}, J.~S., {Zmuidzinas}, J., {Harris}, A.~I., \& {Isaak}, K.~G. 2003, \apj,
  587, 171

\bibitem[{{Wei{\ss}} {et~al.}(2005{\natexlab{a}}){Wei{\ss}}, {Downes},
  {Walter}, \& {Henkel}}]{weiss2005c}
{Wei{\ss}}, A., {Downes}, D., {Walter}, F., \& {Henkel}, C. 2005{\natexlab{a}},
  \aap, 440, L45

\bibitem[{{Wei{\ss}} {et~al.}(2007){Wei{\ss}}, {Downes}, {Walter}, \&
  {Henkel}}]{weiss2007}
{Wei{\ss}}, A., {Downes}, D., {Walter}, F., \& {Henkel}, C. 2007, in
  Astronomical Society of the Pacific Conference Series, Vol. 375, From
  Z-Machines to ALMA: (Sub)Millimeter Spectroscopy of Galaxies, ed.
  {A.~J.~Baker, J.~Glenn, A.~I.~Harris, J.~G.~Mangum, \& M.~S.~Yun }, 25

\bibitem[{{Wei{\ss}} {et~al.}(2003){Wei{\ss}}, {Henkel}, {Downes}, \&
  {Walter}}]{weiss2003}
{Wei{\ss}}, A., {Henkel}, C., {Downes}, D., \& {Walter}, F. 2003, \aap, 409,
  L41

\bibitem[{{Wei{\ss}} {et~al.}(2005{\natexlab{b}}){Wei{\ss}}, {Walter}, \&
  {Scoville}}]{weiss2005b}
{Wei{\ss}}, A., {Walter}, F., \& {Scoville}, N.~Z. 2005{\natexlab{b}}, \aap,
  438, 533

\bibitem[{{Wei{\ss}} {et~al.}(2009){Wei{\ss}}, {Kov{\'a}cs}, {Coppin}, {Greve},
  {Walter}, {Smail}, {Dunlop}, {Knudsen}, {Alexander}, {Bertoldi}, {Brandt},
  {Chapman}, {Cox}, {Dannerbauer}, {De Breuck}, {Gawiser}, {Ivison}, {Lutz},
  {Menten}, {Koekemoer}, {Kreysa}, {Kurczynski}, {Rix}, {Schinnerer}, \& {van
  der Werf}}]{weiss2009}
{Wei{\ss}}, A., {Kov{\'a}cs}, A., {Coppin}, K., {et~al.} 2009, \apj, 707, 1201

\bibitem[{{Wei{\ss}} {et~al.}(2013){Wei{\ss}}, {De Breuck}, {Marrone},
  {Vieira}, {Aguirre}, {Aird}, {Aravena}, {Ashby}, {Bayliss}, {Benson},
  {B{\'e}thermin}, {Biggs}, {Bleem}, {Bock}, {Bothwell}, {Bradford}, {Brodwin},
  {Carlstrom}, {Chang}, {Chapman}, {Crawford}, {Crites}, {de Haan}, {Dobbs},
  {Downes}, {Fassnacht}, {George}, {Gladders}, {Gonzalez}, {Greve},
  {Halverson}, {Hezaveh}, {High}, {Holder}, {Holzapfel}, {Hoover}, {Hrubes},
  {Husband}, {Keisler}, {Lee}, {Leitch}, {Lueker}, {Luong-Van}, {Malkan},
  {McIntyre}, {McMahon}, {Mehl}, {Menten}, {Meyer}, {Murphy}, {Padin},
  {Plagge}, {Reichardt}, {Rest}, {Rosenman}, {Ruel}, {Ruhl}, {Schaffer},
  {Shirokoff}, {Spilker}, {Stalder}, {Staniszewski}, {Stark}, {Story},
  {Vanderlinde}, {Welikala}, \& {Williamson}}]{weiss2013}
{Wei{\ss}}, A., {De Breuck}, C., {Marrone}, D.~P., {et~al.} 2013, \apj, 767, 88

\bibitem[{{Wild} {et~al.}(1992){Wild}, {Harris}, {Eckart}, {Genzel}, {Graf},
  {Jackson}, {Russell}, \& {Stutzki}}]{wild1992}
{Wild}, W., {Harris}, A.~I., {Eckart}, A., {et~al.} 1992, \aap, 265, 447

\bibitem[{{Yang} {et~al.}(2010){Yang}, {Stancil}, {Balakrishnan}, \&
  {Forrey}}]{yang2010}
{Yang}, B., {Stancil}, P.~C., {Balakrishnan}, N., \& {Forrey}, R.~C. 2010,
  \apj, 718, 1062

\bibitem[{{Yao} {et~al.}(2003){Yao}, {Seaquist}, {Kuno}, \& {Dunne}}]{yao2003}
{Yao}, L., {Seaquist}, E.~R., {Kuno}, N., \& {Dunne}, L. 2003, \apj, 588, 771

\bibitem[{{Young} \& {Scoville}(1991)}]{young1991}
{Young}, J.~S., \& {Scoville}, N.~Z. 1991, \araa, 29, 581

\bibitem[{{Zemcov} {et~al.}(2010){Zemcov}, {Blain}, {Halpern}, \&
  {Levenson}}]{zemcov2010}
{Zemcov}, M., {Blain}, A., {Halpern}, M., \& {Levenson}, L. 2010, \apj, 721,
  424

\bibitem[{{Zitrin}(2013)}]{zitrin2013}
{Zitrin}, A. 2013, Private communication

\bibitem[{{Zitrin} {et~al.}(2009){Zitrin}, {Broadhurst}, {Umetsu}, {Coe},
  {Ben{\'{\i}}tez}, {Ascaso}, {Bradley}, {Ford}, {Jee}, {Medezinski},
  {Rephaeli}, \& {Zheng}}]{zitrin2009}
{Zitrin}, A., {Broadhurst}, T., {Umetsu}, K., {et~al.} 2009, \mnras, 396, 1985

\end{thebibliography}
\end{document}